\documentclass[a4paper,11pt]{article}
\usepackage{jheppub}

\usepackage{graphicx}
\usepackage{dcolumn}
\usepackage{bm}
\usepackage{bbm}
\usepackage[dvipsnames]{xcolor}
\usepackage{microtype}
\usepackage{IEEEtrantools}
\usepackage{amsthm}
\usepackage{enumitem}
\usepackage[normalem]{ulem}
\usepackage{hyperref}
\hypersetup{breaklinks=true}
\usepackage[T1]{fontenc}
\usepackage[scr = rsfs, cal = dutchcal]{mathalfa} 
\usepackage[capitalise]{cleveref}

\definecolor{darkgreen}{rgb}{0,0.5,0}

\renewcommand{\P}{\mathcal{P}}
\renewcommand{\O}{\mathcal{O}}

\newcommand{\e}{\varepsilon}

\newcommand{\scri}[0]{$\mathscr{I}^+$}
\newcommand{\order}[1]{\mathcal{O}(#1)}

\font\ec=ecrm0800 at 12pt

\def\edthp{\hbox{\ec\char'360}'}

\def\mb{{\bar{m}}}

\def\g{g}
\newcommand{\MB}[1][]{{M_{\mathcal{B}}^{#1}}}
\newcommand{\MBdot}[1][]{{\dot{M}_{\mathcal{B}}^{#1}}}

\newcommand{\Lie}{{\mathcal L}}

\newcommand{\bsu}{{u}}
\newcommand{\bsr}{{\hat r}}
\newcommand{\bst}{{\hat \theta}}
\newcommand{\bsp}{{\hat \phi}}

\newcommand{\KNu}{{u_{\rm KN}}}
\newcommand{\KNp}{{\phi_*}}

\newcommand{\ro}{{\{1\}}}
\newcommand{\rr}{{\{2\}}}

\newcommand{\bssub}[1]{{\vphantom{\theta}\smash{#1}}}

\font\ec=ecrm0800 at 12pt

\def\edthp{\hbox{\ec\char'360}'}

\def\mb{{\bar{m}}}

\def\m{m}
\def\l{l}

\def\circl{{\circ}}

\makeatletter
\def\@fpheader{\relax}
\makeatother

\begin{document}

\title{The Bondi--Sachs gauge, BMS frames, and memory in black hole perturbation theory}

\author[a,b,c]{Andrew Spiers\footnote{\href{mailto:andrew.spiers@ucd.ie}{Email: andrew.spiers@ucd.ie}}}
\author[c]{Adam Pound}
\author[d]{Jordan Moxon}
\affiliation[a]{
School of Mathematics \& Statistics, University College Dublin, Belfield, Dublin 4, Ireland, D04 V1W8
}
\affiliation[b]{%
 School of Mathematical Sciences \& School of Physics and Astronomy,
University of Nottingham, University Park, Nottingham, NG7 2RD, UK
}%
\affiliation[c]{%
 School of Mathematical Sciences and STAG Research Centre, University of Southampton, Southampton, SO17 1BJ, United Kingdom
}
 \affiliation[d]{TAPIR, California Institute of Technology, Pasadena, CA 91125, USA
}%

\date{\today}

\abstract{
As LISA and other next-generation detectors demand increasingly accurate waveform models, there is a growing need for these models to precisely control gauge freedoms that had previously been inconsequential. One such intrinsic freedom is the choice of the asymptotic Bondi--Metzner--Sachs (BMS) frame. The need to control the BMS frame is particularly pronounced in black hole perturbation theory, where there has been little work to this end---most glaringly in gravitational self-force calculations, which are in an unknown frame and encounter infrared, far-zone gauge singularities at second perturbative order.
Here we present a framework for iteratively transforming to the Bondi--Sachs gauge and fixing the BMS frame on a Kerr background. This includes an extension of the Bondi--Sachs formalism to the multiscale expansions that underpin most self-force-based waveforms, introducing soft hair and a concept of ``forgetful gauges'' in the process. Our framework evades infrared divergences and naturally incorporates memory effects that had previously only ever been added ``after the fact'' in self-force waveforms,  including the recently discovered ``memory distortion''. Our formalism could also be used for ringdown analysis, and we expect it to be vital for comparisons with numerical relativity and post-Newtonian theory.
}

\maketitle

\section{\label{sec:level1}Introduction}

Gravitational-wave astronomy is an advancing field whose expanding experimental reach continues to uncover new subtleties in our theoretical understanding of gravitational-wave sources. With the improved sensitivity of LIGO/Virgo/KAGRA detectors~\cite{abbott2023gwtc} and the planned next-generation detectors (such as LISA~\cite{colpi2024lisa, arun2022new}, Cosmic Explorer~\cite{reitze2019cosmic}, the Einstein Telescope~\cite{ET:2025xjr}, TianQin~\cite{luo2016tianqin}, and Taiji~\cite{ruan2020taiji}), the number of gravitational-wave sources and their signal-to-noise ratios will increase by multiple orders of magnitude over the next two decades~\cite{iacovelli2022forecasting}. Extracting these gravitational-wave signals from detector data requires matched filtering~\cite{owen1999matched, abbott2016characterization} with accurate waveform templates. To avoid biases in parameter measurements, substantial improvements in modelling techniques will be required~\cite{Purrer:2019jcp,bailes2021gravitational,Hu:2022rjq,LISAConsortiumWaveformWorkingGroup:2023arg,Dhani:2024jja, Chandramouli:2024vhw, ET:2025xjr,Mahapatra:2026wsp}. 

Increasing the accuracy of waveform models, and therefore the accuracy of parameter estimation, also requires a sharper understanding of what model parameters represent. One obvious example is orbital eccentricity: a parameter of great astrophysical interest, but one which is gauge dependent and can vary wildly in meaning between one model and another; overcoming this ambiguity has led to a definition of eccentricity that only uses properties of the waveform itself~\cite{Shaikh:2023ypz,Shaikh:2025tae}. But even more broadly, the relationship between a waveform and a system's parameters depends on which asymptotic frame is used to describe future null infinity (\scri), where the waveform is computed~\cite{penrose1963assyptotic,waldbook}. The frame can be altered by Poincaré transformations---translations, rotations, and boosts---and substantial effort has gone into ensuring waveform models describe the waveform in the centre-of-mass-frame, for example (see Refs.~\cite{OShaughnessy:2011pmr,Ochsner:2012dj, Schmidt:2012rh,Boyle:2013nka,Woodford:2019tlo,Sun:2025quw} for a sample of this work). Moreover, an asymptotic frame can actually be transformed in an infinite number of ways, corresponding to the 10 Poincaré transformations together with an infinite-dimensional group of \emph{super}translations. This is the Bondi-Metzner-Sachs (BMS) group, the symmetry group of \scri~\cite{bondi1962gravitational,sachs1962BMS-fav,flanagan2017BMS,compere2020poincare}. Recent work has highlighted the importance of specifying the full BMS frame, not only the Poincaré frame, when comparing between waveform models describing the same physical system~\cite{Boyle:2015nqa,mitman2021fixing,magana2022high,mitman2024review,Khairnar:2026mtm}.

One area of gravitational wave modelling advancement that is increasingly susceptible to these ambiguities is black hole perturbation theory (BHPT)~\cite{chandrabook, Pound-Wardell:2021, pani2013advanced}. Interest in extending calculations from first to second perturbative order has recently intensified, both in calculating quadratic quasi-normal modes~\cite{Nakano:2007cj,Ioka:2007ak,okuzumi2008possible,Loutrel-etal:2020,Ripley-etal:2020,Lagos:2022otp,ma2024excitation,Redondo-Yuste:2023seq,Bourg:2024jme,Bucciotti:2024zyp,Khera:2024yrk,Khera:2024bjs,Kehagias:2024sgh,Zhu:2024rej,Fransen:2025cgv,Bourg:2025lpd} and in second-order self-force theory~\cite{rosenthal2006secondorder,Detweiler:2011tt,pound2012second,gralla2012,pound2012field,pound2017,Miller-Pound:2020,pound2020second,Warburton:2021kwk,Wardell:2021fyy,Spiers:2023mor,spiers2023second,Miller:2023ers,Cunningham:2024dog,Upton:2025bja,Mathews:2025txc}. These steps to second order are essential for next-generation detectors, as part of the black hole spectroscopy programme in the quasi-normal mode case~\cite{Berti:2025hly} and modelling extreme-mass-ratio inspirals in the self-force case~\cite{hindereretal2008,LISAConsortiumWaveformWorkingGroup:2023arg,Burke:2023lno}. Such second-order calculations introduce new challenges not present at first order, many of them related to gauge choices. This paper addresses three such challenges: providing a framework to fix the asymptotic gauge (including the BMS frame) in second-order calculations~\cite{bruni1997perturbations,pound2015gaugeandmotion,l-c}; avoiding infrared divergences that arise from asymptotic behaviour, particularly in second-order self-force calculations~\cite{PoundLargeScales,Cunningham:2024dog}; and naturally incorporating gravitational-wave memory content, which is closely related to the choice of supertranslation frame~\cite{zel1974radiation,braginskii1985kinematic,favata2010gravitational,christodoulou1991nonlinear,gleiser2003gravitational,Cunningham:2024dog}.


In BHPT, gauge dependence grows more complex as one extends from first to second order. General relativity admits four degrees of gauge freedom~\cite{waldbook}. In BHPT, four additional gauge degrees of freedom manifest themselves at each perturbative order, and quantities that are invariant at lower orders typically become gauge dependent at higher orders. A key example is the perturbations to the Weyl scalars $\psi_0$ and $\psi_4$~\cite{newman1962behavior}, the latter of which contains the emitted gravitational-wave content of the spacetime. These quantities are gauge invariant at first order but are gauge dependent at second order~\cite{l-c,spiers2023second}. Similarly, the first-order waveform is insensitive to first-order choices of Poincaré frame, but the second-order waveform is sensitive to them: if the waveform itself is ${\cal O}(\varepsilon)$, where $\varepsilon$ is the small parameter, then a small, order-$\varepsilon$ translation will induce an order-$\varepsilon^2$ change in the waveform.  

To analyse (and ultimately fix) this gauge freedom, we employ the Bondi--Sachs (BS) formalism~\cite{bondi1962gravitational,sachs1962BMS,madler2016bondisachs,flanagan2017BMS,compere2020poincare}, which was first formulated in the 1960s to analyse gravitational radiation. This framework uses coordinates adapted to outgoing null cones, and the residual freedom in these BS coordinates is precisely the BMS symmetry group, corresponding to deformations that leave \scri invariant. In this paper, we define the BS perturbative gauge on Kerr spacetime, and we find the gauge transformation from any initial gauge to the BS gauge. To fully fix the gauge, we also prescribe a scheme for fixing the BMS frame. Specifically, we formulate this gauge-fixing scheme at first perturbative order. This suffices for our purposes because the second-order perturbation of $\psi_4$ is invariant under second-order transformations, meaning it becomes gauge invariant once the first-order gauge is fixed. Hence, our first-order gauge fixing will allow gauge-invariant comparisons of second-order waveforms in a consistent way.

All of this is particularly relevant in self-force calculations for asymmetric binaries, in which one object (the secondary) orbits a much larger black hole (the primary). In this case, the perturbative parameter $\varepsilon$ is the small mass ratio~\cite{barackpound18,Pound-Wardell:2021}. Although there are now advanced waveform models based on self-force theory both at leading order in $\varepsilon$~\cite{Chapman-Bird:2025xtd} and at first subleading order~\cite{Honet:2025lmk}, very little work has been done to understand which asymptotic frame they are in, except in very limited cases~\cite{Detweiler:2003ci,Barack:2019agd} relying on Newtonian intuition and never analysing the full specification of the BMS frame. For systems involving spin-orbit precession, this lack of clarity already complicates comparison with numerical relativity (NR)~\cite{Mathews:2025txc}. Since comparisons with NR and post-Newtonian (PN) theory are often probing minuscule differences, it is also sometimes unclear whether disagreements merely represent a different choice of frame~\cite{Warburton:2021kwk,Warburton:2024xnr}. Our framework should resolve this uncertainty.


A further challenge in second-order BHPT, which is particularly pronounced in self-force calculations, is the appearance of infrared divergences. These arise from second-order source terms that are quadratic combinations of first-order quantities. Such sources can exhibit slow falloff, or even blowup, toward \scri. They can be particularly badly behaved in some of the most commonly used gauges, such as the Regge-Wheeler-Zerilli gauge~\cite{brizuela2009complete,Bucciotti:2024zyp,okuzumi2008possible}. But even in seemingly well-behaved gauges such as the Lorenz gauge, this source generically falls off sufficiently slowly to cause the integral of it against a retarded Green's function to diverge~\cite{PoundLargeScales,Cunningham:2024dog}. In most cases, one can change to a better-behaved field variable by subtracting a suitable counter-term. However, in the self-force case, the divergence stems from a breakdown between near-zone behaviour (described by a multiscale expansion in which all spacetime evolution is expressed in terms of orbital dynamics~\cite{hindereretal2008,Miller-Pound:2020,Pound-Wardell:2021,Mathews:2025nyb,Lewis:2025ydo}) and far-zone behaviour (where gravitational self-interactions can no longer be described in terms of instantaneous functions of the two-body system). The necessary counter-terms in this case involve integrals over the entire past history, which could not have been discerned from the form of the field equation alone~\cite{PoundLargeScales,Cunningham:2024dog}. 

To overcome these infrared divergences and correctly account for far-zone behaviour, one of us introduced a matched-expansions approach, in which the near-zone (multiscale) solution is matched to a far-zone ``self-force multipolar post-Minkowskian'' (SF-MPM) expansion~\cite{Cunningham:2024dog}. This approach, modelled on the traditional MPM expansion used in PN theory~\cite{blanchet1986radiative, Blanchet:1987wq, blanchet1992hereditary, blanchet2014gravitational}, was successful. Yet it has been shown~\cite{ma2024excitation,Khera:2024yrk,Loutrel-etal:2020,Ripley-etal:2020} that infrared divergences at second order can be avoided far more simply by working in an appropriate gauge at first order. In Ref.~\cite{spiers2023second}, we showed that this is the case even for the more pernicious divergences in self-force theory: if one adopts the BS gauge at first order, then the field equation for the second-order perturbation of $\psi_4$ is well behaved. In this paper, we expand on that conclusion, showing how the BS gauge effectively eliminates the distinction between near and far zones. We then develop a practical iterative procedure for transforming to the BS gauge, evading infrared divergences and the need for an SF-MPM expansion.

As discussed in Ref.~\cite{Cunningham:2024dog}, the infinite-time integrals that are associated with the near-versus-far divide are closely tied to gravitational-wave memory: a nonlinear self-interaction between gravitational waves that causes a permanent displacement of test masses~\cite{christodoulou1991nonlinear}. This link to memory also links them to the BMS symmetry group and to the soft graviton theorem~\cite{he2015bms,strominger2016gravitational}. Memory effects are expected to be detectable by next-generation detectors~\cite{grant2023outlook,inchauspe2025measuring,gasparotto2023can,ghosh2023detection}, and they are of exceptional theoretical interest due to their interpretation as ``soft hair'' on black hole spacetimes~\cite{Hawking:2016msc} and their connection to the celestial holography programme~\cite{Raclariu:2021zjz}. Our framework in this paper allows us to show that while memory is ``forgotten'' in the traditional near-zone gauges of first-order self-force theory, it naturally enters the metric through a slowly evolving supertranslation when transforming to the BS gauge; this effectively installs ``soft hair'' onto the two-body system. We also show that this then naturally leads to the ``memory distortion'' effect found in Ref.~\cite{Cunningham:2024dog}, through which the slowly evolving memory distorts the emitted gravitational waves. 

The paper is organized as follows. Section~\ref{sec:Asy-gauges} reviews gauge freedom in BHPT up to second order, our definition of asymptotic flatness, the BS gauge, and the BMS group. Section~\ref{sec:second-order} reviews second-order perturbation theory, how to construct a second-order gauge invariant, and infrared divergences. Section~\ref{sec:KerrBS} describes the Kerr metric in BS form, following Ref.~\cite{bai2007KerrBS}'s description. Section~\ref{sec:BSgaugevector} presents our formalism for transforming to the perturbative BS gauge from any asymptotically flat initial gauge. Section~\ref{sec:ConstrInfBMS} gives our scheme for fixing the BMS frame in regular perturbation theory. Section~\ref{sec:BS gauge in SF theory} extends our analysis from regular perturbation theory to the multiscale expansions used in self-force theory. Here, we introduce the concept of a forgetful gauge; slowly evolving BMS transformations; our iterative approach to transforming to the BS gauge; and how memory and memory distortion arise naturally in this framework. Section~\ref{sec:CompareNR} compares our BMS fixing scheme to those used by the SXS collaboration to fix the BMS frame in NR calculations~\cite{mitman2024review,Khairnar:2026mtm}. Finally, in \cref{sec:conclusions} we present our conclusions and remark on how our formalism could be used in future BHPT calculations. Some important technical details are relegated to the appendices.

We adopt geometric coordinates with $G=c=1$. Greek letters $\alpha,\beta,\gamma,\ldots$ are used for 4-dimensional spacetime indices, lowercase Latin letters $i,j,k,\ldots$ for 3-dimensional indices, and uppercase Latin letters $A,B,C,\ldots$ for 2-dimensional indices. Spacetime indices are lowered and raised with the background metric $g^{(0)}_{\mu\nu}$ and its inverse $g^{\mu\nu}_{(0)}$; 3-dimensional indices, with the Euclidean metric $\delta_{ij}$ and its inverse $\delta^{ij}$; and 2-dimensional indices of tensors on a sphere, with the unit-2-sphere metric $\Omega_{AB}$ and its inverse $\Omega^{AB}$. We adopt Mathematica notation in using square brackets rather than round parentheses to display the arguments of functions: $f[x]$ rather than $f(x)$. Angular brackets around indices denote symmetric trace-free combinations, as in $T_{\langle AB\rangle}:=T_{(AB)} - \frac{1}{2}\Omega_{AB}\Omega^{CD}T_{CD}$.


\section{Asymptotics and gauge}~\label{sec:Asy-gauges}

The four gauge degrees of freedom in General Relativity correspond to diffeomorphisms on the spacetime manifold. A perturbative expansion results in gauge freedoms manifesting themselves at each perturbative order, representing the freedom to choose how one maps points in the perturbed spacetime to points in the background spacetime~\cite{geroch1969limits, stewartbook,bruni1997perturbations,pound2015gaugeandmotion}. Alternatively, one can consider the freedoms as corresponding to the choice of coordinate system (in the so-called \textit{passive view} \cite{pound2015gaugeandmotion}), with additional small coordinate transformations allowed at each perturbative order. Adopting a coordinate system on the background spacetime represents a zeroth-order choice of gauge. Perturbative gauge transformations then alter how tensor perturbations are expressed in the given background coordinates. 

Concretely, a change in choice of perturbative gauge corresponds to the following gauge transformations of metric perturbations~\cite{pound2015gaugeandmotion, bruni1997perturbations}:
\begin{align}\label{eq:1gaguetransform}
h^{(1)}_{\mu\nu} &\rightarrow h^{(1)}_{\mu\nu}+\mathcal{L}_{\vec{\xi}_{(1)}} g^{(0)}_{\mu\nu}, \\
h^{(2)}_{\mu\nu} &\rightarrow h^{(2)}_{\mu\nu}+\mathcal{L}_{\vec{\xi}_{(2)}} g^{(0)}_{\mu\nu} +\mathcal{L}_{\vec{\xi}_{(1)}} h^{(1)}_{\mu\nu} + \frac{1}{2}\mathcal{L}_{\vec{\xi}_{(1)}}\mathcal{L}_{\vec{\xi}_{(1)}} g^{(0)}_{\mu\nu},\label{eq:2gaguetransform} \\
h^{(3)}_{\mu\nu} &\rightarrow ... ,
\end{align}
\noindent where $\vec{\xi}_{(1)}$ and $\vec{\xi}_{(2)}$ are the first- and second-order gauge vectors, respectively. Precisely analogous transformations apply to perturbations of any other tensor field. 

\subsection{Asymptotically flat gauges and far-zone expansions}
\label{sec:AssymFlat}


Gauge freedom in general relativity allows for solutions to the EFE which exhibit divergences. A simple example is the divergence at the Schwarzschild event horizon in Schwarzschild coordinates~\cite{schwarzschild1916gravity}. Such divergences are spurious, due to a poor choice of coordinates. Similarly, it is common to encounter divergences in BHPT caused by pathologies in the perturbative gauge, particularly in the most commonly used gauges: the Regge--Wheeler and radiation gauges~\cite{gleiser1996second,Keidl:2010pm,pound2014gravitational,Thompson:2018lgb,barackpound18}. Such pathologies can cause point-particle singularities to extend away from the particle's location~\cite{pound2014gravitational,Thompson:2018lgb,barackpound18}, for example, or cause the metric to appear singular at the black hole horizon or toward asymptotic infinity~\cite{gleiser1996second,Keidl:2010pm}. In this paper, we are specifically interested in eliminating gauge divergences that can appear towards future null infinity, $\mathscr{I}^+$, the boundary of spacetime which outgoing null geodesics approach. 

One way to examine a spacetime's asymptotic behaviour is to conformally compactify it. Following a conformal transformation $\bar g_{\mu\nu}=\Omega^2 g_{\mu\nu}$, $\mathscr{I}^+$ is attached as a boundary of the conformally transformed spacetime. Asymptotic flatness then corresponds to appropriate conditions on the conformal factor and on the transformed metric~\cite{penrose1963assyptotic,waldbook, madler2016bondisachs}.
However, following the standard BS approach, we will find it more convenient to impose conditions directly on metric components in a retarded polar coordinate system $(u,r,\theta^A)$ where asymptotic flatness is manifest. 

We say a metric is asymptotically flat at $\mathscr{I}^+$ 
if there exist coordinates $(u,r,\theta^A)$ in which the metric appropriately asymptotes to the Minkowski metric when $r\rightarrow \infty$ at fixed $(u,\theta^A)$. Concretely, 
given the Minkowski metric
\begin{align}\label{eq:AF-metric}
g^\mathrm{M}_{\mu\nu}= \left( \begin{array}{cccc}
 -1  & -1 & 0& 0\\
 -1 &  0  &0 &0  \\
0 &  0 &  r^2&  0 \\
0 &  0 &  0 &r^2 \sin\theta  \end{array} \right) ,
\end{align}
an asymptotically flat metric must satisfy
\begin{align}\label{eq:AF-metric-2}
\g_{\mu\nu}-g^\mathrm{M}_{\mu\nu}= \left( \begin{array}{cccc}
 \mathcal{O}(r^{-1})  & \mathcal{O}(r^{-1}) & \mathcal{O}(r^0)& \mathcal{O}(r^0)\\
 \mathcal{O}(r^{-1})  &  \mathcal{O}(r^{-1})  &  \mathcal{O}(r^{0}) &  \mathcal{O}(r^{0})  \\
\mathcal{O}(r^{0})  &  \mathcal{O}(r^{0})  &  \mathcal{O}(r^{1}) &  \mathcal{O}(r^{1})  \\
\mathcal{O}(r^{0})  &  \mathcal{O}(r^{0})  &  \mathcal{O}(r^{1}) &  \mathcal{O}(r^{1})  \end{array} \right)  ,
\end{align}
and an $n$th radial derivative $\partial^n_r\g_{\mu\nu}$ must decay at least $n$ orders more rapidly with $r$ than $\g_{\mu\nu}$, while derivatives with respect to $u$ and $\theta^A$ do not increase the rate of decay with $r$. 

Correspondingly, we say a metric perturbation $h^{(n)}_{\mu\nu}$ is asymptotically flat if there exists a perturbative gauge in which the perturbation's components in $(u,r,\theta^A)$ coordinates satisfy
\begin{align}\label{eq:AF-metric-pert}
h^{(n)}_{\mu\nu}= \left( \begin{array}{cccc}
 \mathcal{O}(r^{-1})  & \mathcal{O}(r^{-1}) & \mathcal{O}(r^0)& \mathcal{O}(r^0)\\
 \mathcal{O}(r^{-1})  &  \mathcal{O}(r^{-1})  &  \mathcal{O}(r^{0}) &  \mathcal{O}(r^{0})  \\
\mathcal{O}(r^{0})  &  \mathcal{O}(r^{0})  &  \mathcal{O}(r^{1}) &  \mathcal{O}(r^{1})  \\
\mathcal{O}(r^{0})  &  \mathcal{O}(r^{0})  &  \mathcal{O}(r^{1}) &  \mathcal{O}(r^{1})  \end{array} \right)  ,
\end{align}
with the same conditions as above on derivatives. We refer to a perturbative gauge in which this is satisfied as an asymptotically flat gauge.

To denote orders in $1/r$ relative to the Minkowski metric, we introduce the following order symbol:
\begin{align}\label{eq:below-Minkowski}
\mathcal{O}(r^{\mathrm{M}-n})= \left( \begin{array}{cccc}
 \mathcal{O}(r^{-n})  & \mathcal{O}(r^{-n}) & \mathcal{O}(r^{1-n})& \mathcal{O}(r^{1-n})\\
 \mathcal{O}(r^{-n})  &  \mathcal{O}(r^{-n})  &  \mathcal{O}(r^{1-n}) &  \mathcal{O}(r^{1-n})  \\
\mathcal{O}(r^{1-n})  &  \mathcal{O}(r^{1-n})  &  \mathcal{O}(r^{2-n}) &  \mathcal{O}(r^{2-n})  \\
\mathcal{O}(r^{1-n})  &  \mathcal{O}(r^{1-n})  &  \mathcal{O}(r^{2-n}) &  \mathcal{O}(r^{2-n})  \end{array} \right)  .
\end{align}
We indicate coefficients of $r^{{\rm M}-n}$ using curly brackets, as in $g^{\{n\}}_{\mu\nu}$.

\subsection{\label{sec:BSIntro}The Bondi-Sachs formalism}

A gauge which is manifestly asymptotically flat is that of the BS formalism, which was formulated to analyse gravitational radiation at $\mathscr{I}^+$~\cite{madler2016bondisachs} with respect to an inertial observer. Bondi-Sachs coordinates, $\hat x^{\gamma}=(u,\bsr,\hat{\theta}^A)$,  foliate spacetime with outgoing null hypersurfaces of constant $u$. Additionally, the angular coordinates are chosen such that $\hat{\theta}^A$ labels the generators of these hypersurfaces (i.e., each curve of constant $u$ and constant $\hat\theta^A$ is an outgoing null geodesic). The radial coordinate, a parameter along each of these null geodesics, is chosen such that the surface element of the 2-surface of constant $u$ and $\hat{r}$ is the same as the surface element on a geometrical sphere of radius $\hat{r}$~\cite{bishop2006kerrBS,flanagan2017BMS}. 

The four BS gauge conditions, which enforce these geometrical features, can be expressed as  $g^{uu}=0$, $g^{uA}=0$, and $  \det[\gamma^{AB}]=\det[\Omega^{AB}]$, where $\Omega_{AB}$ is the unit 2-sphere metric and $\gamma_{AB}$ is the angular metric, per unit $\hat{r}^2$, on surfaces of constant $u$ and $\hat{r}$. The equivalent gauge conditions in index-down form are
\begin{align}
    g_{\hat{r}\hat{r}}=g_{\hat{r}A}&=0, \notag \\
    \det[\gamma_{AB}]&=\det[\Omega_{AB}].\label{eq:BSconditions}
\end{align}
Applying these gauge conditions, along with asymptotic flatness as in \cref{eq:AF-metric-2}, one finds the metric in the BS gauge takes the form
\begin{align}  \label{eq:BSMetric}
\g_{\mu\nu} d\hat x^\mu d\hat x^\nu =& - \frac{V}{\hat{r}} e^{2\beta} du^2 -2 e^{2\beta} du d\hat{r} + \hat{r}^2 \gamma_{AB}(d\hat{\theta}^A -U^A du)(d\hat{\theta}^B -U^B du).
\end{align}
Here $V$, $\beta$, $U^A$, and $\gamma_{AB}$ are functions of $(u,\hat{r},\hat{\theta}^A)$ representing the metric's six physical degrees of freedom (accounting for $\gamma_{AB}=\gamma_{BA}$ and the gauge condition on ${\rm det}[\gamma_{AB}$]). In Sec.~\ref{sec:BMSIntro} we will illustrate how the BS gauge conditions constrain all four gauge degrees of freedom up to the BMS transformations~\cite{sachs1962BMS-fav,bondi1962gravitational}.

For physical solutions, one must also enforce the EFE with a stress-energy tensor with an appropriate fall off. Assuming the spacetime is vacuum outside some finite region, the EFE constrains the metric functions' large-$\hat{r}$ expansions to be~\cite{flanagan2017BMS}
\begin{equation}\label{eq:BSfalloffs}
\begin{split}
V &= \hat{r} - 2\MB -\frac{2Z}{\hat{r}} + \mathcal{O}(\hat{r}^{-2}), \\ 
\beta &= -\frac{1}{32}\frac{C_{AB}C^{AB}}{\bsr^2} +\mathcal{O}(\hat{r}^{-3}),  \\
\gamma_{AB}&= \Omega_{AB}+\frac{C_{AB}}{\hat{r}} + \frac{1}{4}\frac{\Omega_{AB}C_{CD}C^{CD}}{\hat{r}^2} + \mathcal{O}(\hat{r}^{-3}),  \\
   \ \  U^A&= \frac{-D^BC^A_{\ \ B}}{2\hat{r}^2} + \frac{1}{\hat{r}^3}\bigg[ -\frac{2}{3}N^A + \frac{1}{16}D^A(C_{BC}C^{BC})   +\frac{1}{2}C^{AB}D^CC_{BC} \bigg] + \mathcal{O}(\hat{r}^{-4}), 
\end{split}
\end{equation}
where $D_A$ is the 2-dimensional covariant derivative corresponding to $\Omega_{AB}$, and we recall we use $\Omega^{AB}$ to raise uppercase Latin indices. $\MB$ is the Bondi mass aspect, $Z$ is a function of $[u,\hat{\theta}^A]$, $N^A$ is the angular momentum aspect, and $C_{AB}$ is the Bondi shear, which is traceless ($\Omega^{AB}C_{AB}=0$) and is related to the Bondi news tensor by $N_{AB}:=\partial_u C_{AB}$. The Bondi shear represents the gravitational waveform, and the news tensor contains the information about the gravitational wave fluxes emitted to $\mathscr{I}^+$. 

It will be convenient to decompose vectors and tensors on the 2-sphere into their even- and odd-parity pieces. For example, 
\begin{equation}
N_A=N^{\rm even}_A+N^{\rm odd}_A,    
\end{equation}
where $N^{\rm even}_A=D_A \rho$ and $N^{\rm odd}_A=-\epsilon_A{}^B D_B \zeta$  for some scalars $\rho$ and $\zeta$. Here $\epsilon_{AB}$ is Levi–Civita tensor of the unit 2-sphere. Similarly, the shear can be decomposed as~\cite{flanagan2017BMS}
\begin{equation}\label{C=Ce+Co}
    C_{AB} = C^{\rm even}_{AB}+C^{\rm odd}_{AB},
\end{equation}
with $C^{\rm even}_{AB} = \left(D_AD_B-\frac{1}{2}\Omega_{AB}D^CD_C\right)\Phi_{\rm even}$ and $C^{\rm odd}_{AB} = \epsilon_{C(A}D_{B)}D^C\Phi_{\rm odd}$ for some scalars $\Phi_{\rm even/odd}$.

From the EFE, it can be shown~\cite{flanagan2017BMS} that, in stationary regions of spacetime, the function $Z$ relates to the Bondi shear and to the even-parity part of the angular momentum aspect,
\begin{align}\label{eq:Z}
    Z=-\frac16\Big( D_AN^A  +\frac{3}{16}C_{AB}C^{AB} +\frac{3}{4}D_A C^{AB}D^CC_{CB}  \Big).
\end{align}
We note that $D_AN^A_{\rm odd}=0$, picking out the even-parity piece of $N^A$. The EFE also implies evolution equations for the Bondi mass and angular momentum aspects~\cite{flanagan2017BMS}:
\begin{align}
    \frac{d\MB}{du} &= -\frac{1}{8}N_{AB}N^{AB} + \frac{1}{4}D_A D_B N^{AB},\label{eq:MBdot}\\
    \frac{dN_A}{du} &= D_A \MB + \frac{1}{4}D_BD_AD_C C^{BC} - \frac{1}{4}D_BD^BD^CC_{CA}\notag\\
    &\quad +\frac{1}{4}D_B(N^{BC}C_{CA}) + \frac{1}{2}D_BN^{BC}C_{CA}.\label{eq:Ndot}
\end{align}
The linear terms on the right-hand sides of these equations are referred to as ``soft'' terms, as they relate to the soft theorems in quantum field theory~\cite{compere2020poincare}; we will recall how they relate to gravitational-wave memory in later sections. Taking the divergence of the last equation yields
\begin{equation}
    \frac{d}{du}D^AN^{\rm even}_A = D^AD_A \MB +\frac{1}{4}D^AD_B(N^{BC}C_{CA}) + \frac{1}{2}D^A(D_BN^{BC}C_{CA})\label{eq:Ndoteven}.
\end{equation}
These are effectively energy and momentum balance laws, and the quantity $\frac{1}{8}N_{AB}N^{AB}$ represents the angular distribution of instantaneous gravitational-wave energy flux to $\mathscr{I}^+$.

\subsection{\label{sec:BMSIntro}The BMS symmetry group}

The BS gauge conditions do not fully constrain the gauge freedom, coordinate system, or form of the metric (these are equivalent statements). The residual freedom, corresponding to transformations of $\mathscr{I}^+$ onto itself, are the diffeomorphisms of the BMS symmetry group~\cite{bondi1962gravitational,sachs1962BMS, sachs1962BMS-fav}. The BMS group contains not only the Poincaré symmetry group but also an infinite-dimensional subgroup: the \emph{supertranslations} (and the superrotations when singular coordinates are considered~\cite{banks2003critique,barnich2010symmetries,barnich2011supertranslations,flanagan2017BMS}, which we neglect in this paper). Here, we give a brief overview of the BMS symmetries; for more detailed, recent reviews, see Refs.~\cite{flanagan2017BMS,madler2016bondisachs,compere2019advancedGR-BMS}.

Explicitly, the generator of the BMS transformations is~\cite{sachs1962BMS-fav}\footnote{Note this vector is tangent to $\mathscr{I}^+$ and acts only on the intrinsic metric of $\mathscr{I}^+$; it would not preserve the form~\eqref{eq:BSfalloffs} of the full metric. The extension of the transformation off $\mathscr{I}^+$, which does preserve the form~\eqref{eq:BSfalloffs}, is given in Eq.~\eqref{eq:BMSvecExt}.}
\begin{align}\label{eq:BMSVec}
\vec{\xi}=f[u,\bst^A]\partial_u +Y^A[\hat\theta^B]\partial_A, 
\end{align}
where 
\begin{equation}\label{eq:f def}
    f[\bsu,\bst^A]:=\alpha[\hat\theta^A]+\frac{1}{2}u D_A Y^A[\hat\theta^A]
\end{equation}
and $Y^A[\hat\theta^A]$ satisfies
\begin{align}\label{eq:Y Killing}
2D_{(A} Y_{B)}-D_{C}Y^C \Omega_{AB}=0.
\end{align}
The last equation, which can equivalently be written as $D_{\langle A} Y_{B\rangle}=0$, has the general solution
\begin{align}\label{eq:evenoddY}
Y^A=D^A\chi -\epsilon^{AB} D_B \kappa,
\end{align}
where $\chi$ and $\kappa$ are any linear combinations of $l=1$ spherical harmonics~\cite{flanagan2017BMS}. The $\chi$ and $\kappa$ parts of $Y^A$ thus correspond to linear combinations of $l=1$ even- and odd-parity vector harmonics~\cite{martellpoisson2005SVH}, respectively. 

The six degrees of freedom in $Y^A$ represent the Lorentz transformations: the Lorentz boosts are associated with $\chi$, and the rotations are associated with $\kappa$. 
The form of $\alpha[\hat\theta^A]$ in Eq.~\eqref{eq:f def} is then what distinguishes the asymptotic BMS symmetry from the global Poincaré symmetry of Minkowski spacetime. In retarded polar coordinates, the generator of a Poincaré transformation is 
\begin{equation}\label{eq:Poincare}
f\partial_u + \Big(Y^A-\frac{1}{\hat{r}}D^Af\Big) \partial_A - \frac{1}{2}\Big(\hat{r} D_A Y^A - D^2f\Big)\partial_{\hat{r}},
\end{equation}
where $\alpha[\hat\theta^A]$ is a linear combination of $l \in \{0,1\}$ spherical harmonics, corresponding to translations in time and space, respectively. In Eq.~\eqref{eq:f def}, $\alpha[\hat\theta^A]$ can instead be any smooth function on the 2-sphere~\cite{sachs1962BMS-fav}, meaning it can be a linear combination of spherical harmonics for all $\ell\geq0$. The transformations generated by the $\ell\geq2$ piece of $\alpha$ represent  the infinite set of supertranslations.







The asymptotic BMS symmetries are associated with BMS charges via Noether's theorem~\cite{noether1918invariante,dray1984angular,wald2000general}. The charges corresponding to the Poincaré group are the mass (associated with the $\ell=0$ term in $\alpha$), linear momentum (associated with the $\ell=1$ terms in $\alpha$), angular momentum (associated with $\kappa$), and boost charges (associated with $\chi$). Written as Cartesian vectors, these are given by the following integrals over cuts of $\mathscr{I}^+$~\cite{compere2020poincare}:
\begin{align}
    M[u] &:= \frac{1}{4\pi} \int_{S^2} \MB  d\Omega,\label{eq:Pt}\\
    P_i[u] &:= \frac{1}{4\pi} \int_{S^2} \MB n_i d\Omega, \label{eq:PT}\\
   J_i[u] &:= -\frac{1}{8\pi} \int_{S^2} \hat{N}_A\epsilon^{AB} D_B n_i d\Omega,\label{eq:JT} \\
    K_i[u] &:= \frac{1}{8\pi} \int_{S^2}\hat{N}_A  D^A n_i d\Omega.\label{eq:KT}
\end{align}
where $\hat{N}_A$ is the Wald--Zoupas~\cite{wald2000general} angular momentum aspect,\footnote{Note the final two terms in Eq.~(3.2) of Ref.~\cite{compere2020poincare} identically cancel, reducing to our expression, in agreement with Eq.~(3.11) of Ref.~\cite{flanagan2017BMS}. See Ref.~\cite{compere2020poincare} for discussion of other definitions of angular momentum aspect.}
\begin{align}\label{eq:Nhat}
    \hat{N}_A&=N_A - uD_A\MB -\frac{1}{4}C_{AB}D_C C^{BC}-\frac{1}{16}D_A(C_{BC}C^{BC}).
\end{align}
Here $n_i=(\sin\bst\cos\bsp,\sin\bst\sin\bsp, \cos\bst)$ is the Cartesian outward unit normal from the sphere, and the four types of charges $M$, $P_i$, $J_i$, and $K_i$ are generated by Eq.~\eqref{eq:BMSVec} with $\alpha=1$, $\alpha=n_i$, $\kappa=n_i$, and $\chi=n_i$, respectively. From a physical viewpoint, it is often useful to also consider the centre-of-mass charge, 
\begin{align}
    G_i[u]=(K_i+uP_i).\label{eq:centre-of-mass-charge}
\end{align}
$K_i$ and $P_i$ are both constant in nonradiative regions of $\mathscr{I}^+$ (i.e., regions with vanishing news), and $G_i$ then behaves like a Newtonian center of mass, with $K_i$ corresponding to its initial position and $P_i$ corresponding to its initial velocity.

The supertranslations also have associated charges, the supermomentum. There are multiple definitions of supermomentum. For example, \cref{eq:PT} can be generalized by replacing $n_i$ with any $\l\geq2$ spherical harmonic~\cite{mitman2024review, compere2020poincare}; up to a minus sign, this definition is known as the Geroch supermomentum~\cite{Geroch1977,dray1984angular},
\begin{align}
    \P_G[u,\hat\theta^A]
    &:= -\MB .
\end{align}
Geroch suppermomentum is supertranslation-invariant in nonradiative regions of the spacetime, meaning it cannot be used to fix supertranslation freedom. A form of supermomentum that does not have this adverse property is the Moreschi supermomentum~\cite{mitman2022fixing, dray1984angular, moreschi1988supercentre, moreschi1998rest},
\begin{align}
    \P_M[u,\hat\theta^A]:=-\MB +\frac{1}{4}D_AD_BC^{AB}.
    \label{eq:MoreschiSupermomentum} 
\end{align}
We can see another desirable property of this supermomentum from Eq.~\eqref{eq:MBdot}: its rate of change is given by the gravitational energy flux,
\begin{equation}\label{PMdot}
    \frac{d\P_M}{du} = \frac{1}{8}N_{AB}N^{AB}.
\end{equation}

There have also been generalisations of the rotation and boost charges to super-Lorentz charges~\cite{campiglia2014asymptotic,campiglia2015new,compere2020poincare,barnich2011supertranslations,flanagan2017BMS}.
We omit these superrotations and superboosts as they introduce singularities into the asymptotic metric.

All of the BMS charges satisfy flux-balance laws derivable from Eqs.~\eqref{eq:MBdot} and~\eqref{eq:Ndot}~\cite{compere2020poincare}. We will make use of these when needed.


\section{Second-order perturbation theory}\label{sec:second-order}

Before specializing to the BS formalism, in this section we review several generic aspects of second-order BHPT. We particularly highlight the second-order Teukolsky formalism, the construction of a gauge-fixed second-order Teukolsky variable, and the infrared divergences that arise in extant second-order self-force calculations. 

\subsection{Field equations}

The basic equations of BHPT are the perturbative EFEs. These are obtained by expanding the Einstein tensor of the total metric $\g_{\alpha\beta} = g^{(0)}_{\alpha\beta}+\varepsilon h^{(1)}_{\alpha\beta} + \varepsilon^2 h^{(2)}_{\alpha\beta} + \order{\varepsilon^3}$ as 
\begin{equation}
   G_{\alpha\beta}[\g] = G_{\alpha\beta}[g^{(0)}] + \e G^{(1)}_{\alpha\beta}[h^{(1)}] 
   + \e^2\! \left(G^{(1)}_{\alpha\beta}[h^{(2)}] 
   + G^{(2)}_{\alpha\beta}[h^{(1)},h^{(1)}]\right)  + {\cal O}(\e^3),\label{G expansion}
\end{equation}
where $G^{(1)}_{\alpha\beta}[h^{(n)}]$ is the linearized Einstein tensor constructed from $h^{(n)}_{\mu\nu}$, $G^{(2)}_{\alpha\beta}$ is quadratic (bilinear in its arguments), and so on to higher order. Given an expansion of the stress-energy tensor $T_{\alpha\beta}$ in powers of~$\varepsilon$, the EFE then reduces to a sequence of linear equations,
\begin{align}
    G^{(1)}_{\alpha\beta}[h^{(1)}] &= 8\pi T^{(1)}_{\alpha\beta},\label{EFE1}\\
    G^{(1)}_{\alpha\beta}[h^{(2)}] &= 8\pi T^{(2)}_{\alpha\beta}- G^{(2)}_{\alpha\beta}[h^{(1)},h^{(1)}],\label{EFE2}
\end{align}
and so on. Here we suppress tensor indices on functional arguments for visual clarity.

From any linear EFE, $G^{(1)}_{\mu\nu}[h]=S_{\mu\nu}$, we can construct an associated Teukolsky equation, ${\cal O}[\psi_4]={\cal S}[S_{\mu\nu}]$, where $\psi_4$ is a linear perturbation of the spin $-2$ Weyl scalar and ${\cal O}$ and ${\cal S}$ are second-order linear differential operators~\cite{Pound-Wardell:2021}.\footnote{Here we follow the notation of Ref.~\cite{spiers2023second}. The operators would instead be denoted ${\cal O}'$ and ${\cal S}_4$ in Ref.~\cite{Pound-Wardell:2021}.} This equation can be written as the operator identity $\mathcal{O}\mathcal{T}[h]=\mathcal{S}\mathcal{E}[h]$, valid for any $h_{\mu\nu}$, where the linear operator $\mathcal{T}$ acts on a metric perturbation to return a linearized Weyl scalar, and $\mathcal{E}$ acts on a metric perturbation to return the corresponding $G^{(1)}_{\mu\nu}$~\cite{wald1978}. Applying this identity to the first- and second-order EFEs above, one obtains~\cite{green2020teukolsky,Spiers:GR22,spiers2023second}  
\begin{align}
   {\cal O}[\psi^{(1)}_{4}] &= {\cal S}\!\bigl[8\pi T^{(1)}_{\mu\nu}\bigr],\label{Teukolsky1}\\
\mathcal{O}[\psi_{4L}^{(2)}]&=\mathcal{S}\big[8\pi T^{(2)}_{\mu\nu} - G^{(2)}_{\mu\nu}[h^{(1)},h^{(1)}]\big],\label{eq:pure-2nd-Teuk}
\end{align}
where $\psi^{(1)}_4:={\cal T}[h^{(1)}_{\mu\nu}]$ and $\psi^{(2)}_{4L}:={\cal T}[h^{(2)}_{\mu\nu}]$.

The field variable in Eq.~\eqref{eq:pure-2nd-Teuk}, $\psi_{4L}^{(2)}$, corresponds to only a piece of the second-order perturbation of the spin $-2$ Weyl scalar. If the Weyl scalar is expanded in powers of $\varepsilon$, as in $\psi_4=\varepsilon\psi^{(1)}_4 + \varepsilon^2\psi^{(2)}_4 + \order{\varepsilon^3}$, then the second-order term is given by~\cite{l-c}
\begin{align}\label{eq:psi42deconstructed}
\psi_4^{(2)}&=\psi_{4L}^{(2)}+\psi_{4Q}^{(2)},
\end{align}
where $\psi_{4L}^{(2)}$ is the field defined above, linear in $h^{(2)}_{\mu\nu}$, and  $\psi_{4Q}^{(2)}$ is a quadratic function of $h^{(1)}_{\mu\nu}$ and of the perturbations $(l^\mu_{(1)},n^\mu_{(1)},m^\mu_{(1)},\bar m^\mu_{(1)})$ to the Newman--Penrose tetrad. Because of this, in Ref.~\cite{spiers2023second} we referred to Eq.~\eqref{eq:pure-2nd-Teuk} as the \emph{reduced} second-order Teukolsky equation.

In Ref.~\cite{spiers2023second} we also detailed the relative merits of working with $\psi_{4L}^{(2)}$ or $\psi_4^{(2)}$. Importantly, since $\psi_{4Q}^{(2)}$ decays like $1/\bsr^2$ at large $\bsr$, it cannot contribute to the gravitational waveform. This implies the waveform can be extracted from either variable:\footnote{We refer to Appendix~C of Ref.~\cite{Bourg:2025lpd} for elaboration of some subtleties related to this point.}
\begin{equation}
    \lim_{r\to\infty}\Bigl(\bsr\psi^{(2)}_{4L}\Bigr) = \lim_{\bsr\to\infty}\Bigl(\bsr\psi^{(2)}_{4}\Bigr) = - \frac{1}{2}\bar m^A\bar m^B\partial^2_u C^{(2)}_{AB}.
\end{equation}
where $C^{(2)}_{AB}$ is the second-order shear tensor and $\bar m^A = \frac{1}{\sqrt{2}}(1,-i\csc\theta)$ is the usual Newman--Penrose complex tetrad leg restricted to the sphere at $\mathscr{I}^+$ and scaled by $\bsr$.


 \subsection{\label{sec:gauge-fixing} Gauge-invariant second-order variable via gauge fixing}

In first-order BHPT calculations, $\psi^{(1)}_4$ is very often preferred  over $h^{(1)}_{\alpha\beta}$ for several reasons: $\psi^{(1)}_4$ is a single (complex) scalar rather than 10 coupled components; the Teukolsky equation is fully separable in a Kerr background (while the linearized EFE is not known to be); and $\psi^{(1)}_4$ is gauge invariant. The last point follows immediately from the fact that the background Weyl scalar vanishes: analogously to Eq.~\eqref{eq:1gaguetransform}, we have 
\begin{equation}
\psi^{(1)}_4\to \psi^{(1)}_4 + \Delta\psi^{(1)}_4    
\end{equation}
with 
\begin{equation}
\Delta\psi^{(1)}_4 = {\cal L}_{\vec\xi_{(1)}}\psi^{(0)}_4=0.    
\end{equation}
Consequently, we can also write
\begin{equation}\label{T[Lie g]=0}
{\cal T}\bigl[{\cal L}_{\vec\xi_{(1)}}g^{(0)}_{\mu\nu}\bigr] = 0    
\end{equation}
for any vector $\xi^\mu_{(1)}$, since $\Delta \psi^{(1)}_4 = {\cal T}[\Delta h^{(1)}_{\mu\nu}] = {\cal T}\bigl[{\cal L}_{\vec\xi_{(1)}}g^{(0)}_{\mu\nu}\bigr]$.

However, at second order, both $\psi_{4}^{(2)}$ and  $\psi_{4L}^{(2)}$ are gauge dependent. For example, analogously to Eq.~\eqref{eq:2gaguetransform}, 
\begin{align}
\Delta\psi_4^{(2)} &=\mathcal{L}_{\vec\xi_{(2)}}\psi_4^{(0)} + \frac{1}{2}\mathcal{L}_{\vec\xi_{(1)}}\mathcal{L}_{\vec\xi_{(1)}}\psi_4^{(0)} + \mathcal{L}_{\vec\xi_{(1)}}\psi_4^{(1)}\notag\\
&=\mathcal{L}_{\vec\xi_{(1)}}\psi_4^{(1)}.\label{eq:Delta psi42}
\end{align}
Similarly,
\begin{align}
\Delta\psi_{4L}^{(2)}&=\mathcal{T}[\Delta h^{(2)}_{\mu\nu}] \notag\\
&=\mathcal{T}\!\left[\mathcal{L}_{\vec\xi_{(1)}} h^{(1)}_{\mu\nu} + \tfrac{1}{2}\mathcal{L}_{\vec\xi_{(1)}}\mathcal{L}_{\vec\xi_{(1)}} g_{\mu\nu}\right],\label{eq:phi42GT}
\end{align}
where the ${\cal L}_{\vec\xi_{(2)}}g^{(0)}_{\mu\nu}$ term dropped out by virtue of Eq.~\eqref{T[Lie g]=0}.

Significantly, no second-order gauge vector appears in \cref{eq:Delta psi42} or \cref{eq:phi42GT}. Hence, $\psi_{4}^{(2)}$ and $\psi_{4L}^{(2)}$ are both invariant under a purely second-order gauge transformation; they only depend on the choice of gauge for $h^{(1)}_{\mu\nu}$. ($\psi_{4}^{(2)}$ additionally depends on the choice of tetrad perturbations, but $\psi_{4L}^{(2)}$ does not have that adverse property.) 
Therefore, we can fix the gauge of the second-order Weyl scalar variables by fixing the gauge of $h^{(1)}_{\mu\nu}$. This first-order gauge fixing requires a prescription for calculating the gauge vector, $\xi^{\mu}_{(1)}$, from an arbitrary initial gauge, with a given metric perturbation $h_{\mu\nu}^{(1)}$, to a specific, fixed, final gauge; we denote the metric perturbation in this fixed gauge as $\hat h_{\mu\nu}^{(1)}$.

In practice, this gauge fixing only manifests itself through the source term for the second-order Teukolsky equations; for example, with the gauge of $h^{(1)}_{\mu\nu}$ fixed,
\begin{align}
 \mathcal{O}[\hat\psi^{(2)}_{4L}] &= \mathcal{S}\Big[8\pi\hat T^{(2)}_{\mu\nu} - G^{(2)}_{\mu\nu}[\hat h_{\alpha\beta},\hat h_{\alpha\beta}]\Big].\label{eq:Gauge-Fixed-pure-2nd-Teuk}
\end{align}
Here $\hat h^{(1)}_{\mu\nu}:=h^{(1)}_{\mu\nu}+{\cal L}_{\vec\xi_{(1)}}g^{(0)}_{\mu\nu}$ and $\hat T^{(2)}_{\mu\nu}:=T^{(2)}_{\mu\nu}+{\cal L}_{\vec\xi_{(1)}}T^{(1)}_{\mu\nu}$ are effectively functions of $h^{(1)}_{\mu\nu}$ in the initial gauge because the gauge vector will itself be expressed in terms of $h^{(1)}_{\mu\nu}$, meaning  $\xi^{\gamma}_{(1)}=\xi^{\gamma}_{(1)}\Big[h_{\mu\nu}^{(1)}\Big]$. If the target gauge is fully specified, with no residual freedom, then $\xi^{\mu}_{(1)}$ will be uniquely determined by $h^{(1)}_{\mu\nu}$, with no free parameters associated with residual freedom. One can then consider $\hat h^{(1)}_{\mu\nu}$ as a gauge-invariant function of $h^{(1)}_{\mu\nu}$, and $\hat\psi^{(2)}_{4L}$ as a gauge-invariant variable. Appendix~A of Ref.~\cite{Spiers:2023mor}, for example, expounds on this point. Alternatively, one can simply consider this procedure as a way of working in the desired, fixed target gauge. 

In Secs.~\ref{sec:BSgaugevector} and~\ref{sec:ConstrInfBMS}, we will give such a gauge fixing formalism to the perturbative BS gauge, in a fixed BMS frame, near \scri from any asymptotically flat initial gauge.

\subsection{Infrared divergences in second-order perturbation theory}\label{sec:goodgaugesEMRIs}

In Ref.~\cite{spiers2023second}, we discussed the asymptotic behaviour of the reduced second-order Teukolsky equation, \cref{eq:pure-2nd-Teuk}, near \scri. Our conclusion was that in generic gauges, infrared divergences appear in the solution to  \cref{eq:pure-2nd-Teuk} due to the slow falloff of the source term ${\cal S}[G^{(2)}_{\mu\nu}]$. However, we also showed that these divergences can be evaded by fixing to a BS gauge. Here, for clarity, we summarize parts of that discussion as well as discussions of the EFE in Refs.~\cite{PoundLargeScales,Cunningham:2024dog}. 

As a first example, consider the second-order EFE~\eqref{EFE2} and impose the Lorenz gauge condition: 
\begin{equation}
\nabla^\alpha\bar h^{(2)}_{\alpha\beta}=0,
\end{equation}
with
\begin{align}
    \bar h^{(2)}_{\alpha\beta}:=h^{(n)}_{\alpha\beta}-\frac{1}{2}g_{\alpha\beta}g^{\mu\nu}h^{(n)}_{\mu\nu}.
\end{align}
At large distances from the source, the linearized Einstein tensor $G^{(1)}_{\alpha\beta}$ in this gauge reduces to the Minkowski-space d'Alembertian (up to a factor of $-1/2$). The analysis is then simplest if we work in asymptotically Cartesian coordinates $(t,\vec x)$ and transform to the frequency domain, such that the EFE~\eqref{EFE2} reduces to
\begin{equation}\label{eq:h2 Lorenz large r}
    -\frac{1}{2}(\vec\nabla^2+\omega^2)\widehat h^{(2)}_{\alpha\beta} = -\widehat G^{(2)}_{\alpha\beta}.
\end{equation}
Here we use wide hats to denote Fourier transformed quantities, with the metric perturbation written as an integral over modes, $\bar h^{(n)}_{\alpha\beta}(t,\vec x) = \int d\omega\, e^{-i\omega t}\widehat h^{(n)}_{\alpha\beta}(\omega,\vec x) $, as is~$G^{(2)}_{\alpha\beta}=\int d\omega\, e^{-i\omega t}\widehat G^{(2)}_{\alpha\beta}$, and we assume the stress-energy tensor vanishes outside some finite range of $|\vec x|$. At first order, the metric perturbation behaves like a sum of outgoing waves $\sim\frac{A_{\alpha\beta}(\omega,\theta^A)e^{i\omega u}}{r}$ in the time domain, meaning
\begin{equation}
\widehat h^{(1)}_{\alpha\beta}\sim \frac{A_{\alpha\beta}(\omega,\theta^A)e^{i\omega r}}{r}
\end{equation}
in the frequency domain, with $r:=|\vec x|$ and angles $\theta^A$ defined in the natural way from $\vec x$. The second-order Einstein tensor, which is made up of terms with the form $\nabla h^{(1)}\nabla h^{(1)}$ or $ h^{(1)}\nabla\nabla h^{(1)}$, then behaves as\footnote{Terms proportional to $(\omega_1)^2$ and $(\omega_2)^2$ also appear, but we suppress them for simplicity.}
\begin{equation}\label{eq:G2 large r}
    \widehat G^{(2)}_{\alpha\beta}(\omega,\vec x) \sim \int d\omega_2
    \,\omega_1\omega_2 P_{\alpha\beta}^{\gamma\delta\mu\nu}A_{\gamma\delta}(\omega_1)A_{\mu\nu}(\omega_2)\frac{e^{i(\omega_1+\omega_2)r}}{r^2},
\end{equation}
with $\omega_1+\omega_2=\omega$ and where $P_{\alpha\beta}^{\gamma\delta\mu\nu}$ is made up of algebraic combinations of the Minkowski metric. This $e^{i\omega r}/r^2$ structure of this source is what leads to problems at large distances. 

To see the nature of the problem, we can attempt to solve Eq.~\eqref{eq:h2 Lorenz large r} using the retarded Green's function, 
\begin{equation}
    \widehat G(\vec x,\vec x') = -\frac{e^{i\omega|\vec x-\vec x'|}}{4\pi |\vec x-\vec x'|}.
\end{equation}
Since the problem arises from the large-$r$ behaviour of the source, we can use the large-$r'$ approximation for the Green's function, $G(\vec x,\vec x')\approx -\frac{e^{i\omega r'}}{4\pi r'}$. The contribution to the second-order field from the large-$r'$ region is then
\begin{align}
    \widehat h^{(2)}_{\alpha\beta} &= \int_{r'>r^+} G(\vec x,\vec x')\widehat G^{(2)}_{\alpha\beta}(\vec x') d^3x'\\
     &\sim C_{\alpha\beta}(\omega)
     \int_{r'>r^+} \frac{e^{2i\omega r'}}{4\pi r'} dr',
\end{align}
where $r^+\gg r$ is some large radius, and $C_{\alpha\beta}(\omega) = P_{\alpha\beta}^{\gamma\delta\mu\nu}\int d\omega_2d\Omega' A_{\gamma\delta}(\omega_1,\theta'^A)A_{\mu\nu}(\omega_2,\theta'^A)$. For $\omega=0$, we see immediately that the integral over $r'$ is logarithmically divergent. In other words, when two first-order modes beat against one another with opposite frequencies ($\omega_2=-\omega_1$), they create a stationary, $\omega=0$ source, and the retarded integral of that source fails to converge. 

This analysis assumes $r\ll r'$, but the same conclusion is reached for any $r$: the retarded integral diverges for $\omega=0$. A more detailed analysis without the restriction $r\ll r'$, in Sec.~5.1 of Ref.~\cite{Cunningham:2024dog}, shows that the retarded integral \emph{does} converge for $\omega\neq0$, but it converges to an asymptotically ill-behaved solution $\sim\displaystyle\frac{\log r}{r}e^{i\omega r}$. Reference~\cite{Cunningham:2024dog} also explains how the divergent retarded integral for $\omega=0$ specifically relates to gravitational-wave memory. We return to that relationship in Sec.~\ref{sec:forgetful gauges} below.

Applying this Green's-function analysis to the second-order Teukolsky equation is not straightforward because of well-known pathologies of the Teukolsky retarded Green's function~\cite{Poisson:1996ya}. But we can, nevertheless, analyse the behaviour of the equation and of its solutions at large distances. To tie this more directly to the methods of the present paper, we adopt BS background coordinates rather than the asymptotically Cartesian coordinates we used above. At large $\bsr$ we can neglect the effects of the background mass and spin and decompose $\psi^{(2)}_{4L}$ in spin-weighted spherical harmonics and retarded Fourier modes $e^{-i\omega u}$: 
\begin{equation}
    \psi^{(2)}_{4L} = \frac{1}{\bsr}\int d\omega\, e^{-i\omega u} R^{(2)}_{lm\omega}(r){}_{-2}Y_{lm}(\bst^A).
\end{equation}
One then quickly finds the Teukolsky equation~\eqref{eq:pure-2nd-Teuk} reduces to
\begin{equation}\label{eq:teuk2 omega}
    \left[\partial_\bsr^2+\left(\frac{4}{\bsr}+2 i \omega \right)\partial_\bsr -\frac{(l+2)(l-1)}{\bsr^2}\right]\!R^{(2)}_{lm\omega} = -2\bsr S^{(2)}_{lm\omega},
\end{equation}
starting from  Eq.~(212b) of Ref.~\cite{Spiers:2023mor}, for example. Clearly the dominant terms at large $\bsr$ are even more simple: 
\begin{equation}\label{eq:teuk2 large r}
    2 i \omega\partial_\bsr R^{(2)}_{lm\omega} = -2\bsr S^{(2)}_{lm\omega},
\end{equation}
Here 
\begin{equation}\label{S2}
    S^{(2)}  = \int d\omega\, e^{-i\omega u}S^{(2)}_{lm\omega}(r){}_{-2}Y_{lm}(\bst^A)
\end{equation}
represents the decomposition of the right-hand side of Eq.~\eqref{eq:pure-2nd-Teuk}, ${\cal S}\bigl[8\pi T^{(2)}_{\mu\nu}-G^{(2)}_{\mu\nu}[h^{(1)},h^{(1)}]\bigr]$. 

To assess the falloff of the source, we assume $T^{(2)}_{\mu\nu}$ vanishes at large $\bsr$ and that $G^{(2)}_{\mu\nu}$ has the same falloff as in the Lorenz-gauge analysis above,
\begin{equation}
    G^{(2)}_{\mu\nu} = \bsr^{\rm M}\left[ \frac{s_{\mu\nu}(u,\bst^A)}{\bsr^2} + \order{\bsr^{-3}}\right]
\end{equation}
for some function $s_{\mu\nu}$, where the $\bsr^{\sf M}$ appears due to the change to polar coordinates. In the special case of a Lorenz-gauge $h^{(1)}_{\alpha\beta}$, we can more concretely quote Eq.~(140) of Ref.~\cite{Cunningham:2024dog}:
\begin{equation}\label{eq:G2 in Lorenz}
    s_{\alpha\beta} = \left[-\frac{1}{2}N^{(1)}_{AB}N^{AB}_{(1)} +\frac{\partial}{\partial u}\!\left(C^{(1)}_{AB}N^{AB}_{(1)}\right) \right]l_\alpha l_\beta + 4 M^{(1)}\partial_u N^{(1)}_{AB}e^A_\alpha e^B_\beta, 
\end{equation}
where $M^{(1)}$, $C_{AB}^{(1)}$, and $N_{AB}^{(1)}=\partial_u C^{(1)}_{AB}$ are the first-order Bondi mass, shear, and news, respectively, and $l_\alpha:=-\partial_\alpha u$ and $e^A_\alpha:=\partial_\alpha \bst^A$ form the coordinate basis on $\mathscr{I}^+$. Now, $\mathcal{S}$ is a second-order differential operator, and at large $\bsr$, $S^{(2)}$ is dominated by terms in $\mathcal{S}[G^{(2)}]_{\mu\nu}$ in which both the derivatives are in the $u$ direction; other derivatives always reduce the order by one power of $1/\bsr$. More explicitly, from (59b) of Ref.~\cite{Pound-Wardell:2021}, for example, we can quickly read off
\begin{equation}
    S^{(2)} = -\frac{\partial^2_u s_{\bar m\bar m}}{2\bsr^2} + {\cal O}(\bsr^{-3}).
\end{equation}
where $s_{\bar m\bar m} = s_{\alpha\beta}\bar m^\alpha \bar m^\beta$. Here $\bar m^\beta$ is the standard Newman-Penrose complex null tetrad leg, with angular components $\bar m^A=\frac{1}{\sqrt{2}\bsr}(1,-i\csc\bst)$ at large $\bsr$. The essential feature here is simply that the Teukolsky source, like the source in the EFE, decays like $1/\bsr^2$. 

Returning to Eq.~\eqref{eq:teuk2 large r}, we infer the source 
\begin{equation}
    S^{(2)}_{lm\omega} = \frac{\omega^2 s^{lm\omega}_{\bar m\bar m}}{2\bsr^2} + {\cal O}(\bsr^{-3}),
\end{equation}
and we immediately read off that each mode of the solution has the form 
\begin{equation}
 \frac{R^{(2)}_{lm\omega}e^{-i\omega u}}{\bsr} = \frac{\bigl[a_{lm\omega}+i\omega s^{lm\omega}_{\bar m\bar m}\log\bsr\bigr]e^{-i\omega u}}{2\bsr} + {\cal O}(\bsr^{-2}),
\end{equation}
for some constant $a_{lm\omega}$. Like the Lorenz-gauge $h^{(2)}_{\alpha\beta}$, the second-order Weyl scalar has a logarithmic irregularity in its $\omega\neq0$ modes. 

On the other hand, the reduced second-order Teukolsky equation \emph{does not} exhibit quite the same extreme pathologies as the Lorenz-gauge field equations for $\omega=0$ modes. Intuitively, this is because at large $\bsr$, $\psi^{(2)}_{4L}$ reduces to $-\frac{1}{2}\partial^2_u h^{(2)}_{\bar m \bar m}$~\cite{Pound-Wardell:2021}; the time derivatives kill off the misbehaving stationary modes. Since no $u$ derivatives contribute, the stationary part of the Teukolsky source is suppressed by two powers of $\bsr$ relative to the source in the EFE. Explicitly, again referring to (59b) of Ref.~\cite{Pound-Wardell:2021}, we read off
\begin{equation}
    \bigl\langle S^{(2)}\bigr\rangle = -\frac{\edthp^2\langle s_{uu}\rangle}{4\bsr^4} + {\cal O}(\bsr^{-5}),
\end{equation}
where $\edthp$ is a spin-lowering operator, normalized as in Eq. (79b) of Ref.~\cite{Spiers:2023mor}, and we use angular brackets to denote the time-averaged, $\omega=0$ piece of a function. With this source, we can readily check that $\omega=0$ solutions of Eq.~\eqref{eq:teuk2 omega} exist and decay at large $\bsr$. However, they do not decay as rapidly as a vacuum solution should, behaving as $1/\bsr$ rather than $1/\bsr^{l+2}$.

As we highlighted in the Introduction, to overcome these problems, particularly the divergent Lorenz-gauge integrals, Refs.~\cite{PoundLargeScales,Cunningham:2024dog} developed a post-Minkowskian expansion similar to the MPM construction in PN theory~\cite{blanchet1986radiative, blanchet1987radiative, blanchet1998multipole}. 
This method involves baroque analytical calculations, motivating us to explore an alternative method for removing the infrared divergences. 
In Ref.~\cite{spiers2023second}, we proposed a conceptually simple alternative: transform to a better-behaved gauge. We showed that in sufficiently regular gauges, the infrared divergences are avoided, in particular for the reduced second-order Teukolsky equation, \cref{eq:pure-2nd-Teuk}. Such a gauge satisfies the following falloff conditions: 
\begin{align}
    h^{(1)}_{\mu\nu}= \O(r^{-1}), \ \ h^{(1)}_{l\mu}= \O(r^{-2}), \ \ h^{(1)}_{m\mb}= \O(r^{-2}), \ \ 
    h^{(1)}_{ll}= \O(r^{-3}),\label{eq:gaugeCondsh}
\end{align}
where $(l^\mu,n^\mu,m^\mu,\bar m^\mu)$ is a (zeroth-order) Newman--Penrose null tetrad aligned with the Kerr principal null directions. Multiple gauges satisfy these conditions, including the standard outgoing radiation gauge~\cite{cohen1975space, chrzanowski1975vector} popular in second-order QNM calculations~\cite{ma2024excitation, Khera:2024yrk, Bourg:2024jme}. The ingoing radiation gauge of metric reconstruction of Refs.~\cite{Loutrel-etal:2020, Ripley-etal:2020} also satisfy \cref{eq:gaugeCondsh}; the traditional metric reconstruction scheme leads to metric perturbations in the ingoing radiation gauge that are \emph{not} asymptotically flat~\cite{Keidl:2010pm}, but an additional gauge transformation of the form in Ref.~\cite{hollands2024metric} can put the metric perturbation in an ingoing radiation gauge satisfying~\cref{eq:gaugeCondsh}. In this paper we prioritize the BS gauge over those alternatives in order to connect to the large body of literature on BMS frames.

\section{\label{sec:KerrBS}Kerr spacetime in Bondi-Sachs coordinates and its associated BMS frame}

Before addressing the gauge transformation to the perturbative BS gauge, we analyse how the background spacetime, the Kerr black hole, can be expressed in BS coordinates. In this section, for simplicity we omit ``$(0)$'' labels on the Kerr mass and spin parameters $M$ and $a$.

There has been a substantial body of literature on the Kerr metric in BS coordinates: 
\begin{itemize}
    \item Fletcher and Lun~\cite{fletcher2003kerrgenBS} express the Kerr metric in generalised BS form, where the angular metric gauge condition in \cref{eq:BSconditions} is not imposed. Their coordinate system differs from BS coordinates only in its choice of radial coordinate.  
    \item The asymptotic expansion of Fletcher and Lun's metric has been transformed to a BS gauge by Barnich and Troessaert~\cite{Barnich:2011mi} and Gaur~\cite{Gaur:2024oms}. However, the result is in a BMS frame with nonvanishing shear and with coordinate singularities at the poles\footnote{This approach has also been generalised to express Kerr--Newman spacetime in BS coordinates~\cite{Galoppo:2024vww}.}.
    \item Hollands and Toomani~\cite{hollands2024metric} transform Fletcher and Lun's generalised BS metric to Newman-Unti form~\cite{newman1962behavior} without using an asymptotic expansion. One could similarly transform Fletcher and Lun's metric to a BS gauge without an asymptotic expansion, extending Ref.~\cite{Gaur:2024oms}, but the result would again be in a BS gauge with nonvanishing shear and singularities at the poles.
    \item Bishop and Venter~\cite{bishop2006kerrBS} express the Kerr metric analytically in a BS gauge with vanishing shear and without singularities at the poles. However, their expression for the metric involves highly nontrivial analytical functions, containing integrals which must be evaluated numerically\footnote{There also appear to be errors in Eqs.~(35) and (36) of Ref.~\cite{bishop2006kerrBS}, which become apparent upon evaluating their respective asymptotic limits. One would expect $j_{11}=1+\mathcal{O}(\hat{r}^{-1})$ and $j_{1}=\mathcal{O}(\hat{r}^{-1})$, but evaluating their expression gives $j_{11}=\frac{1}{2}+\mathcal{O}(\hat{r}^{-1})$ and $j_{1}=\mathcal{O}(\hat{r}^{3})$, respectively.}.
    \item Arganaraz and Moreschi~\cite{Arganaraz:2021fpm} find BS null coordinates for the Kerr metric, with vanishing shear and free of coordinate singularities at the poles, expressed in terms of radial integrals. 
    \item Reference~\cite{chrusciel2001hamiltonian} presents a gauge transformation from Kerr--Newman coordinates to coordinates that asymptotically approach BS coordinates.
    \item Bai et al.~\cite{bai2007KerrBS} calculate an asymptotic expansion for the Kerr metric in the BS gauge; while the expansion is incomplete, it can be evaluated to arbitrarily high order, and the expansion they provide is to a sufficient order for our analysis. Additionally, the Kerr BS coordinates in Bai et al.~\cite{bai2007KerrBS} are regular and shear-free. We expect that the singular BS coordinates derived from Fletcher and Lun can be recovered by performing a singular BMS transformation on the coordinates in Bai et al.~\cite{bai2007KerrBS} (and vice versa).
\end{itemize}

We find it most convenient to follow Bai et al.~\cite{bai2007KerrBS}, who derive a BS form of the Kerr metric by deriving a coordinate transformation from Boyer--Lindquist coordinates $(t,r,\theta,\phi)$ to BS coordinates $(\bsu,\bsr,\hat \theta,\hat \phi)$, such that the BS gauge conditions~\eqref{eq:BSconditions} are satisfied. The coordinate transformation is
\begin{align}\label{eq:BStoBLtransform}
\bsu&=t-r-2M\log\left[\frac{r}{2M} \right]+\frac{4M^2-\frac{1}{2}a^2\sin^2\theta}{r} +\frac{4M^3-Ma^2}{r^2} +\mathcal{O}(r^{-3}), \notag \\
\hat{r}&=r+\frac{a^2\sin^2\theta}{2r}+\frac{a^2M\sin^2\theta}{2r^2} +\mathcal{O}(r^{-3}), \notag \\
\hat{\theta} &= \theta+\frac{a^2\cos\theta\sin\theta}{2r^2} +\mathcal{O}(r^{-4}),\notag \\
\hat{\phi} &=\phi+\frac{Ma}{r^2}+\frac{4M^2a}{3r^3}+\mathcal{O}(r^{-4}).
\end{align}
We invert \cref{eq:BStoBLtransform} to find the transformation from BS to Boyer--Lindquist coordinates,
\begin{align}
t&=\bsu+\hat{r}+2M\log\left[\frac{\hat{r}}{2M}\right]-\frac{4M^2}{\hat{r}}-\frac{4M^3-Ma^2+\frac{3}{2}Ma^2\sin^2\hat{\theta}}{\hat{r}^2}+\mathcal{O}(\hat{r}^{-3}), \notag \\
r&=\hat{r}-\frac{a^2\sin^2\hat{\theta}}{2\hat{r}}-\frac{Ma^2\sin^2\hat{\theta}}{2\hat{r}^2}+\mathcal{O}(\hat{r}^{-3}), \notag \\
\theta &= \hat{\theta}-\frac{a^2\cos \hat{\theta}\sin\hat{\theta}}{2\hat{r}^2}+\mathcal{O}(\hat{r}^{-4}), \notag \\
\phi &= \hat{\phi} -\frac{Ma}{\hat{r}^2}-\frac{4M^2a}{3\hat{r}^3}+\mathcal{O}(\hat{r}^{-4}).
\end{align}
Implementing the gauge transformation~\eqref{eq:BStoBLtransform} on the Boyer--Lindquist Kerr metric gives
\begin{align}\label{eq:KerrBS}
g^{(0)}_{\mu\nu} d\hat{x}^\mu d\hat{x}^\nu =&    -\bigg(1-\frac{2M}{\hat{r}}+\frac{Ma^2(2-3\sin^2\hat{\theta})}{\hat{r}^3}\bigg)d\bsu ^2 - 2d\bsu d\hat{r} \nonumber\\ 
& + \frac{6Ma^2\sin\hat{\theta}\cos\hat{\theta}}{\hat{r}^2}d\bsu d\hat{\theta} - \frac{4Ma\sin^2\hat{\theta}}{\hat{r}}d\bsu d\hat{\phi} \notag \\ 
&+ \bigg(\hat{r}^2-\frac{Ma^2 \sin^2\hat{\theta}}{\hat{r}} \bigg)d\hat{\theta}^2 + \bigg(\hat{r}^2 \sin^2\hat{\theta} + \frac{Ma^2\sin^4\hat{\theta}}{\hat{r}} \bigg)d\hat{\phi}^2 + \mathcal{O}(\hat{r}^{\mathrm{M}-4}).
\end{align}
We have corrected some typos in the metric presented in Bai et al.~\cite{bai2007KerrBS}, and we have given each component in \cref{eq:KerrBS} to a consistent order up to $ \mathcal{O}(\hat{r}^{\mathrm{M}-4})$; see \cref{eq:below-Minkowski}. We compare the BS coordinates to the more traditional  retarded coordinates of Kerr spacetime (Kerr--Newman coordinates) in Appendix~\ref{app:Kerr-Newman}.



Equation~\eqref{eq:KerrBS} is not a unique form of the Kerr metric in the BS gauge; it is transformable under the BMS freedoms. The BS metric quantities in this BMS frame are given by
\begin{align}\label{KerrBSQuantities}
        \MB[(0)]=M,\ Z^{(0)}=0,\ C^{(0)}_{AB}=0,\ N_{(0)}^A=(0,-3Ma),
\end{align}
which are the properties of a \emph{canonical} BMS frame for stationary spacetimes~\cite{flanagan2017BMS}. See also the discussion around Eq.~(5.8) of Ref.~\cite{compere2020poincare}. Additionally, from Eq.~\eqref{eq:KerrBS} one can identify that the BS functions \eqref{eq:BSfalloffs} take the form
\begin{align}
    V^{(0)}&=  \bsr-2M+\frac{Ma^2(2-3\sin^2\hat{\theta})}{\hat{r}^2}+\O(\bsr^{-3}),  \notag\\ 
    \beta^{(0)}&=\O(\bsr^{-4}),\notag \\
    \gamma^{(0)}_{AB}&= \Omega_{AB}+\mathcal{O}(\hat{r}^{-3}),  \notag \\
    U_{(0)}^A&=   -\frac{2}{3}\frac{N_{(0)}^A}{\hat{r}^3} + \mathcal{O}(\hat{r}^{-4}). 
\end{align}

Equation~\eqref{eq:KerrBS} is simple and near Boyer--Lindquist in nature: all super charges vanish; the Bondi mass aspect is equal to the ADM mass; the angular momentum aspect is an $l=1$ odd-parity vector harmonic (meaning, in particular, the even-parity, center-of-mass charge vanishes); and the shear vanishes. Up to order $\bsr^{{\rm M}-3}$ terms, the nonzero components of the metric have the simple form
\begin{align}
    g^{(0)}_{uu} &= -1+\frac{2M}{\bsr} + \order{\bsr^{-3}},\label{g0uu}\\
    g^{(0)}_{u\bsr} &= -1 + \order{\bsr^{-4}},\label{g0ur}\\
    g^{(0)}_{uA} &= \frac{2}{3}\frac{N^{(0)}_A}{\bsr} +\order{\bsr^{-2}},\label{g0uA}\\
    g^{(0)}_{AB} &= \bsr^2 \Omega_{AB} + \order{\bsr^{-1}}.\label{g0AB}
\end{align}
We will also require the coefficients of $\bsr^{{\rm M}-3}$ in some components, which we denote $g^{\{3\}}_{\mu\nu}$. With some algebra, we obtain
\begin{align}
    g^{\{3\}}_{uA} &= \frac{1}{3M}N^B D_A N_B^{(0)} ,\label{g3uA}\\
    g^{\{3\}}_{AB} &= \frac{2}{9M}\epsilon_{C\langle A} \epsilon_{B\rangle D}N^C_{(0)}N^D_{(0)}\label{g3AB}
\end{align}
from Eq.~\eqref{eq:KerrBS}.

As we are concerned with BMS frame fixing, it is necessary to ask whether there are residual BMS transformations that preserve Eq.~\eqref{eq:KerrBS}; that is, are there BMS transformations that leave Eq.~\eqref{eq:KerrBS} unchanged, or does Eq.~\eqref{eq:KerrBS} correspond to a fully fixed BMS frame? The answer presents itself if we consider the symmetries of Kerr spacetime: stationarity and axial symmetry. Time translations and axial rotations preserve Eq.~\eqref{eq:KerrBS}; therefore, the frame is not fully fixed by the form of the metric in Eq.~\eqref{eq:KerrBS}. In Sec.~\ref{sec:ConstrInfBMS}, we prove that the other BMS degrees of freedom (two further rotations, three Lorentz boosts, spatial translations, and supertranslations) are fully constrained by the form of the Kerr metric in Eq.~\eqref{eq:KerrBS}.

In BHPT, calculations are often executed in tetrad basis formalisms, such as the Newman--Penrose~\cite{np1962} and GHP~\cite{geroch1973space} formalisms. For convenience in those formalisms, we express the Kinnersley tetrad~\cite{kinnersley1969type} in the Kerr BS coordinates of \cref{eq:BStoBLtransform}:
\begin{align}
    l^{\gamma}&=\bigg(   \frac{a^2\sin^2\hat\theta}{2\bsr^2},  1 - \frac{a^2 \sin^2\hat \theta}{2 \bsr^2} ,   -\frac{a^2\cos\hat\theta\sin\hat\theta}{\bsr^3} , \frac{a}{\bsr^2}  \bigg),\label{eq:Kinnersley-l}  \\
    n^{\gamma}&=\bigg(   1+\frac{3a^2\sin^2\hat\theta}{4\bsr^2}, -\frac{1}{2}+\frac{M}{\bsr} - \frac{a^2 \sin^2\hat\theta}{4\bsr^2}, \frac{a^2 \sin[2\hat\theta]}{4\bsr^3} , \frac{a}{2\bsr^2} +\frac{aM}{\bsr^3}  \bigg),\label{eq:Kinnersley-n} \\
    m^{\gamma}&=\Bigg(   \frac{i a\sin\hat\theta}{\sqrt2 \bsr}, \frac{ a^2\cos\hat\theta\sin\hat\theta}{\sqrt2 \bsr^2}, \frac{1}{\sqrt2 \bsr}-\frac{ia\cos\hat\theta}{\sqrt2 \bsr^2} - \frac{a^2\cos^2\hat\theta}{2\sqrt2 \bsr^3} , 
\notag \\
    & \ \ \ \ \ \ \ \frac{i\csc[\hat\theta]}{\sqrt2 \bsr}\left[ 1-\frac{ia\cos\hat\theta}{\bsr} -\frac{a^2\cos[2\hat\theta]}{2\bsr^2}  \right] \Bigg), \label{eq:Kinnersley-m}
\end{align}
up to and not including $\O(\bsr^{{\rm M}-4})$ contributions.

\section{\label{sec:BSgaugevector}Transformation to the perturbative Bondi–Sachs gauge at linear order}

In this section, we derive the transformation to the perturbative BS gauge, from an arbitrary initial asymptotically flat gauge, at first order in perturbation theory. Like in the last section, we omit ``$(0)$'' labels on background parameters $(M,a)$; similarly, we omit the ``$(1)$'' label on the first-order metric perturbation, simply writing it as~$h_{\mu\nu}$.


\subsection{The perturbative Bondi--Sachs gauge}\label{sec:pertBSgauge}

Applying the BS gauge conditions, \cref{eq:BSconditions}, to the metric perturbation gives the perturbative BS gauge conditions at linear order in $\varepsilon$,
\begin{align}\label{eq:pertBSconditions}
    \hat h_{\hat{r}\hat{r}}=\hat h_{\hat{r}A}=0 \quad\text{and} \quad\Omega^{AB}\hat h_{AB}=\order{\bsr^{-2}},
\end{align}
where we use a hat to denote metric perturbations in the the BS gauge. The condition on the angular trace derives from the condition on the determinant in Eq.~\eqref{eq:BSconditions}; see Appendix~\ref{sec:perturbative det condition}.

Perturbatively expanding \cref{eq:BSMetric} in powers of $\varepsilon$, we find the first-order metric perturbation in the BS gauge takes the form
\begin{align}  \label{eq:perturbedBSMetric}
\hat h_{\mu\nu} dx^\mu dx^\nu =& - \left(\frac{V^{(1)}}{\hat{r}} e^{2\beta^{(0)}} +\frac{2\beta^{(1)}V^{(0)}}{\hat{r}}\right)du^2 -4\beta^{(1)} du d\hat{r} \notag \\
&+ \hat{r}^2 \gamma^{(1)}_{AB}\left(d\hat{\theta}^A -U_{(0)}^A du\right)\left(d\hat{\theta}^B -U_{(0)}^B du\right)\notag \\
&- \hat{r}^2 \gamma^{(0)}_{AB}\left(d\hat{\theta}^AU_{(1)}^B du+U_{(1)}^A dud\hat{\theta}^B\right).
\end{align}
For a Kerr background satisfying \cref{KerrBSQuantities}, the perturbative expansion of \cref{eq:BSfalloffs} gives
\begin{align}
V^{(1)} &= - 2\MB[(1)] -\frac{2Z^{(1)}}{\hat{r}} + \mathcal{O}(\hat{r}^{-2}), \notag \\ \beta^{(1)} &= \mathcal{O}(\hat{r}^{-3}), \notag \\
\gamma^{(1)}_{AB} &= \frac{C_{AB}^{(1)}}{\hat{r}}
+\mathcal{O}(\hat{r}^{-3}), \notag \\
   \ \  U_{(1)}^A &= \frac{-D_BC^{AB}_{(1)}}{2\hat{r}^2}  -\frac{2}{3}\frac{N^A_{(1)}}{\hat{r}^3} + \mathcal{O}(\hat{r}^{-4}). \label{eq:pertBSfalloffs}
\end{align}
Here we again assume the spacetime is vacuum outside some finite region, a consistent assumption for first-order self-force and QNM calculations. Under this condition, one can also see from expressions in Ref.~\cite{flanagan2017BMS} that $\beta^{(1)}$ exhibits the even stronger falloff $\beta^{(1)} = \mathcal{O}(\hat{r}^{-4})$. 

To fix the BMS frame of $\hat h_{\mu\nu}$, we will set several of the Poincaré charges and the even-parity piece of the shear to zero at some reference time. The linear Poincaré charges $M^{(1)}$, $P^{(1)}_i$, $J^{(1)}_i$, $K^{(1)}_i$ are constructed from $\MB[(1)]$ and $N^A_{(1)}$ via \cref{eq:PT,eq:JT,eq:KT}. Several features of these charges simplify at linear order. First, the linear term in the the Wald--Zoupas angular momentum aspect, \cref{eq:Nhat}, is simply
\begin{align}\label{eq:WaldZoupasN1}
    \hat{N}^{(1)}_A&=N^{(1)}_A - u D_A\MB[{(1)}].
\end{align}
In stationary regions of spacetime, \cref{eq:Z} becomes
\begin{align}\label{eq:ZN}
    Z^{(1)}=-\frac{D_A N^A_{(1)}}{6},
\end{align}
given our shear-free background coordinates [\cref{KerrBSQuantities}]. More importantly, the charges' time dependence becomes trivial by virtue of Eqs.~\eqref{eq:MBdot}--\eqref{eq:Ndoteven}, which reduce to
\begin{align}
    \frac{d\MB[(1)]}{du} &= \frac{1}{4}D_A D_B \partial_u C_{(1)}^{AB},\label{eq:MBdot1 regular}\\
    \frac{d N^{(1)\rm even}_A}{du} &= -\frac{1}{4}D_AD_B D_C \partial_u C_{(1)}^{BC},\label{eq:Ndot1 even regular}\\ 
    \frac{d\hat N^{(1)\rm odd}_A}{du} &=  \frac{1}{4}D_BD_AD_C C_{(1)}^{BC} - \frac{1}{4}D_BD^BD^CC^{(1)}_{CA}.\label{eq:Ndot1 odd regular}
\end{align}
Here we have used Eq.~\eqref{eq:WaldZoupasN1} and its consequence, $\hat N^{(1)\rm odd}_A = N^{(1)\rm odd}_A$. Since the Poincaré charges are $l=0$ and $l=1$ pieces of $\MB[(1)]$ and $\hat N^{(1)}_A$, and the shear only contains $l\geq2$ modes, we immediately infer
\begin{equation}\label{constant Poincare}
    \frac{dM^{(1)}}{du}=\frac{dP^{(1)}_i}{du} = \frac{dJ^{(1)}_i}{du} = \frac{dK^{(1)}_i}{du} = 0.
\end{equation}
Hence, all the Poincaré charges are constant at first order. This is the expected result because gravitational fluxes are quadratic.


Finally, to confirm that the BS gauge achieves our desired goal removing infrared divergences, we can assess how the tetrad contractions of the perturbed metric behave near \scri. Contracting \cref{eq:perturbedBSMetric} with \cref{eq:Kinnersley-l,eq:Kinnersley-n,eq:Kinnersley-m} gives
\begin{align}
    \hat h_{ll}&=\O(\bsr^{-3}), \ \hat h_{ln}=\O(\bsr^{-2}), \notag \\
    \hat h_{lm}&=\O(\bsr^{-2}), \ \hat h_{l\mb}=\O(\bsr^{-2}), \notag\\ 
    \hat h_{m\mb}&=\O(\bsr^{-2}),\label{eq:tetradh}
\end{align}
and the other components are $\O(\bsr^{-1})$. 
Therefore, the metric perturbation in the BS gauge satisfies the requirements~\eqref{eq:gaugeCondsh} for the second-order Teukolsky equation to avoid infrared divergences, as desired.

\subsection{Gauge transformation to the perturbative Bondi--Sachs gauge}\label{sec:BSgaugevector-coord-form}
To derive the gauge transformation from a generic gauge (with metric perturbation $h_{\mu\nu}$) to the perturbative BS gauge, we begin with \cref{eq:1gaguetransform}, written in terms of the first-order gauge vector, $\xi^{\gamma}$,
\begin{align}\label{eq:gaugetransform}
    \hat h_{\mu\nu}= h_{\mu\nu} + \mathcal{L}_{\vec\xi} \,g^{(0)}_{\mu\nu},
\end{align}
where $h_{\mu\nu}$ is in any initial gauge and $\hat h_{\mu\nu}$ is in the perturbative BS gauge. In this section we leave the form of $h_{\mu\nu}$ unspecified, before assuming a series in $1/\bsr$ in the next section.

Referring to the form of the Kerr metric in Eqs.~\eqref{g0uu}--\eqref{g3AB}, we find the gauge transformation can be written as
\begin{align}
    \mathcal{L}_{\vec\xi} \,g^{(0)}_{uu} &= -2\partial_\bsu\xi^\bsr -2\biggl(1-\frac{2M^{(0)}}{\bsr}\biggr)\partial_\bsu\xi^\bsu  -\frac{2M^{(0)}}{\bsr^2}\xi^\bsr +\frac{4}{3}\frac{N^{(0)}_A}{\bsr}\partial_\bsu\xi^A  +{\cal O}(\bsr^{-3}),\label{Liexig0uu}\\
    \mathcal{L}_{\vec\xi} \,g^{(0)}_{u\bsr} &= -\partial_\bsu\xi^\bsu -\partial_\bsr\xi^r - \biggl(1-\frac{2M^{(0)}}{\bsr}\biggr)\partial_\bsr\xi^\bsu  +\frac{2}{3}\frac{N^{(0)}_A}{\bsr}\partial_\bsr\xi^A +{\cal O}(\bsr^{-3}),\label{Liexig0ur}\\
    \mathcal{L}_{\vec\xi} \,g^{(0)}_{uA} &= \bsr^2\partial_\bsu\xi_A - D_A\xi^r -\biggl(1-\frac{2M^{(0)}}{\bsr}\biggr)D_A\xi^\bsu -\frac{2}{3}\frac{N^{(0)}_A}{\bsr^2}\xi^\bsr + \frac{2}{3}\frac{D_BN^{(0)}_A}{\bsr}\xi^B \notag\\
    &\quad + \frac{2}{3}\frac{N^{(0)}_A}{\bsr}\partial_\bsu\xi^\bsu  +\frac{2}{3} \frac{N^{(0)}_B}{\bsr} D_A\xi^B  + \frac{g^{\{3\}}_{AB}}{\bsr}\partial_u\xi^B  + {\cal O}(\bsr^{-2}),\label{Liexig0uA}\\
    \mathcal{L}_{\vec\xi} \,g^{(0)}_{\bsr\bsr} &= -2\partial_\bsr\xi^\bsu + {\cal O}(\bsr^{-4}),\label{Liexig0rr}\\
    \mathcal{L}_{\vec\xi} \,g^{(0)}_{\bsr A} &= \bsr^2\partial_\bsr\xi_A - D_A\xi^\bsu + \frac{g^{\{3\}}_{AB}}{\bsr}\partial_\bsr\xi^B + \frac{2}{3}\frac{N^{(0)}_A}{\bsr}\partial_\bsr\xi^u 
    +{\cal O}(\bsr^{-3}),\label{Liexig0rA}\\
    \mathcal{L}_{\vec\xi} \,g^{(0)}_{AB} &= 2\bsr\xi^\bsr \Omega_{AB} +2 \bsr^2 D_{(A}\xi_{B)} - \frac{g^{\{3\}}_{AB}}{\bsr^2}\xi^\bsr +\frac{4}{3}\frac{N^{(0)}_{(A}D_{B)}\xi^u}{\bsr} + \frac{1}{\bsr}{\cal L}_{\xi^C} g^{\{3\}}_{AB}+ {\cal O}(\bsr^{-2}).\label{Liexig0AB}
\end{align}
Here we assumed the scalings $\xi^\bsu={\cal O}(\bsr^0)$, $\xi^\bsr={\cal O}(\bsr)$, and $\xi^A={\cal O}(\bsr^0)$, consistent with~Eq.~\eqref{eq:Poincare}. We also carried the $\bsr\bsr$, $\bsr A$, and $AB$ components to $\order{\bsr^{{\rm M}-3}}$ inclusive, one order higher than the other components, for reasons explained below. 

From the expressions~\eqref{Liexig0uu}--\eqref{Liexig0AB}, we draw several conclusions: 
\begin{itemize}
    \item First, the components we wish to impose conditions on ($\hat h_{\bsr\bsr}$, $\hat h_{\bsr A}$, and $\Omega^{AB}\hat h_{AB}$) involve no $u$ derivatives of $\xi^\alpha$, meaning the gauge conditions can be enforced along each null slice $u=\text{constant}$ regardless of the metric's $u$ dependence.
    \item Those three components ($rr$, $rA$, and the angular trace) of ${\cal L}_{\vec \xi}\,g^{(0)}_{\mu\nu}$ are identical to what they would be in flat spacetime up to order $\bsr^{{\rm M}-3}$ terms, meaning the first three orders of the gauge transformation will be insensitive to the background Kerr parameters.
    \item Controlling the mass and linear momentum, which appear at order $1/\bsr$ in $\hat h_{uu}$, requires calculating $\xi^\alpha$ to higher order in $1/\bsr$ than would be required to simply enforce the falloff conditions~\eqref{eq:tetradh}. From Eq.~\eqref{Liexig0uu} we see that to correctly extract the mass aspect $\MB[(1)]$ (and therefore the mass and linear momentum), we must enforce the gauge conditions to sufficiently high order to fix the $\bsr^0$ and $1/\bsr$ terms in $\xi^u$; the $\bsr^1$, $\bsr^0$, and $1/\bsr$ terms in $\xi^\bsr$; and the $\bsr^0$ term in $\xi^A$. 
    \item Controlling the angular momentum and boost charges, which appear at order $1/\bsr$ in $\hat h_{uA}$, similarly requires enforcing the gauge condition to higher order. From Eq.~\eqref{Liexig0uA} we see that to correctly extract $N^{(1)}_A$, we must fix the $\bsr^0$ and $1/\bsr$ terms in $\xi^u$; the $\bsr^1$, $\bsr^0$, and $1/\bsr$ terms in $\xi^\bsr$; and the $\bsr^0$, $1/\bsr$, $1/\bsr^2$, and $1/\bsr^3$ terms in~$\xi^A$.
\end{itemize}

We find $\xi^\alpha$ to the required order in three steps: 
\begin{enumerate}
    \item Solve $\hat h_{\bsr\bsr}=0$ for $\xi^u$ through order $1/\bsr$. From Eq.~\eqref{Liexig0rr}, we see this requires controlling $h_{\bsr\bsr}$ through order~$1/\bsr^2$.
    \item Solve $\hat h_{\bsr A}=0$ for $\xi^A$ through order $1/\bsr^3$. From Eq.~\eqref{Liexig0rA}, we see this requires controlling $h_{\bsr A}$ and terms in Eq.~\eqref{Liexig0rA} through order $1/\bsr^2$.
    \item Solve $\Omega^{AB}\hat h_{AB}=\order{\bsr^{-2}}$ for $\xi^\bsr$ through order $1/\bsr$. From Eq.~\eqref{Liexig0AB}, we see this requires controlling $h_{AB}$ and terms in Eq.~\eqref{Liexig0AB} through order $\bsr^0$. 
\end{enumerate}
Note the final step requires one less order in $1/\bsr$ than given in Eq.~\eqref{Liexig0AB}; we provide that higher order only to be consistent across the $\bsr\bsr$, $\bsr A$, and $AB$ components.

In the first step, the condition $\hat h_{\hat{r}\hat{r}}=0$ with Eq.~\eqref{Liexig0rr} gives
\begin{align}
  h_{\bsr \bsr} - 2\partial_{\bsr}\xi^\bsu=  \O(\bsr^{-4}).\label{eq:hrr}
\end{align}
Solving for $\xi^u$ yields
\begin{align}\label{eq:xiu}
    \xi^\bsu=\int \frac{h_{\bsr \bsr}}{2} d\bsr + \xi^\bsu_\circl + \O(\bsr^{-3}),
\end{align}
where quantities labelled with a $\circl$ sub/superscript are independent of $\bsr$.

In the second step, we impose $\hat h_{\hat{r}A}=0$. Equation~\eqref{Liexig0rA} implies
\begin{equation}
h_{\bsr A} +\bsr^2\partial_\bsr\xi_A - D_A\xi^\bsu + \frac{g^{\{3\}}_{AB}}{\bsr}\partial_\bsr\xi^B + \frac{2}{3}\frac{N^{(0)}_A}{\bsr}\partial_\bsr\xi^u 
    = {\cal O}(\bsr^{-3}).\label{hrA eqn}    
\end{equation}
Solving for $\xi^A$ gives
\begin{equation}
    \xi^A = \int \frac{\Omega^{AB}}{\bsr^2}
    \left(D_B\xi^u - h_{\bsr B} - \frac{2}{3}\frac{N^{(0)}_B}{\bsr}\partial_\bsr\xi^u\right)d\bsr + \xi^A_\circl +\order{\bsr^{-4}}.\label{eq:xiA}
\end{equation}
Note that $g^{\{3\}}_{AB}$ has not contributed in the end. This is because it can only contribute at order $1/\bsr^2$ in Eq.~\eqref{hrA eqn} if $\xi^A=\order{\ln\bsr}$, which would require $h_{\bsr B}$ to grow like $\bsr$ in the integrand in Eq.~\eqref{eq:xiA}.


Finally, we impose $ \Omega^{AB}\hat h_{AB}=\order{\bsr^{-1}}$, noting again we do not require $1/\bsr$ terms. From Eq.~\eqref{Liexig0AB} we read off
\begin{equation}\label{Omega.hAB eqn}
    \Omega^{AB}h_{AB} + 4\bsr\xi^\bsr  +2 \bsr^2 D_{A}\xi^{A}  
    = {\cal O}(\bsr^{-1}).
\end{equation}
Rearranging for $\xi^\bsr$, we obtain
\begin{equation}\label{eq:xir}
    \xi^\bsr = -\frac{1}{4\bsr}\left(\Omega^{AB}h_{AB} +2 \bsr^2 D_{A}\xi^{A} 
    \right) + \order{\bsr^{-2}}.
\end{equation}



\subsection{Radial expansion of gauge transformation to the Bondi--Sachs gauge}

Our results in the previous section are valid for any initial $h_{\mu\nu}$---even one that is in a non-asymptotically-flat gauge, though the results are valid to lower order than advertised in that case.  
We now assume the initial gauge is asymptotically flat and adopt the following expansion in powers of $1/\bsr$:
\begin{align}
    h_{\mu\nu} = \bsr^{\rm M}\left[\frac{h^\ro_{\mu\nu}[u,\bst^A]}{\bsr}+\frac{h^\rr_{\mu\nu}[u,\bst^A]}{\bsr^2}+\frac{h^{\{3\}}_{\mu\nu}[u,\bst^A]}{\bsr^3}+\O(\bsr^{-4})\right].\label{eq:metricPertExpansion}
\end{align}
This expansion is related to the analogous one in Kerr--Newman coordinates (which are more likely to be used in practice) in Appendix~\ref{App:pBSinBL}. One can readily extend the calculations to allow for $\ln\bsr$ terms in $h^{\{n\}}_{\mu\nu}$, but we exclude them here for simplicity. 

Substituting the expansion~\eqref{eq:metricPertExpansion} into \cref{eq:xiu} and evaluating the integral, we obtain
\begin{align}\label{eq:xiu3}
    \xi^\bsu = \xi^\bsu_\circl + \frac12\left(h^\ro_{\bsr \bsr}\ln\bsr-\frac{h^\rr_{\bsr \bsr}}{\bsr}\right) + \O(\bsr^{-2}).
\end{align}
Note any length scale in the logarithm can be absorbed into the integration constant $\xi^\bsu_\circl$. Applying the same procedure to \cref{eq:xiA} yields
\begin{align}
    \xi^A&=\xi^A_\circl + \Omega^{AB}\Bigg(\frac{2h_\bssub{\bsr B}^\ro-2D_{B}\xi^\bsu_{\circl}-(1+\ln\bsr)D_Bh_\bssub{\bsr\bsr}^{\ro}}{2\bsr}+\frac{2h_\bssub{\bsr B}^{\rr}+D_Bh_\bssub{\bsr\bsr}^{\rr}}{4\bsr^2}\Bigg)\notag\\
    &\quad + \Omega^{AB}\Bigg(\frac{3h_{\bsr B}^{\{3\}}+N^{(0)}_B h_{\bsr\bsr}^{\{1\}}}{9 \bsr^3}\Bigg) +\O(\bsr^{-4}).\label{eq:xiA3}
\end{align}
Finally, we expand \cref{eq:xir} using \cref{eq:metricPertExpansion,eq:xiu3,eq:xiA3}:
\begin{align}
    \xi^\bsr&= -\frac{\bsr}{2}D_A\xi^A_{\circl} +\frac{1}{4}(1+\ln\bsr)D^2 h^{\{1\}}_{\bsr\bsr} -\frac{1}{4}\Bigl(\Omega^{AB}h^{\{1\}}_{AB}+2D^Ah^{\ro}_\bssub{\bsr A}  -2D^2\xi^\bsu_{\circl}\Bigr) \notag \\
    &\quad -\frac{1}{8\bsr}\Bigl(2\Omega^{AB}h^{\{2\}}_{AB}+2D^Ah^{\{2\}}_\bssub{\bsr A}-D^2 h^{\{2\}}_{\bsr\bsr}\Bigr)+\O(\bsr^{-2}),\label{eq:xir3}
\end{align}
where $D^2:=D^AD_A$.

This completes our calculation of the gauge vector to the perturbative BS gauge. For the sake of practical implementations, we provide the vector's tetrad components, expressed in terms of Kerr--Newman coordinates, in Appendix~\ref{App:pBSinBL}.

The gauge vector in \cref{eq:xiu3,eq:xiA3,eq:xir3} may appear to produce non-asymptotically-flat $uu$, $u\bsr$, $uA$ components of $\hat h_{\mu\nu}$, via \cref{Liexig0uu,Liexig0ur,Liexig0uA}. However, if $h_{\mu\nu}$ satisfies the linearised Einstein field equations with $T^{(1)}_{\bsr \bsr}=\O(r^{-3})$, $T^{(1)}_{\bsr A}=\O(r^{-2})$, and $T_{uA}=\O(r^{-3})$, then the offending $h_{\alpha\beta}$ terms in the gauge vector defined by \cref{eq:xiu3,eq:xiA3,eq:xir3} will necessarily conspire to cancel in ${\cal L}_{\vec\xi}\, g^{(0)}_{u\alpha}$; this follows from the fact that every asymptotically flat metric can be written in BS gauge, and we have no remaining freedom in the transformation. The exception to this is the contribution from the integration constants $\xi^u_\circ$ and $\xi^A_\circ$, which will violate asymptotic flatness unless their dependence on $u$ and $\bst^A$ is restricted. Imposing asymptotic flatness forces $\xi^u_\circ\partial_u + \xi^A_\circ\partial_A$ to be the generator of a BMS transformation. To make this clear, we note the following:
\begin{itemize}
    \item Eliminating the $\bsr^0$ term in Eq.~\eqref{Liexig0ur} requires $\partial_u\xi^u_\circl = \frac{1}{2}D_A\xi^A_\circl$. 
    \item Eliminating the $\bsr^2$ term in Eq.~\eqref{Liexig0uA} requires $\xi^A_\circl=\xi^A_\circl[\bst^A]$.
    \item Eliminating the $\bsr^2$ traceless term in Eq.~\eqref{Liexig0AB} requires $D_{\langle A}\xi^{\circl}_{ B\rangle}=0$.
\end{itemize}
Together these imply the BMS form
\begin{equation}\label{xicirc}
    \xi^u_\circl = f[u,\bst^A] \quad\text{and}\quad \xi^A_\circl=Y^A[\bst^B],
\end{equation}
with $f$ and $Y^A$ as in Eq.~\eqref{eq:BMSVec}. We address this residual BMS freedom in the next section.

\section{\label{sec:ConstrInfBMS}Fixing the BMS frame at first perturbative order}

Next, we address the BMS frame in the perturbative BS gauge. As reviewed in Sec.~\ref{sec:BMSIntro}, the BMS freedoms correspond to the residual gauge freedom within the BS formalism; this is the only remaining freedom in our gauge vector~\eqref{eq:xiu3}--\eqref{eq:xir3}, represented by $\xi^\alpha_\circl$. 
In this section we present a method of fixing the (first-order) BMS frame by fixing the residual BMS gauge vector. 

We now take our initial metric perturbation ($\hat{h}_{\mu\nu}$) to already be in a BS gauge---as it is after applying the transformation of the last section, for any choice of~$\xi^\alpha_\circl$. We then impose conditions on $\hat{h}^\prime_{\mu\nu}=\hat{h}_{\mu\nu}+\mathcal{L}_{\vec\xi}\, g^{(0)}_{\mu\nu}$ to fix the BMS frame, where $\xi^{\gamma}$ is a BMS gauge vector.

In this section, we will find it convenient to restore explicit labels ``(1)'' on first-order quantities.

\subsection{Linearized BMS transformations and BMS charges}\label{sec:BMS fixing strategy}

For any asymptotically flat spacetime in BS coordinates, 
we can write $\xi^{\gamma}$ as an expansion in orders of $1/\bsr$ by extending the BMS vector~\eqref{eq:BMSVec} into the interior of the spacetime~\cite{flanagan2017BMS} in such a way that it preserves the BS form of the metric: 
\begin{align}\label{eq:BMSvecExt}
\vec{\xi}&= f\partial_u + \Big[Y^A-\frac{1}{\hat{r}}D^Af + \frac{1}{2\hat{r}^2}C^{AB}D_B f +\mathcal{O}(\hat{r}^{-3})\Big] \partial_A \notag \\
&\quad - \Big[ \frac{1}{2}\hat{r} D_A Y^A -\frac{1}{2}D^2f+\frac{1}{4\hat{r}}D_A C^{AB} D_B f + \frac{1}{4\hat{r}}D_A(C^{AB}D_Bf)+\mathcal{O}(\hat{r}^{-2}) \Big]\partial_{\hat{r}}.
\end{align}
This can be compared with the Poincaré transformation~\eqref{eq:Poincare} and the $\xi^\alpha_\circl$ terms in Eqs.~\eqref{eq:xiu3}--\eqref{eq:xir3}. Under such a transformation, the changes in the BS metric functions $\MB$, $N_A$, and $C_{AB}$ are~\cite{flanagan2017BMS}\footnote{In Eq.~\eqref{DeltaC} we have replaced Flanagan and Nichols' $D_{\langle A}D_{B\rangle}f$ with $D_{\langle A}D_{B\rangle}\alpha$ using the fact that $D_{\langle A}D_{B\rangle}D_CY^C=0$ because $D_CY^C$ is an $l=1$ scalar.}
\begin{align}
    \Delta\MB &= 
    \frac{1}{4}N^{AB}D_A D_B f +\frac{1}{2}D_AfD_BN^{AB} + \frac{3}{2}\MB D_AY^A+Y^AD_A\MB \notag\\
    &\quad +\frac{1}{8}C^{AB}D_AD_BD_CY^C,\label{DeltaMB}\\
    \Delta N_A &= (f\partial_u + {\Lie}_{\vec Y}+D_BY^B)N_A+ 3M D_A f - \frac{3}{4}D_Bf(D^BD^C C_{CA} - D_AD_CC^{BC}),\label{DeltaN} \\
    \Delta C_{AB} &= fN_{AB} -2D_{\langle A}D_{B\rangle}\alpha -\frac{1}{2}D_CY^CC_{AB} + \Lie_{\vec Y}C_{AB}.\label{DeltaC}
\end{align}
The corresponding change in the Poincaré charges can be extracted from Eqs.~\eqref{eq:Pt}--\eqref{eq:KT}, and the change in the Moreschi supermomentum from Eq.~\eqref{eq:MoreschiSupermomentum}.

Specialised to linear order in $\varepsilon$ and to the Kerr values of BS quantities in Eq.~\eqref{KerrBSQuantities}, Eq.~\eqref{eq:BMSvecExt} simplifies to 
\begin{align}
        \vec{\xi}_{(1)}=&f_{(1)}\partial_u+\Big[Y_{(1)}^A-\frac{1}{\hat{r}}D^A f_{(1)}  +\mathcal{O}(\hat{r}^{-3}) \Big] \partial_A \notag \\
        &-\Big[ \frac{1}{2}\hat{r}D_AY_{(1)}^A -\frac{1}{2}D^2f_{(1)} +  \mathcal{O}(\hat{r}^{-2})\Big]\partial_{\hat{r}} , \label{eq:simplified-extended-bms-vector}
    \end{align}
with $f_{(1)}=\alpha_{(1)}+\frac{1}{2}u D_AY^A_{(1)}$. Similarly, from Eqs.~\eqref{DeltaMB}--\eqref{DeltaC}, we find the corresponding changes in $\MB[(1)]$, in the Wald--Zoupas angular momentum aspect~\eqref{eq:WaldZoupasN1}, and in the shear:
\begin{align}\label{eq:DeltaM}
        \Delta \MB[(1)] &= \frac{3M}{2} D_AY_{(1)}^A , \\
        \Delta \hat{N}^{(1)}_A &= 3MD_A\alpha_{(1)} +\mathcal{L}_{\vec{Y}_{(1)}} N^{(0)}_A +D_BY_{(1)}^B N^{(0)}_A, \label{eq:DeltaNhat}\\
        \Delta C^{(1)}_{AB} &= -2D_{\langle A}D_{B\rangle}\alpha^{(1)},\label{DeltaC1}
\end{align}
where $\Delta \MB[(1)] := \MB[\prime(1)]-\MB[(1)]$, for example.

We now recall that all the Poincaré charges are constant in time at first order, as shown in Eq.~\eqref{constant Poincare}. This allows us to fix the BMS frame as follows:
\begin{itemize}
    \item We fix the boosts (corresponding to even-parity modes of $Y^A_{(1)}$) by setting the linear momentum  $P_i^{\prime(1)}$ to zero. Equivalently, we set the $l=1$ terms in $\MB[\prime(1)]$ to zero.
    \item We fix the translations (corresponding to $l=1$ modes of $\alpha^{(1)}$) by setting the boost charges $K_i^{\prime(1)}$ to zero. Equivalently, we set the even-parity, $l=1$ terms in $\hat{N}_A^{\prime(1)}$ to zero. When combined with $P_i^{\prime(1)}=0$, this selects the centre-of-mass frame.
    \item We fix the rotations around the $x$ and $y$ axes (corresponding to odd-parity $m=\pm1$ modes of $Y^A_{(1)}$) by setting the angular momentum components $J_x^{\prime(1)}=J_y^{\prime(1)}=0$. Equivalently, we set the odd-parity, $l=1, m=\pm1$ terms in $\hat{N}_A^{\prime(1)}$ to zero. This aligns the perturbed spacetime's angular momentum with the $z$ axis. 
    \item We fix the supertranslations (corresponding to $l\geq2$ modes of $\alpha^{(1)}$) by setting the even-parity part of the shear $C^{\prime(1)}_{AB}$ to zero at a chosen reference time.
\end{itemize}
This leaves two transformations unfixed: time translations and rotations around the $z$ axis. Since these two transformations are isometries of the background spacetime, they do not contribute to ${\cal L}_{\vec\xi}\,g^{(0)}_{\mu\nu}$, making it impossible to fix them through conditions on $h^{\prime(1)}_{\mu\nu}$. We discuss this further below.
    


\subsection{Fixing to the comoving frame}

Equation~\eqref{eq:DeltaM} is independent of all BMS transformations except for boosts, as $D_AY_{(1)}^A=D_AD^A\chi_{(1)}=-2\chi_{(1)}$; see Eq.~\eqref{eq:evenoddY}. Hence, we can fix the boost freedom by eliminating the dipole part of $\MB[(1)]$. Doing so eliminates the linear momentum, making the frame comoving with the system. 

To achieve this, we express $Y_{(1)}^A$ in terms of $l=1$ spherical harmonics, such that Eq.~\eqref{eq:evenoddY} becomes
 \begin{align}\label{YA as harmonics}
 Y_{(1)}^A = \sum_{\m=-1}^{1} \left(\chi^{(1)}_{(1,\m)} D^AY_{1,m} -\kappa^{(1)}_{(1,\m)}\epsilon^{AB} D_B  Y_{1,m}\right).
 \end{align} 
Similarly, we express $\Delta \MB[(1)]$ as a sum of harmonics,
\begin{align}
    \Delta \MB[(1)]= \sum_{l=0}^\infty\sum_{m=-l}^{l}\Delta \MB[(1)]_{\!\!\!\!(l,\m)}Y_{lm}.
\end{align}
Equation~\eqref{eq:DeltaM} can then be rearranged for $\chi^{(1)}_{(1,\m)}$:  
\begin{align}
   D^2\chi^{(1)}_{(1,\m)}Y_{1,\m}  &= \frac{ 2\Delta \MB[(1)]_{\!\!\!\!(1,\m)}Y_{1,\m}}{3 M} , \notag \\
   \Rightarrow \chi^{(1)}_{(1,\m)}  &= -\frac{ \Delta \MB[(1)]_{\!\!\!\!(1,\m)}}{3 M},
\end{align}
where in the second line we have used the eigenvalue equation for the spherical harmonics, $D^2Y_{\l\m}=-\l(\l+1)Y_{\l\m}$. Setting $\Delta \MB[(1)]_{\!\!\!\!(1,\m)} = - \MB[(1)]_{\!\!\!\!(1,\m)}$ to enforce $\MB[\prime(1)]_{\!\!\!\!\!(1,\m)}=0$, we obtain
\begin{equation}
\chi^{(1)}_{(1,\m)} = \frac{\MB[(1)]_{\!\!\!\!(1,\m)}}{3 M}.\label{eq:boost-transform}    
\end{equation}

\subsection{Fixing to the centre-of-mass frame and fixing the rotation axis}


Next, we fix the rotations and translations by eliminating the $l=1$ modes of $\hat{N}^{\prime(1)}_A$ (excluding the odd-parity $m=0$ mode). Isolating each transformation is straightforward when a decomposition into vector spherical harmonics is applied~\cite{hecht2000decomp, shiraishi2013probing}. Following the conventions of Martel and Poisson~\cite{martellpoisson2005SVH}\footnote{We denote the even harmonics with $Z$ rather than $Y$ to avoid confusion with the boost and rotation generator $Y^A$.}, even ($Z^{lm}_A$) and odd ($X^{lm}_A$) vector spherical harmonics are defined as 
\begin{align}
	Z^{lm}_A &:= D_A Y_{lm}, \label{eq:Z_A}\\
    X^{lm}_A &:=-\epsilon_{A}^{\ \ B} D_B Y_{lm}.\label{eq:X_A}
\end{align}

In terms of these harmonics, the even and odd-parity pieces of Eq.~\eqref{YA as harmonics} read 
\begin{align}
Y^A_{{\rm e}(1)}&=D^A\chi^{(1)} = \sum_{m=-1}^1 \chi^{(1)}_{1,m}Z_{1,m}^A ,\\
Y^A_{{\rm o}(1)} &=-\epsilon_{A}^{\ \ B} D_B\kappa^{(1)} = \sum_{m=-1}^1 \kappa^{(1)}_{(1,m)}X^A_{1,m},
\end{align}
where we abbreviate the ``even'' and ``odd'' labels as ``e'' and ``o'' for brevity. Then, by using
\begin{align}
    N_A^{(0)}&\sim X_A^{1,0},\\
    Y^A_{{\rm e}(1)} &\sim Z^A_{1,m}, \\
    Y^A_{{\rm o}(1)} &\sim X^A_{1,m},
\end{align}
 one finds the constituent pieces of \cref{eq:DeltaNhat} decompose into the following vector harmonics:
\begin{align}
D_A\alpha_{(1)} &\sim  Z_A^{\l\m} ;\label{eq:decomp1}\\
\mathcal{L}_{\vec{Y}_{{\rm o}(1)}} N^{(0)}_A &\sim X_A^{1,\pm 1}; \label{eq:decomp2} \\
\mathcal{L}_{\vec{Y}_{{\rm e}(1)}} N^{(0)}_A &\sim X_A^{2, \pm 1},X_A^{2,0};\\
D_BY_{(1)}^B N^{(0)}_A &\sim Z_A^{1,\pm 1}, X_A^{2,\pm 1},X_A^{2,0} ;\label{eq:decomp4}\\
\Delta \hat{N}^{(1)}_A &= \sum_{\l=1}^\infty\sum_{\m=-l}^l \left(\Delta N^{{\rm e}(\l,\m)}_{(1)} Z_A^{\l\m}+\Delta N^{{\rm o}(\l,\m)}_{(1)}X_A^{\l\m}\right).\label{eq:decomp5}
\end{align}

From Eqs.~\eqref{eq:decomp1} to ~\eqref{eq:decomp5}, we can identify that the only pieces of \cref{eq:DeltaNhat} that are proportional to $X_A^{1,\pm 1}$ are Eqs.~\eqref{eq:decomp2} and~\eqref{eq:decomp5}; therefore,
we have
\begin{align}
    \Delta \hat{N}^{{\rm o}(1,\pm 1)}_{(1)}X_A^{1,\pm 1}&=\mathcal{L}_{\vec{Y}^{\ (1,\pm 1)}_{o(1)}} N^{(0)}_A,\\
    &=\pm  i\sqrt{\frac{3}{4\pi}}\kappa^{(1)}_{(1,\pm 1)}N^{(0)}  X^{1,\pm1}_A,\label{eq:Nodd}
\end{align}
where $N^{(0)}:=2\sqrt{3\pi}Ma$ arises from a vector harmonic decomposition of Eq.~\eqref{eq:KerrBS} (with $N^A_{(0)}=N^{(0)}X^A_{1,0}$). We solve \cref{eq:Nodd} for $\kappa^{(1)}_{(1,\pm 1)}$ and set $\Delta \hat{N}^{(1)}_{{\rm o}(1,\pm 1)}=-\hat{N}^{(1)}_{{\rm o}(1,\pm 1)}$, giving
\begin{align}\label{eq:BMSRotationconst}
\kappa^{(1)}_{(1,\pm 1)}= \pm \frac{i \hat{N}^{(1)}_{{\rm o}(1,\pm 1)}}{3 M a}.
\end{align}

Next, we fix the three spatial translations, which correspond to the $\l=1$ part of \cref{eq:decomp1}. For this, we need the $\propto Z_A^{1,\pm 1}$ piece of \cref{eq:decomp1,eq:decomp4}, which we determine to be
\begin{align}
    \Delta \hat{N}^{{\rm e}(1,\m )}_{(1)}Z_A^{1,\m}= 3M\alpha^{(1)}_{(1,\m)}D_AY_{1,\m} - 3Ma\m i \chi^{(1)}_{(1,\m)}D_A Y_{1,\m}.\label{eq:decompNhat}
\end{align}
Using \cref{eq:Z_A}, we can solve \cref{eq:decompNhat} for $\alpha^{(1)}_{(1,\m)}$, finding
\begin{align}
    \alpha^{(1)}_{(1,\m)}&=-\frac{1}{3M}\hat{N}^{{\rm e}(1,\m )}_{(1)}+a\m i \chi^{(1)}_{(1,\m)},\\
    &=\frac{1}{3M}\left(-\hat N^{{\rm e}(1,\m )}_{(1)}+a\m i  \MB[(1)]_{\!\!\!\!(1,\m)}\right),\label{eq:translation-transform}
\end{align}
where we set $\Delta N^{{\rm e}(1,\m )}_{(1)}= -N^{{\rm e}(1,\m )}_{(1)}$ and used \cref{eq:boost-transform} in the second line.

\subsection{Fixing to the odd-parity-shear frame}

Finally, we fix the supertranslations by eliminating the even-parity part of the shear at some chosen reference time $u_0$. To easily do so, we write Eq.~\eqref{C=Ce+Co} as
\begin{equation}
    C^{(1)}_{AB} =  D_{\langle A}D_{B\rangle}\Phi^{(1)}_{\rm e} + \epsilon_{C(A}D_{B)}D^C\Phi^{(1)}_{\rm o}.\label{C1=Phie+Phio}
\end{equation}
Referring to Eq.~\eqref{DeltaC1} and setting $\Delta C^{(1)}_{AB}=-C^{{\rm e}(1)}_{AB}$, we can immediately read off
\begin{equation}\label{eq:supertranslation-transform}
    \alpha^{(1)}_{(l,m)} = \frac{1}{2}\Phi^{(1)KN}_{{\rm e}(l,m)}[u_0] \quad (l\geq2),
\end{equation}
where we noted that $l<2$ modes are in the kernel of the operators in Eq.~\eqref{C1=Phie+Phio}. With this frame-fixing, the waveform has purely odd parity at time $u_0$.

Unlike for the $l<2$ modes of the BMS transformation, here we have had to choose a reference time, $u_0$, because $\Phi^{(1)}_{\rm e}$ evolves with $u$. The reference time could be, for example, the infinite past, the infinite future, or the time at which the waveform attains some reference frequency. This might seem disadvantageous compared to setting the Moreschi supermomentum~\eqref{eq:MoreschiSupermomentum} to zero, for example, since Eq.~\eqref{PMdot} implies ${\cal P}^{(1)}_{M}$ is constant. However, a condition on the shear is by far the most easily imposed in practice, since one might have access to the waveform without having access to the higher moments of the mass aspect. Moreover, in Sec.~\ref{sec:BS gauge in SF theory} we will find that a reference time is also required for the $l<2$ modes when we extend our method to self-forced inspirals. We return to this question of how best to fix the supertranslations in Sec.~\ref{sec:CompareNR}.

\subsection{Summary and discussion}\label{sec:BMS fixing summary}

In summary, the transformation from a generic asymptotically flat gauge to a maximally fixed BS gauge is given by Eqs.~\eqref{eq:xiu3}--\eqref{eq:xir3} with the BMS generator
\begin{equation}
\xi^\alpha_\circl = \left(\alpha^{(1)}+\frac{1}{2}uD_AY^A_{(1)}\right)\partial_u + Y^A_{(1)}\partial_A 
\end{equation}
now fixed by Eqs.~\eqref{eq:translation-transform} (for the $l=1$ piece of $\alpha^{(1)}$), \eqref{eq:supertranslation-transform} (for the $l\geq2$ pieces of $\alpha^{(1)}$), \eqref{eq:boost-transform} (for the even-parity piece of $Y^A_{(1)}$, and \eqref{eq:BMSRotationconst} (for the odd-parity piece of $Y^A_{(1)}$).
%
%
Concretely,
\begin{align}
    \xi^\bsu_\circl[u_0,\bsu,\hat{\theta}^A]
    &= \alpha^{(1)}_{(0,0)}Y_{00} + \frac{1}{3M}\sum_{m=-1}^{+1}\left\{-\hat{N}^{{\rm e}(1,\m)}_{(1)} + (a\m i  -  u)\MB[(1)]_{\!\!\!\!(1,\m)}\right\}Y_{1,\m}[\hat{\theta}^A]\notag\\
    &\quad +\frac{1}{2}\sum_{l\geq2}^\infty\Phi^{(1)}_{{\rm e}(l,m)}[u_0]Y_{\ell m}[\bst^A],\label{eq:xiuBMS}
\end{align}
and
\begin{align}
    \xi^A_\circl [u_0,\hat{\theta}^A]
    &= \kappa^{(1)}_{(1,0)} X^A_{1,0}[\hat{\theta}^B]+ \frac{1}{3M}\sum_{m=-1}^1\left\{ \MB[(1)]_{\!\!\!\!(1,\m)}Z^A_{1,\m}[\hat{\theta}^B] + \frac{i \m}{a}\hat{N}^{(1)}_{{\rm o}(1, \m)} X^A_{1,\m}[\hat{\theta}^B]\right\}
    \label{eq:xiABMS}
\end{align}
in terms of the $l=1$ multipoles of the mass and angular momentum aspects, the even-parity part of the shear~\eqref{C1=Phie+Phio}, and the vector harmonics $Z^A_{lm}$ and $Z^A_{lm}$ defined in Eqs.~\eqref{eq:Z_A} and~\eqref{eq:X_A}. As outlined in Sec.~\ref{sec:BMS fixing strategy}, this transformation (i) puts the system in the centre-of-mass frame by setting the linear momentum $P^{(1)}_i$ and boost charge $K^{(1)}_i$ to zero (and therefore also the centre of mass $G_i^{(1)} = K_i^{(1)} + u P^{(1)}_i$), (ii) aligns the $z$ axis with the total angular momentum, and (iii) eliminates the even-parity part of the shear $C^{(1)}_{AB}$ at a reference time $u_0$.

In Eqs.~\eqref{eq:xiuBMS} and \eqref{eq:xiABMS}, the time translation $\alpha^{(1)}_{(0,0)}Y_{00}$, the axial rotation $\kappa^{(1)}_{(1,0)} X^A_{1,0}$, and the reference time $u_0$ are all arbitrary. The time translation and axial rotation cannot be fixed through conditions on the first-order metric perturbation because they are isometries of the background Kerr spacetime. In some scenarios with nonstationary, non-axially-symmetric matter distributions, one could fix this remaining freedom by imposing conditions on the stress-energy tensor at second-order, $T_{\mu\nu}^{(2)}$, which would nontrivially transforms under these background Killing symmetries:
\begin{align}
    \Delta T_{\mu\nu}^{(2)}=\Lie_{\vec \xi}\, T_{\mu\nu}^{(1)}.
\end{align}
However, in practice, the remaining degrees of freedom are treated as physically irrelevant. When comparing two waveforms, one fixes the time and axial shifts by aligning the time and phase of the two waveforms at some reference criterion (e.g., at the peak of the waveform or when the waveform reaches a particular reference frequency)~\cite{mitman2024review}. Similarly, $u_0$ can be chosen to be the time at that reference criterion. Alternatively, one could choose $u_0$ to be the infinite future, putting the system in the super-rest frame of the final black hole, for example.

A subtler issue is that our fixing of the BMS frame has made use of the fact that all the first-order Poincaré charges are constant in time (in any BMS frame). This feature of regular perturbation theory does not hold in an inspiral scenario, where the duration of the inspiral scales as $1/\varepsilon$. Once the charges evolve, the BMS frame alignment can only fix their values at a single reference time $u_0$, just as we have done here for the shear. We explore this in the next section.

\section{Bondi--Sachs gauge in self-force theory}\label{sec:BS gauge in SF theory}

In this section, we outline the application of our results to the problem of asymmetric-mass binaries modelled using self-force theory. As mentioned in the Introduction and Sec.~\ref{sec:second-order}, this problem requires an enlarged class of slowly evolving BMS-like transformations that relate ``forgetful'' gauges to BS gauges. Such transformations, we show, introduce slowly evolving soft hair on the two-body system---additional degrees of freedom beyond the system's slowly evolving masses, spins, and orbital parameters.

We review the multiscale expansion in Sec.~\ref{sec:multiscale GSF} and describe how typical first-order gauge choices ``forget'' memory in Sec.~\ref{sec:forgetful gauges}. This is a symptom of the general breakdown of the multiscale expansion at very large distances in these gauges. In Sec.~\ref{sec:iterative transformation}, we outline the iterative transformation scheme suggested by our earlier Ref.~\cite{spiers2023second}, which bypasses the complicated SF-MPM matching procedure from Refs.~\cite{PoundLargeScales,Cunningham:2024dog} and eliminates the distinction between near and far zones. In Sec.~\ref{sec:multiscale BMS} we show how the transformations in this iterative scheme, from forgetful gauges to the BS gauge, extends the BMS frame-fixing from the previous section and requires the enlarged class of transformations mentioned above. In Sec.~\ref{sec:memory in SF}, we describe how these transformations naturally introduce both memory (soft hair) and memory distortion (the interaction of this hair with oscillatory modes).

In this section, we consistently include ``$(0)$'', ``(1)'', and ``(2)'' labels to indicate coefficients of powers of $\varepsilon$.

\subsection{Self-force theory in the multiscale framework}\label{sec:multiscale GSF}

Self-force calculations and waveforms are typically formulated in a multiscale expansion that differs significantly from regular perturbation theory~\cite{hindereretal2008,Miller-Pound:2020,Pound-Wardell:2021,Mathews:2025nyb,Lewis:2025ydo}.  

Suppose we work in some coordinates $x^\mu=(u,x^i)$ in a neighbourhood of $\mathscr{I}^+$. In regular perturbation theory, the metric in these coordinates is expanded in the form $g^{(0)}_{\alpha\beta}[x^i] + \sum_{n\geq1} \varepsilon^n h^{(n)}_{\alpha\beta}[u,x^i]$, where we recall that $\varepsilon$ denotes the mass ratio in this context. In the multiscale expansion, on the other hand, each coefficient $h^{(n)}_{\alpha\beta}$ depends on $\varepsilon$ in addition to $x^\mu$, and the dependence on time $u$ takes a very particular form: all time dependence is encoded in a dependence on the mechanical variables describing the binary, and these mechanical variables are treated as independent of one another.

This more intricate expansion is motivated by the quasi-periodicity of the problem. In an asymmetric-mass binary, the smaller, secondary object executes approximately tri-periodic orbits around the larger, primary black hole, with independent frequencies $\Omega^i=(\Omega^r,\Omega^\theta,\Omega^\phi)$ of radial, polar, and azimuthal motion~\cite{Schmidt:2002qk}. Each frequency has an associated phase $\varphi^i$ satisfying $d\varphi^i/dt=\Omega^i$, which evolves by $2\pi$ in a given radial, polar, or azimuthal period. The frequencies are determined by the mass and spin of the primary black hole and by orbital variables $\pi_i$ that are constants on geodesic orbits: energy $E$, azimuthal angular momentum $L_z$, and Carter constant $K$; or quasi-Keplerian parameters semi-latus rectum $p$, eccentricity $e$, and inclination $\iota$~\cite{Schmidt:2002qk,Pound-Wardell:2021}. Due to GW emission and the self-force acting on the secondary, the constants of geodesic motion (as well as the mass and spin of the primary) slowly evolve, on a long time scale of order $1/\varepsilon$. 

The spacetime metric, via the EFE, inherits this structure of periodic phases and slowly evolving frequencies. More concretely, we first foliate the spacetime with slices of constant time $s$, where $s$ reduces to advanced time $v$ in a region extending to the primary's horizon, $t$ in a region including the secondary's position, and $u$ in a region extending to $\mathscr{I}^+$. With the metric written as $g_{\alpha\beta}+h_{\alpha\beta}$, the metric perturbation then takes the form~\cite{Lewis:2025ydo}
\begin{equation}\label{multiscale h}
h_{\alpha\beta} = \sum_{n\geq1}\varepsilon^n h^{(n)}_{\alpha\beta}[\varphi^i,\pi_i,M_K,x^i],   
\end{equation}
where $h^{(n)}_{\alpha\beta}$ is $2\pi$-periodic in each of the phases, and all dependence on time $s$ is encoded in the dependence on the orbital phase-space variables $(\varphi^i,\pi_i)$ and the primary's mass and spin $M_K=(M,J)$. The orbital variables $(\varphi^i,\pi_i)$ are extended off the secondary's worldline, and $M_K$ is extended off the primary's horizon, as constants on each slice of constant $s$. $h^{(n)}_{\alpha\beta}$ is then treated as a function on the 11-dimensional space with coordinates $(\varphi^i,\pi_i,M_K,x^i)$.

In this evolving system, we can choose $(\varphi^i,\pi_i,M_K)$ such that they satisfy structurally simple ordinary differential equations~\cite{Pound-Wardell:2021,Mathews:2025nyb,Lewis:2025ydo} (see also~\cite{VanDeMeent:2018cgn,Lynch:2023gpu}),
\begin{align}
    \frac{d\varphi^i}{dt} &= \Omega^i[\pi_j,M_K],\label{eq:phidot}\\
    \frac{d\pi_i}{dt} &= \varepsilon \left(F^{(0)}_i[\pi_j,M_K]+\varepsilon F^{(1)}_i[\pi_j,M_K]+{\cal O}(\varepsilon^2)\right)\!,\label{eq:pidot}
\end{align}
along with 
\begin{equation}\label{eq:MKdot}
\frac{dM_K}{dt} = \varepsilon^2{\cal F}^{(1)}_K[\pi_i,M_L] + {\cal O}(\varepsilon^3),
\end{equation}
where ${\cal F}^{(1)}_K$ are the flux of energy and (azimuthal) angular momentum carried by $h^{(1)}_{\alpha\beta}$ down the primary's horizon. 
An approximation that only includes $F^{(0)}_i$ is referred to as adiabatic (0PA); one that includes also $F^{(1)}_i$ is referred to as first post-adiabatic order (1PA). Here we work in a ``fixed frequencies'' gauge~\cite{Pound-Wardell:2021,Mathews:2025nyb,Lewis:2025ydo}, in which $(\pi_j,M_K)$ are related to the orbital frequencies by the Kerr-geodesic relationships, such that Eq.~\eqref{eq:phidot} contains no ${\cal O}(\varepsilon^n)$ corrections.

Rather than working with a slowly evolving background metric, it will be convenient to write $M_K$ as 
\begin{equation}
M_K = M_K^{(0)} + \varepsilon \delta M_K,
\end{equation}
where $M^{(0)}_K=(M^{(0)},J^{(0)})$ are constants and the slowly evolving corrections satisfy
\begin{equation}\label{eq:dMKdot}
\frac{d\delta M_K}{dt} = \varepsilon{\cal F}^{(1)}_K[\pi_i,M^{(0)}_L] + {\cal O}(\varepsilon^2).
\end{equation}
We can then work with a vacuum background Kerr metric with parameters $M^{(0)}_K$; the corrections $\delta M_K$ enter the perturbations $h^{(n)}_{\alpha\beta}$ beginning at $n=1$, and in the 1PA (and higher) forcing functions $F^{(1)}_i$.

For our purposes, the important consequences of all this are that in the region extending to $\mathscr{I}^+$, the metric is $2\pi$-periodic in the phase variables $\varphi^i(u,\varepsilon)$ and depends on slowly evolving variables $\pi_i(u,\varepsilon)$ and $\delta M_K(u,\varepsilon)$; here we recall that since the mechanical variables are constant on slices of constant $s$, they become functions of $u$ when evaluated near $\mathscr{I}^+$. When a time derivative acts on a metric perturbation, we apply the chain rule:
\begin{equation}
    \partial_u = \Omega^i\partial_{\varphi^i} + \varepsilon \left(F^{(0)}_i\partial_{\pi_i}  +{\cal F}^{(1)}_K\partial_{\delta M_K} \right) + {\cal O}(\varepsilon^2).
\end{equation}
In what follows, to keep expressions compact, we write functions of $(\pi_i,M_K)$ as functions of  ``slow time'' $\tilde u = \varepsilon u$. This is motivated by the fact that solutions to Eqs.~\eqref{eq:pidot} and \eqref{eq:dMKdot} can be written as $\pi_i = \pi^{(0)}_i(\varepsilon t) + \varepsilon\pi^{(1)}_i(\varepsilon t) + \ldots$ and $\delta M_K = M^{(1)}_K(\varepsilon t) + \varepsilon M^{(2)}_K(\varepsilon t)+\ldots$, for example. We then apply the chain rule
\begin{equation}\label{eq:chain rule simple}
    \partial_u = \Omega^i\partial_{\varphi^i} + \varepsilon\partial_{\tilde u} + {\cal O}(\varepsilon^2).
\end{equation}

We will use an overdot to denote $\Omega^i\partial_{\varphi^i}$, as in $\MBdot := \Omega^i\partial_{\varphi^i}\MB$. Angular brackets denote the average over the 3-torus with angles $\varphi^i$,  $\langle \cdot\rangle := \frac{1}{(2\pi)^3}\oint \cdot \,d^3\varphi$. An important consequence of these definitions is that
\begin{equation}\label{<fdot>}
    \langle \dot f \rangle = 0 \qquad\text{and}\qquad \langle \partial_{\tilde u} f \rangle =  \partial_{\tilde u}\langle  f \rangle
\end{equation}
for all functions $f$ that are $2\pi$-periodic in $\varphi^i$.

\subsection{BMS charges in the multiscale expansion: forgetful gauges and soft hair}\label{sec:forgetful gauges}

We now consider the following problem. Suppose we solve the linearized EFE for $h^{(1)}_{\alpha\beta}$, with binary parameters $(\pi_i,M_K)$, using any of the usual gauges of black hole perturbation theory: the Lorenz gauge, radiation gauge, or Regge-Wheeler-Zerilli gauge. Does memory content enter this $h^{(1)}_{\alpha\beta}$? The answer, which we know from SF-MPM calculations, is that in traditional self-force gauges, nonlinear memory does enter $h^{(1)}_{\alpha\beta}$, but \emph{only in the far zone}, at distances $r\gtrsim M/\varepsilon$; traditional self-force calculations are confined to the near zone, $r\ll M/\varepsilon$, and in this zone they find no sign of memory in $h^{(1)}_{\alpha\beta}$. (Note these statements are insensitive to whether we use Boyer--Lindquist $r$ or BS $\hat r$.) The gauges used in these calculations are the \emph{forgetful gauges} we alluded to in the Introduction. 

To understand how and why memory enters only in the far zone in these calculations, consider a gauge (such as Lorenz) in which the linearized EFE reduces to the flat-spacetime wave equation at large distances. At leading order in the multiscale expansion and at large distances (but still much smaller than $M/\varepsilon$), the linearized EFE then takes the form
\begin{equation}
    -2(\partial_{\bsr}+1/\bsr)\dot h^{(1)}_{\alpha\beta} + \nabla^2 h^{(1)}_{\alpha\beta} = 0
\end{equation}
in Cartesian coordinates $(u,\hat x^i)$ with $\bsr=\sqrt{\delta_{ij}\hat x^i\hat x^j}$. Here $\nabla^2$ denotes the flat-space Laplacian. We can pick out the quasistationary (i.e., $\varphi^i$-independent) part of this equation by averaging over the torus, which yields
\begin{equation}
    \nabla^2 \bigl\langle h^{(1)}_{\alpha\beta}\bigr\rangle = 0
\end{equation}
by virtue of Eq.~\eqref{<fdot>}. If we decompose into spherical harmonics, the asymptotically flat solution to this Poisson equation decays as $1/\bsr^{l+1}$, as in the familiar setting of electrostatics. Hence, the monopole is the only term in the solution that decays like $1/\hat r$; all modes with $l>0$ decay more rapidly. This implies, in particular, that there is no gravitational-wave content (which begins at $l=2$) in $\bigl\langle h^{(1)}_{\alpha\beta}\bigr\rangle$. In other words, this is a gauge in which there is no quasistationary shear $\bigl\langle C^{(1)}_{AB}\bigr\rangle$. The same is true in all gauges used in first-order self-force calculations (and we note again that the analysis is identical regardless of the choice of background coordinates, though we have adopted BS coordinates for concreteness).

Now suppose that, in the far zone outside the binary, we transition to the BS gauge at all orders. The relationships~\eqref{eq:MBdot} and \eqref{eq:Ndoteven} then hold fully nonlinearly, and we can straightforwardly extend the analysis in Eqs.~\eqref{eq:MBdot1 regular}--\eqref{constant Poincare} to the multiscale expansion. If we expand the quantities in this equation in powers of $\varepsilon$ at fixed $(\tilde u,\varphi^i,\bst^A)$, then at first order the relationships~\eqref{eq:MBdot} and \eqref{eq:Ndot} reduce to analogues of Eqs.~\eqref{eq:MBdot1 regular}--\eqref{eq:Ndot1 odd regular}:
\begin{align}
    \MBdot[(1)] &= \frac{1}{4}D_A D_B\dot C_{(1)}^{AB},\label{eq:MBdot1}\\
    \dot N^{{\rm e}(1)}_A &= -\frac{1}{4}D_AD_B D_C \dot C_{(1)}^{BC},\\
    \dot {\hat N}^{{\rm o}(1)}_A &= \frac{1}{4}D_BD_AD_C C_{(1)}^{BC} - \frac{1}{4}D_BD^BD^CC^{(1)}_{CA}.
\end{align}
The $l<2$ parts of the first equation imply the generalization of Eq.~\eqref{constant Poincare}:
\begin{equation}
    \dot M^{(1)} = \dot P_i^{(1)} = \dot J_i^{(1)} = \dot K_i^{(1)} = 0.
\end{equation}
However, this no longer implies the charges are constants; instead, it implies they are functions of $\tilde u$ only. But we can get additional information from Eq.~\eqref{eq:Ndoteven}, which reduces to
\begin{equation}
    D^A\dot N^{{\rm e}(1)}_A = D^AD_A \MB[(1)].
\end{equation}
Averaging this equation and appealing to Eq.~\eqref{<fdot>} yields
\begin{equation}
   D^AD_A \bigl\langle\MB[(1)]\bigr\rangle=0.
\end{equation}
This implies that $\bigl\langle\MB[(1)]\bigr\rangle$ is purely monopolar. In other words, the averaged mass aspect is equal to the averaged mass:
\begin{equation}\label{<MB>=<M>}
\bigl\langle\MB[(1)]\bigr\rangle = \bigl\langle M^{(1)}\bigr\rangle.
\end{equation}
As a consequence, the averaged linear momentum vanishes, and since $\dot P^{(1)}_i=0$, we conclude
\begin{equation}
    P_i^{(1)} = 0.\label{P1=0}
\end{equation}

The result~\eqref{<MB>=<M>} is in agreement with the near-zone analysis above, where we determined that the only quasistationary content of the $1/\bsr$ term in the near-zone perturbation $h^{(1)}_{\alpha\beta}$ is a monopole. But an essential difference from the near-zone analysis, leading to the appearance of memory, arises when we consider the second-order terms in the relationship~\eqref{eq:MBdot}:
\begin{align}\label{MBdot 2}
    \MBdot[(2)] + \partial_{\tilde u}\MB[(1)] &= -\frac{1}{8}\dot C^{(1)}_{AB}\dot C_{(1)}^{AB} + \frac{1}{4}D_A D_B \partial_{\tilde u} C_{(1)}^{AB} + \frac{1}{4}D_A D_B\dot C_{(2)}^{AB}.
\end{align}
Here we stress the appearance of slow-time derivatives of first-order quantities, which enter the second-order equations due to the $\varepsilon$ suppression in Eq.~\eqref{eq:chain rule simple}. Averaging Eq.~\eqref{MBdot 2} over $\varphi^i$ and appealing to Eqs.~\eqref{<fdot>} and \eqref{<MB>=<M>}, we find
\begin{equation}\label{<dMB1>}
    \partial_{\tilde u}\bigl\langle M^{(1)}\bigr\rangle  = -\frac{1}{8}\bigl\langle\dot C^{(1)}_{AB}\dot C_{(1)}^{AB}\bigr\rangle + \frac{1}{4}D_A D_B \partial_{\tilde u}\bigl\langle C_{(1)}^{AB}\bigr\rangle.
\end{equation}
The linear term on the right-hand side is the ``soft'' term that contributes to $\langle M^{(1)}\rangle$ in the zero-frequency limit. Further averaging over a sphere, and noting that $C_{(1)}^{AB}$ only contains $l>1$ modes, we see Eq.~\eqref{<dMB1>} implies the standard averaged mass-loss formula:
\begin{equation}
    \partial_{\tilde u}\bigl\langle M^{(1)}\bigr\rangle = -\frac{1}{32\pi}\int d\Omega\,\bigl\langle\dot C^{(1)}_{AB}\dot C_{(1)}^{AB}\bigr\rangle.
\end{equation}
All of the non-spherically-symmetric content of the energy flux $\bigl\langle\dot C^{(1)}_{AB}\dot C_{(1)}^{AB}\bigr\rangle$ in Eq.~\eqref{MBdot 2} must then be cancelled by $\partial_{\tilde u}\bigl\langle C_{(1)}^{AB}\bigr\rangle$. Since the shear only contains $l\geq2$ modes, we conclude
\begin{equation}\label{eq:memory in h1}
   D_A D_B \bigl\langle C_{(1)}^{AB} \bigr\rangle = \frac{1}{2}\int d\tilde u\,\bigl\langle \dot C^{(1)}_{AB}\dot C_{(1)}^{AB}\bigr\rangle_{l\geq2}.
\end{equation}
This is consistent with the usual definition of nonlinear memory~\cite{blanchet1992hereditary,favata2010gravitational,Frauendiener:1992dmu,mitman2024review,Cunningham:2024dog}:
\begin{equation}\label{eq:nonlinear memory}
   D_A D_B C_{\rm mem}^{AB} = \frac{1}{2}\int du\, (N_{AB}N^{AB})_{l\geq2}.
\end{equation}  
Similar calculations can be carried out to determine the slow evolution of $K^{(1)}_i$ and $J^{(1)}_i$, but we defer this to future work.

From this analysis, we see that a quasistationary, slowly evolving shear $\bigl\langle C_{(1)}^{AB} \bigr\rangle$---a nonlinear memory in the first-order metric perturbation---necessarily arises at $\mathscr{I}^+$. It is a strictly nonlinear, quadratic effect, but it appears in the first-order perturbation due to the long time scale $M/\varepsilon$ over which it accumulates. In a forgetful gauge, this memory is forgotten in the near zone. 

The presence of a nonzero $\bigl\langle C_{(1)}^{AB} \bigr\rangle$ complicates the typical description of the multiscale expansion. Traditionally, one thinks of the state of the system as being fully described by the parameters $(\pi_i,M_{K})$, and the system slowly evolves between those states. When solving the first-order field equations for $h^{(1)}_{\alpha\beta}$ in the near zone, the solution (in a given gauge) is almost fully determined by (i) periodicity in $\varphi^i$ and (ii) regularity conditions at infinity and at the black hole's horizon (implying no incoming radiation from $\mathscr{I}^-$ and $\mathscr{H}^-$). The only freedom in the solution is the freedom to add ``trivial perturbations'' corresponding to corrections $\delta M_K$ to the background black hole's mass and spin $M^{(0)}_K$, as these are the only homogeneous solutions that satisfy the regularity and periodicity conditions.\footnote{Quasinormal mode perturbations would satisfy the regularity conditions but not the periodicity condition.} No other quasistationary perturbations are allowed because they cannot be both asymptotically flat and regular at the horizon; this is a statement of the no-hair theorem. But this is an artefact of working in the near zone. Once we move further from the binary, toward the true geometrical $\mathscr{I}^+$ (as opposed to the large-$\bsr$ limit of the near zone), infinitely many additional degrees of freedom necessarily appear in $\bigl\langle C_{(1)}^{AB} \bigr\rangle$, which evolve independently of $(\pi_i,M_{K})$ and are required in order to fully specify the state of the system. These additional degrees of freedom can be interpreted as soft hair.

\subsection{Iterative procedure for solving the multiscale field equations}\label{sec:iterative transformation}

Our analysis in the preceding section might seem to suggest that memory simply does not arise in the forgetful gauges used in typical self-force calculations. This is misleading. The correct statement is that in these gauges, memory does not appear \emph{in the near zone}. The SF-MPM analysis in Ref.~\cite{Cunningham:2024dog} shows that in the Lorenz gauge, for example, memory does arise in the far zone and recovers Eq.~\eqref{eq:memory in h1} (as was the case in the very first derivation of nonlinear memory, which was done in the MPM framework~\cite{Blanchet:1990twn}). In this sense, the important feature of forgetful gauges is that they require such a split into near and far zones; the emergence of memory is, in fact, closely related to the breakdown of the near-zone multiscale expansion in these gauges, a fact we review below. In this section, we outline our iterative scheme that avoids this breakdown. In the following sections we show how, by eliminating the division between near and far zones, this scheme extends the first-order memory into the near zone.




We first lay out the multiscale expansion of the EFE. In place of the regular expansion~\eqref{G expansion}, the Einstein tensor of the total metric $g^{(0)}_{\alpha\beta}+h_{\alpha\beta}$  is now expanded as 
\begin{multline}
   G_{\alpha\beta}[g^{(0)}+h] = G_{\alpha\beta}[g^{(0)}] + \e G^{(1,0)}_{\alpha\beta}[h^{(1)}] \\
   + \e^2\! \left(G^{(1,0)}_{\alpha\beta}[h^{(2)}] 
   + G^{(2,0)}_{\alpha\beta}[h^{(1)},h^{(1)}] +  G^{(1,1)}_{\alpha\beta}[h^{(1)}]\right)  + {\cal O}(\e^3),\label{G multiscale}
\end{multline}
where $(n,0)$ indicates that we omit $\partial_{\tilde u}$ derivatives, and $(n,1)$ indicates the first subleading correction from $\partial_{\tilde u}$ derivatives. If we expand the stress-energy tensor in the same way and equate coefficients of powers of $\varepsilon$, then the EFE reduces to the hierarchy of linear equations
\begin{align}
    G^{(1,0)}_{\alpha\beta}[h^{(1)}] &= 8\pi T^{(1)}_{\alpha\beta},\label{multiscale EFE1}\\
    G^{(1,0)}_{\alpha\beta}[h^{(2)}] &= 8\pi T^{(2)}_{\alpha\beta}- G^{(2,0)}_{\alpha\beta}[h^{(1)},h^{(1)}]- G^{(1,1)}_{\alpha\beta}[h^{(1)}].\label{multiscale EFE2}
\end{align}
Corresponding to these EFEs, there are associated Teukolsky equations, 
\begin{align}
   {\cal O}^{(0)}[\psi^{(1)}_{4}] &= 8\pi {\cal S}^{(0)}\!\bigl[T^{(1)}_{\alpha\beta}\bigr],\label{multiscale Teukolsky1}\\
   {\cal O}^{(0)}[\psi^{(2)}_{4L}] &= 8\pi {\cal S}^{(0)}\!\bigl[T^{(2)}_{\alpha\beta}\bigr] - {\cal S}^{(0)}\!\Bigl[G^{(2,0)}_{\alpha\beta}[h^{(1)},h^{(1)}]\Bigr] - {\cal S}^{(0)}\!\Bigl[G^{(1,1)}_{\alpha\beta}[h^{(1)}]\Bigr].\label{multiscale Teukolsky2}
\end{align}
On their right-hand sides, Eqs.~\eqref{multiscale EFE2} and~\eqref{multiscale Teukolsky2} differ from the ordinary second-order equations~\eqref{EFE2} and \eqref{eq:pure-2nd-Teuk} through their addition of a source term, $G^{(1,1)}_{\alpha\beta}$ or ${\cal S}^{(0)}\!\bigl[G^{(1,1)}_{\alpha\beta}[h^{(1)}]\bigr]$, generated by the slow evolution of $h^{(1)}_{\alpha\beta}$. On the left-hand side, the equations differ from the ordinary Einstein and Teukolsky equations through their neglect of $\tilde u$ derivatives.

The above hierarchy of equations hold true in any gauge that is compatible with the multiscale expansion (which excludes gauge perturbations with polynomial time dependence, for example, that grow large on the radiation-reaction time scale). However, in forgetful gauges they are valid \emph{only in the near zone}. Their breakdown in the far zone is best understood for the case of the Lorenz gauge, in which $G^{(1,1)}_{\alpha\beta}$ decays as $1/\bsr^3$ at large $\bsr$, while $G^{(2,0)}_{\alpha\beta}$ decays as $1/\bsr^2$~\cite{Miller-Pound:2020,Cunningham:2024dog}. If we define a retarded Green's function $G_{\alpha\beta\mu'\nu'}(x,x')$ for the (Lorenz-gauge) operator $G^{(1,0)}_{\alpha\beta}$, then the integral of $G_{\alpha\beta}{}^{\mu'\nu'}$ against $G^{(2,0)}_{\mu'\nu'}$ diverges due to the slow, $1/\bsr^2$ falloff, as in Sec.~\ref{sec:goodgaugesEMRIs}; see Sec.~5.1 of Ref.~\cite{Cunningham:2024dog}. This divergence is what necessitates a separate (SF-MPM) far-zone treatment.

More fundamentally, the breakdown occurs because of how the multiscale expansion treats the spacetime's slow evolution. In arriving at the hierarchy of equations~\eqref{multiscale EFE1} and~\eqref{multiscale EFE2}, we treated $(\varphi^i,\pi_i,M_K)$ [or simply $(\varphi^i,\tilde u)$ here] as independent variables. That is what justified splitting the EFE according to explicit powers of $\varepsilon$ in Eq.~\eqref{G multiscale}; otherwise such splitting would not be implied, since the coefficients themselves depend on $\varepsilon$ when treated as functions of $(u,\varepsilon)$ rather than as functions of $(\varphi^i,\tilde u)$. Consequently, we treat $(\varphi^i,\tilde u)$ as independent when actually solving the hierarchy of equations. Because there are no $\tilde u$ derivatives on the left-hand side of the equations, this means each equation is solved at a fixed value of $\tilde u$ (i.e., for a ``frozen'' set of binary parameters $\pi_i$ and $M_K$). In other words, when we integrate against a Green's function, we only integrate over $(\varphi^i,x^i)$; the source terms do not slowly evolve inside the integral. While this formulation is valid in the near zone, it breaks down in the far zone, where the evolution on large scales cannot be neglected inside the integrals. The breakdown is curable by matching to an MPM solution because the SF-MPM scheme does not treat the variables as independent and does not freeze $\tilde u$ within the retarded integrals. 


The divergent near-zone integrals are the same ones that give rise to nonlinear memory in the SF-MPM solution. Evaluating those integrals in the SF-MPM scheme and then matching them to the near zone ends up introducing memory into the near zone, but only in the near-zone $h^{(2)}_{\mu\nu}$, not in the in the near-zone $h^{(1)}_{\mu\nu}$, as detailed in Ref.~\cite{Cunningham:2024dog}. This is analogous to how memory effects in PN theory enter the waveform at 0PN~\cite{favata2010gravitational} but are suppressed to 5PN in the near zone~\cite{Porto:2024cwd}. 

All of this contrasts with the behaviour of the multiscale expansion in the BS gauge. In the BS gauge, the second-order source decays rapidly enough to avert divergent integrals in the near zone. An intuitive explanation for this is that the BS gauge ensures the multiscale expansion's time coordinate $s$ reduces to the retarded time in the perturbed spacetime rather than retarded time in the background spacetime. In the near zone, the genuine light cones cannot differ substantially from the background light cones, but the two will differ by $\order{\varepsilon^0}$ in the far zone. By ensuring that the true light cones extend all the way from the near zone to $\mathscr{I}^+$, the BS gauge prevents this breakdown, making the same multiscale expansion valid in both the near and far zones.

Hence, to bypass the need for an SF-MPM expansion, we propose the following scheme. We first solve Eq.~\eqref{multiscale EFE1} in any desired gauge (e.g., the Lorenz gauge). Before proceeding to second order, we transform the solution to the BS gauge, using the transformation detailed in the previous sections:
\begin{equation}
    \hat h^{(1)}_{\alpha\beta} =  h^{(1)}_{\alpha\beta} + {\Lie}^{(0)}_{\vec\xi_{(1)}}g_{\alpha\beta}.
\end{equation}
Here we do not require $\hat h^{(1)}_{\alpha\beta}$ to be in the BS gauge everywhere, nor even in an open neighbourhood of $\mathscr{I}^+$. Instead, we only require the BS gauge conditions to be satisfied to appropriate orders in $1/r$, as explained previously. The second-order field equation~\eqref{multiscale EFE2} then becomes
\begin{equation}\label{multiscale EFE2 BS}
    G^{(1,0)}_{\alpha\beta}[h^{(2)}] = 8\pi T^{(2)}_{\alpha\beta} - G^{(2,0)}_{\alpha\beta}[\hat h^{(1)},\hat h^{(1)}]- G^{(1,1)}_{\alpha\beta}[\hat h^{(1)}].
\end{equation}
Here $h^{(2)}_{\alpha\beta}$ can be in any convenient gauge. The essential feature of this equation, pointed out in Ref.~\cite{spiers2023second}, is that the source now exhibits specific falloff properties: 
\begin{align}
    G^{(2,0)}_{nn}[\hat h^{(1)},\hat h^{(1)}] &={\cal O}(r^{-2}), \\ 
    G^{(2,0)}_{n\mb}[\hat h^{(1)},\hat h^{(1)}] &={\cal O}(r^{-3}), \\ 
    G^{(2,0)}_{\mb\mb}[\hat h^{(1)},\hat h^{(1)}] &={\cal O}(r^{-4}).
\end{align}
As Ref.~\cite{spiers2023second} explained, these properties ensure sufficiently rapid falloff of the source in the associated Teukolsky equation,
\begin{equation}\label{multiscale Teukolsky2 BS}
   {\cal O}^{(0)}[\psi^{(2)}_{4L}] = 8\pi T^{(2)}_{\alpha\beta} -{\cal S}^{(0)}\!\bigl[G^{(2,0)}_{\alpha\beta}[\hat h^{(1)},\hat h^{(1)}]\bigr] - {\cal S}^{(0)}\!\bigl[G^{(1,1)}_{\alpha\beta}[\hat h^{(1)}]\bigr].
\end{equation}
Specifically,
\begin{equation}
    {\cal S}^{(0)}\!\bigl[G^{(2,0)}_{\alpha\beta}[\hat h^{(1)},\hat h^{(1)}]\bigr] = {\cal O}(r^{-4});
\end{equation}
integrating this source against the retarded Green's function for the Teukolsky equation yields a well-behaved $\psi^{(2)}_{4L}$. This contrasts with the behaviour in the Lorenz gauge, in which
\begin{equation}
    {\cal S}^{(0)}\!\bigl[G^{(2,0)}_{\alpha\beta}[h^{(1)},h^{(1)}]\bigr] = {\cal O}(r^{-2})
\end{equation}
and the retarded integral diverges. With the well-behaved BS-gauge source, one can solve for $\psi^{(2)}_{4L}$ and, if needed, reconstruct $h^{(2)}_{\alpha\beta}$ in a convenient gauge.

In summary, the iterative scheme is the following:
\begin{enumerate}
    \item Solve Eq.~\eqref{multiscale EFE1} for $h^{(1)}_{\alpha\beta}$ in a convenient gauge, or solve the corresponding Teukolsky equation~\eqref{multiscale Teukolsky1} and reconstruct $h^{(1)}_{\alpha\beta}$.
    \item Transform $h^{(1)}_{\alpha\beta}$ to the Bondi-Sachs gauge: $h^{(1)}_{\alpha\beta}\to \hat h^{(1)}_{\alpha\beta}$. 
    \item Solve Eq.~\eqref{multiscale EFE2 BS} for $h^{(2)}_{\alpha\beta}$ in a convenient gauge, or solve the corresponding Teukolsky equation~\eqref{multiscale Teukolsky2 BS} and reconstruct $h^{(2)}_{\alpha\beta}$.
    \item Transform $h^{(2)}_{\alpha\beta}$ to the Bondi-Sachs gauge.
    \item Continue on to higher orders.
\end{enumerate}
In many cases, one can halt this scheme after the calculation of the highest-order Teukolsky scalar of interest, without performing a subsequent metric reconstruction and transformation to the BS gauge. For example, to compute the 1SF inputs for 0PA waveform generation, one only requires $\psi^{(1)}_4$, not the whole of $h^{(1)}_{\alpha\beta}$; to compute the 2SF inputs for 1PA waveform generation, one likely only requires $\psi^{(2)}_{4L}$, not the whole of $h^{(2)}_{\alpha\beta}$; and so on.

We explain in the next section how, at each order, when transforming to the BS gauge (i.e., in Steps 2 and 4), we introduce memory (i.e., soft hair) into the metric. 

\subsection{Slowly evolving BMS transformations}\label{sec:multiscale BMS}

We now return to the analysis of the transformation to BS gauge in Secs.~\ref{sec:BSgaugevector} and~\ref{sec:ConstrInfBMS}, extending the transformation to the case of a multiscale expansion. 

As we highlighted in Sec.~\ref{sec:BSgaugevector}, the gauge vector in Eqs.~\eqref{eq:xiu3} , \eqref{eq:xiA3}, and \eqref{eq:xir3} is valid regardless of how the metric perturbation depends on $u$. Hence, it remains valid for coefficients $h^{\{n\}}_{\mu\nu}=h^{\{n\}}_{\mu\nu}[\varphi^i,\tilde u,\bst^A]$. Moreover, this vector does not involve any derivatives of $h^{\{n\}}_{\mu\nu}$ with respect to $u$, meaning we do not need to be concerned with $\partial_{\tilde u}h^{\{n\}}_{\mu\nu}$ terms that would be demoted to second order. 

Therefore, we can immediately focus on the terms involving $\xi^\alpha_\circl$, which we recall was the generator of a BMS transformation in the case of regular perturbation theory. Following the same reasoning as in the final paragraph of Sec.~\ref{sec:BSgaugevector}, we can restrict this vector by requiring asymptotic flatness of $h^{(1)}_{\mu\nu}$:
\begin{itemize}
    \item Eliminating the $\bsr^0$ term in Eq.~\eqref{Liexig0ur} requires $\dot\xi^u_\circl = \frac{1}{2}D_A\xi^A_\circl$. 
    \item Eliminating the $\bsr^2$ term in Eq.~\eqref{Liexig0uA} requires $\xi^A_\circl=\xi^A_\circl[\tilde u,\bst^B]$.
    \item Eliminating the $\bsr^2$ traceless term in Eq.~\eqref{Liexig0AB} requires $D_{\langle A}\xi^{\circl}_{ B\rangle}=0$.
\end{itemize}
Together these imply
\begin{equation}
    \xi^u_\circl = \alpha^{(1)}[\tilde u,\bst^A] \quad\text{and}\quad \xi^A_\circl = Y^A_{(1)}[\tilde u,\bst^B].
\end{equation}
Unlike in regular perturbation theory, here $Y^A_{(1)}$ is not only purely dipolar, but also must satisfy $D_AY^A_{(1)}=0$ because $0=\langle\dot\xi^u_\circl\rangle = \frac{1}{2}D_A\xi^A_\circl$. This additional condition forbids first-order boosts. Hence,
\begin{equation}
    Y^A_{(1)} = -\epsilon^{AB}D_B\kappa^{(1)}.
\end{equation}
$\alpha^{(1)}$ and $\kappa^{(1)}$ are, at this stage, arbitrary functions of $\tilde u$, representing a slowly evolving supertranslation and rotation, respectively.

With these restrictions, the transformation generated by Eqs.~\eqref{eq:xiu3}--\eqref{eq:xir3} 
%
puts the first-order metric perturbation in the BS-gauge form
\begin{align}
    h^{(1)}_{uu} &= \frac{2\MB[(1)][\varphi^i,\tilde u, \theta^A]}{\bsr} - \frac{D_AN^A_{(1)}[\varphi^i,\tilde u, \theta^A]}{3\bsr^2} + {\cal O}(\bsr^{-3}),\\
    h^{(1)}_{u\bsr} &= {\cal O}(\bsr^{-3}),\\
    h^{(1)}_{uA} &= D^B C^{(1)}_{AB}[\varphi^i,\tilde u, \theta^A] +\frac{2}{3}\frac{N^{(1)}_A[\varphi^i,\tilde u, \theta^A]}{\bsr} + {\cal O}(\bsr^{-2}),\\
    h^{(1)}_{AB} &= \bsr\, C^{(1)}_{AB}[\varphi^i,\tilde u, \theta^A] + {\cal O}(\bsr^{-1}),
\end{align}
and $h^{(1)}_{\bsr\bsr}=h^{(1)}_{\bsr A}=\order{\bsr^{-3}}$. Here the mass aspect, angular momentum aspect, and shear must satisfy Eqs.~\eqref{eq:MBdot1}--\eqref{P1=0}.
The specific contribution to this $h^{(1)}_{\alpha\beta}$ from the slowly evolving BMS vector $\xi^\alpha_\circl$ is
\begin{align}\label{eq:Dh1multiscale}
    \Delta h^{(1)}_{\alpha\beta} &= {\Lie}^{(0)}_{\vec\xi_\circl} g^{(0)}_{\alpha\beta},
\end{align}
where the superscript $(0)$ on the Lie derivative indicates that derivatives with respect to $\tilde u$ are neglected when applying Eq.~\eqref{eq:chain rule simple}. Concretely, up to ${\cal O}(\bsr^{{\rm M}-3})$ residuals,
\begin{align}
    \Delta h^{(1)}_{uu} &= -\frac{M^{(0)}}{\bsr^2}D^2f^{(1)}
    + {\cal O}(\bsr^{-3}),\\
    \Delta h^{(1)}_{uA} &= -\frac{1}{2}D_A(D^2+2)\alpha^{(1)} +\frac{2}{\bsr}\left(M^{(0)}D_Af^{(1)} +\frac{1}{3}{\Lie}_{\vec Y_{(1)}}N^{(0)}_A\right) 
    +{\cal O}(\bsr^{-2}),\\
    \Delta h^{(1)}_{AB} &= -2\bsr D_{\langle A}D_{B\rangle}\alpha^{(1)} +{\cal O}(\bsr^{-1}),\label{Dh1AB multiscale}
\end{align}
along with $\Delta h^{(1)}_{\bsr\bsr} =0$ and $\Delta h^{(1)}_{u\bsr} =  \Delta h^{(1)}_{\bsr A} = {\cal O}(\bsr^{{\rm M}-3})$. Here we have used $(D^2+2)Y_{1m}=0=D_{\langle A}D_{B\rangle}Y_{1m}$ and $\chi^{(1)}=0$ to eliminate some $Y^A_{(1)}$ terms. We can immediately read off the corresponding contributions to the mass aspect, angular momentum aspect, and shear:
\begin{align}
    \Delta \MB[(1)] &= 0,\\
    \Delta N^{(1)}_A &= {\Lie}_{\vec Y_{(1)}}N^{(0)}_A+ 3M^{(0)}D_A f^{(1)},\\
    \Delta C^{(1)}_{AB} &= -2D_{\langle A}D_{B\rangle}\alpha^{(1)}.
\end{align}

If we begin from a forgetful gauge, where $\bigl\langle C^{(1)}_{AB}\bigr\rangle = 0$, then the only slowly evolving part of the shear is 
\begin{equation}\label{eq:<CAB>=DDalpha}
   \bigl\langle C^{(1)}_{AB}\bigr\rangle = -2D_{\langle A}D_{B\rangle}\alpha^{(1)}.
\end{equation}
In other words, the slowly evolving supertranslation has induced a slowly evolving shear. This soft hair is now analogous to the mass and spin of the primary: something that appears arbitrary from the perspective of a first-order calculation, but in fact is forced to dynamically evolve on long time scales by virtue of the EFE. In an asymptotically well-behaved BS gauge, the soft hair continually grows due to the EFE; in a forgetful gauge, the near-zone solution continually sheds the soft hair, but such a gauge cannot smoothly extend to $\mathscr{I}^+$. 

We will consider the slow evolution of the supertranslation in the next section, ultimately showing that it  recovers the standard nonlinear memory~\eqref{eq:memory in h1}.

\subsection{Memory and memory distortion}\label{sec:memory in SF}

In the iterative approach above, the first-order metric perturbation $h^{(1)}_{\alpha\beta}$ is first computed, generically, in a forgetful gauge. $h^{(1)}_{\alpha\beta}$ then contains no sign of nonlinear memory, as explained in Sec.~\ref{sec:multiscale BMS}. In the SF-MPM approach, one discovers, after the fact, that the calculation of $h^{(1)}_{\alpha\beta}$ has only been valid in the near zone; the memory content in $h^{(1)}_{\alpha\beta}$ only arises in the far zone, where it must be computed separately using MPM techniques. Using the SF-MPM expansion, Ref.~\cite{Cunningham:2024dog} also discovered that calculations of the second-order (1PA) waveform had omitted an effect, termed ``memory distortion'', that arises from the coupling between slowly varying memory modes and oscillatory modes in the waveform. 

Here we show how, in the iterative approach, (i) the memory content arises directly from the slowly evolving supertranslation that moves from the forgetful gauge to the BS gauge, (ii) the memory content extends across the spacetime, from near to far zone, in the BS gauge, and (iii) our iterative scheme then naturally includes the memory distortion found in Ref.~\cite{Cunningham:2024dog}. 

To obtain these results, we consider the second-order Einstein equation in the multiscale expansion, Eq.~\eqref{multiscale EFE2 BS}, at large $\bsr$. Since the stress-energy tensor vanishes at large $\bsr$, the equation reduces to
\begin{equation}\label{eq:EFE2 vacuum BS}
    G^{(1,0)}_{\alpha\beta}[h^{(2)}] = - G^{(2,0)}_{\alpha\beta}[\hat h^{(1)},\hat h^{(1)}]- G^{(1,1)}_{\alpha\beta}[\hat h^{(1)}].    
\end{equation}
We will recover gravitational-wave memory from the non-oscillatory terms in this equation, while we will infer the memory distortion from the oscillatory terms.

One can straightforwardly calculate the nonlinear source term at large $\bsr$, which evaluates to
\begin{equation}\label{eq:G2 in Bondi}
 G^{(2,0)}_{\alpha\beta}[\hat h^{(1)},\hat h^{(1)}] =  -\frac{l_\alpha l_\beta}{4\bsr^2}\left[\dot C^{AB}_{(1)}\dot C^{(1)}_{AB} - 2\Omega^i\partial_{\varphi^i}\left(C^{AB}_{(1,\rm osc.)}\dot C^{(1)}_{AB}\right) - 2 \bigl\langle C^{AB}_{(1)}\bigr\rangle\ddot C^{(1)}_{AB}\right] + {\cal O}(1/\bsr^3),
\end{equation}
where we have written the shear as a sum of non-oscillatory and oscillatory terms, 
\begin{equation}
    C^{(1)}_{AB} = \bigl\langle C^{(1)}_{AB}\bigr\rangle + C^{(1,\rm osc.)}_{AB}.
\end{equation}
Compare Eq.~\eqref{eq:G2 in Bondi} with the Lorenz-gauge expression~\eqref{eq:G2 in Lorenz}. In both cases, the expression can be derived by following the same steps that led to Eq.~(140) of Ref.~\cite{Cunningham:2024dog}. In the Lorenz gauge, additional components, not directed along $l_{\alpha}l_\beta$, are present, and there would be no non-oscillatory term in $C^{(1)}_{AB}$; recall $\bigl\langle C^{(1)}_{AB}\big\rangle$ only appears from the slowly evolving supertranslation that is required when transforming from the forgetful Lorenz gauge to the BS gauge. 

The second source term in Eq.~\eqref{eq:EFE2 vacuum BS}, arising from slow evolution, is
\begin{equation}\label{eq:G11 BS}
    G^{(1,1)}_{\alpha\beta}[\hat h^{(1)}] = -\frac{1}{2\bsr^2}\Bigl(4\partial_{\tilde u} \MB[(1)]-D^AD^B\partial_{\tilde u}C^{(1)}_{AB}\Bigr)l_\alpha l_\beta  + O(1/\bsr^3).
\end{equation}
If we divide the mass aspect into non-oscillatory and oscillatory pieces, $\MB[(1)]=\bigl\langle\MB[(1)]\bigr\rangle + \MB[(1,\rm osc.)]$, then Eq.~\eqref{eq:MBdot1} implies $\MB[(1,\rm osc.)] = \frac{1}{4}D^A D^B C^{(1,\rm osc.)}_{AB}$. The oscillatory terms in Eq.~\eqref{eq:G11 BS} cancel, leaving
\begin{equation}
    G^{(1,1)}_{\alpha\beta}[\hat h^{(1)}] = -\frac{1}{2\bsr^2}\Bigl(4\partial_{\tilde u} \bigl\langle\MB[(1)]\bigr\rangle + D^A D^B\partial_{\tilde u} \bigl\langle C^{(1)}_{AB}\bigr\rangle\Bigr) l_\alpha l_\beta + O(1/\bsr^3). 
\end{equation}
After combining this with Eq.~\eqref{eq:G2 in Bondi}, we divide the field equation~\eqref{eq:EFE2 vacuum BS} into its non-oscillatory and oscillatory parts:
\begin{align}
    \Bigl\langle G^{(1,0)}_{\alpha\beta}[h^{(2)}]\Bigr\rangle &= \frac{l_\alpha l_\beta}{4\bsr^2}\left[\bigl\langle\dot C^{AB}_{(1)}\dot C^{(1)}_{AB}\bigr\rangle + 8\partial_{\tilde u} \bigl\langle\MB[(1)]\bigr\rangle + 2D^A D^B\partial_{\tilde u} \bigl\langle C^{(1)}_{AB}\bigr\rangle\right] + {\cal O}(1/\bsr^3)\label{eq:EFE2 BS non-oscillatory}\\
    \Bigl(G^{(1,0)}_{\alpha\beta}[h^{(2)}]\Bigr)^{\rm osc.} &= \frac{l_\alpha l_\beta}{4\bsr^2}\left[\Bigl(\dot C^{AB}_{(1)}\dot C^{(1)}_{AB}\Bigr)^{\rm osc.} - 2\Omega^i\partial_{\varphi^i}\left(C^{AB}_{(1,\rm osc.)}\dot C^{(1)}_{AB}\right) - 2 \langle C^{AB}_{(1)}\rangle\ddot C^{(1)}_{AB}\right] \nonumber\\
    &\quad + {\cal O}(1/\bsr^3).\label{eq:EFE2 BS oscillatory}
\end{align}

The non-oscillatory equation quickly yields the formula for the gravitational-wave memory $\bigl\langle C^{(1)}_{AB}\bigr\rangle$. One way to see this is that our analysis of Eq.~\eqref{MBdot 2}, which led to Eq.~\eqref{eq:memory in h1} for $\bigl\langle C^{(1)}_{AB}\bigr\rangle$, is in actuality already an analysis of Eq.~\eqref{eq:EFE2 BS non-oscillatory}---in the special case that $h^{(2)}_{\alpha\beta}$ is in the BS gauge. This follows from the fact that Eq.~\eqref{eq:MBdot}, from which Eq.~\eqref{MBdot 2} was derived, is the exact $uu$ component of the $1/\bsr^2$ term in the fully nonlinear BS-gauge EFE. However, the result for $\bigl\langle C^{(1)}_{AB}\bigr\rangle$ is not restricted to the case that $h^{(2)}_{\alpha\beta}$ is in the BS gauge. If we require that $h^{(2)}_{\alpha\beta}$ to decay at large $\bsr$, then it follows in any gauge that $\Bigl\langle G^{(1,0)}_{\alpha\beta}[h^{(2)}]\Bigr\rangle$ decays more rapidly than $1/\bsr^2$; $\bsr$ and $\bst^A$ derivatives each reduce the order in $\bsr$ by one, and $u$ derivatives (treated as $\Omega^i\partial_{\varphi^i}$) average to zero. Hence, the coefficient of $\bsr^2$ on the right-hand side of Eq.~\eqref{eq:EFE2 BS non-oscillatory} must vanish, implying 
\begin{equation}
    \frac{1}{4}\left\langle \dot C^{(1)}_{AB}\dot C^{AB}_{(1)} \right\rangle = -2\left\langle\partial_{\tilde u} \MB[(1)]\right\rangle - \frac{1}{2}\left(D^2+2\right)\!D^2\partial_{\tilde u}\alpha^{(1)}.\label{slow balance law 1}
\end{equation}
Here we have made use of Eq.~\eqref{eq:<CAB>=DDalpha} to write the equality explicitly in terms of our slowly evolving supertranslation. 
Integrating this equation over the sphere, we recover the Bondi mass-loss formula~\eqref{<dMB1>}. 
If we instead integrate Eq.~\eqref{slow balance law 1} against $Y^*_{lm}$, we obtain a relationship between the supertranslation $\alpha^{(1)}$ and the angular distribution of GW flux: 
\begin{equation}
\partial_{\tilde u}\alpha^{(1)}_{lm} = \frac{1}{2}\frac{(l-2)!}{(l+2)!}\left\langle \dot C^{(1)}_{AB}\dot C^{AB}_{(1)}\right\rangle_{\!lm}.
\end{equation}
Or, re-expressed in term of the shear using Eq.~\eqref{eq:<CAB>=DDalpha},
\begin{equation}
    \left\langle  C^{(1)}_{AB}\right\rangle = -\frac{1}{4}\sum_{lm}\frac{(l-2)!}{(l+2)!}\int d\tilde u\left\langle \dot C^{(1)}_{CD}\dot C^{CD}_{(1)}\right\rangle_{\!lm}Z^{lm}_{AB},
\end{equation}
where $Z_{AB}^{lm}:=D_{\langle A}D_{B\rangle}Y_{lm}$ is an even-parity tensor harmonic. This is the standard expression for the memory~\cite{Cunningham:2024dog}, equivalent to Eq.~\eqref{eq:memory in h1}.

Finally, we turn to the memory distortion. To explain how it arises, we simply note that in Eq.~\eqref{eq:EFE2 BS oscillatory}, coupling between $\bigl\langle C^{(1)}_{AB}\bigr\rangle$ and $C^{(1,\rm osc.)}_{\alpha\beta}$ appears explicitly as a source term for the oscillatory part of $h^{(2)}_{\alpha\beta}$. Hence, this coupling contributes to the waveform  $C_{AB}^{(2,\rm osc.)}$ and therefore to the 1PA energy flux, $\frac{1}{16\pi}\oint\dot C^{(1)}_{AB}\dot C^{AB}_{(2)}d\Omega$. This contribution to $C_{AB}^{(2,\rm osc.)}$ represents a slow distortion of the waveform, created by the slowly evolving supertranslation of the asymptotic frame. It arises naturally in our iterative scheme because the first-order memory extends into the near zone: when solving Eq.~\eqref{eq:EFE2 BS oscillatory} numerically, for example, one would automatically obtain this contribution. If $h^{(1)}_{\alpha\beta}$ is instead in a forgetful gauge, where there is no sign of $\bigl\langle C^{(1)}_{AB}\bigr\rangle$ in the near zone, then the memory distortion can only be obtained from a separate analysis of the far-zone field equations, as in Ref.~\cite{Cunningham:2024dog}.

\section{\label{sec:CompareNR}Comparison to numerical relativity BMS frame fixing}

Alternative BMS frame-fixing techniques have been used to compare NR waveforms, particularly those produced by the SXS collaboration, to waveforms produced using perturbation theory (PN or BHPT)~\cite{Boyle:2015nqa,mitman2021fixing,magana2022high,mitman2022fixing,mitman2024review,DaRe:2025glj,sun2024optimizing,Khairnar:2026mtm}. To compare to PN theory, NR results are transformed to the so-called \textit{PN frame} (e.g., in Refs.~\cite{mitman2022fixing,Khairnar:2026mtm}), in which the linear momentum $P_i$, boost charge $K_i$, and shear $C_{AB}$ all vanish in the infinite past; this is the canonical centre-of-mass-frame of the binary's infinite past. Similarly, to compare to ringdown analysis from BHPT, NR results are transformed to the so-called \textit{superrest frame}~\cite{magana2022high}, in which the linear momentum $P_i$, boost charge $K_i$, and shear $C_{AB}$ all vanish in the infinite future; this is the canonical centre-of-mass frame of the remnant black hole. Note that since the odd-parity part of the shear automatically vanishes in non-radiative regions of $\mathscr{I}^+$, both of these options fall within our gauge-fixing scheme. However, the schemes diverge in how they fix the angular momentum: we directly use the Bondi angular momentum and align it with the preferred spin axis provided by the Kerr background; the NR frame-fixing scheme used in SXS~\cite{mitman2024review} instead makes use of an angular velocity vector that is accessible from the shear alone, without requiring direct access to the Bondi angular momentum.

Here, we delineate the similarities and differences between the two schemes. We must first consider that the SXS frame-fixing scheme works in terms of the full metric, not a perturbative expansion, and it does not have access to a preferred background frame, whereas our formalism fixes the BMS frame only at first perturbative order about the BMS frame associated with the Kerr Bondi--Sachs coordinates given in Bai et al.~\cite{bai2007KerrBS}. 
This means that the SXS frame-fixing must be performed iteratively, for example, until the error between the NR and PN results converge, because each BMS charge they fix is dependent on other BMS transformations nonlinearly. Our formalism allows one to fix all the BMS charges independently, as we have separated all the BMS transformation dependences in our charges. Our formalism only fixes the BMS frame to first perturbative order, but since we can work with a scalar variable, $\psi_{4L}^{(2)}$, that is invariant under second-order transformations, our first-order gauge fixing suffices to determine an invariant waveform through second order; see Sec.~\ref{sec:gauge-fixing}. Additionally, we expect our BMS frame-fixing formalism to be extendable to higher orders. 

Distinctions between our BMS fixing scheme and the SXS scheme also emerge in the specific choice of variables used to fix the BMS frame, most prominently in the case of rotations, but also for the other transformations. We directly utilize the perturbed Poincaré charges, Eqs.~\eqref{eq:Pt}--\eqref{eq:KT}, together with the shear, while Ref.~\cite{mitman2024review} fixes the Moreschi supermomentum~\eqref{eq:MoreschiSupermomentum}, the centre-of-mass charge~\eqref{eq:centre-of-mass-charge}, and an alternative quantifier of rotation we introduce below.



Concretely, 
\begin{itemize}
    \item Our scheme fixes the translations and boosts by eliminating the boost and linear momentum charges $K^{(1)}_i$ and $P^{(1)}_i$, respectively, on a given time slice, using Eqs.~\eqref{eq:translation-transform} and~\eqref{eq:boost-transform}; this then enforces that the center-of-mass charge $G_i^{(1)}$ in Eq.~\eqref{eq:centre-of-mass-charge} also vanishes on that slice. In Ref.~\cite{mitman2024review}, the translations and boosts are instead fixed by directly using $G_i$. To fix the translations and boots simultaneously, the $u$ dependence of $G_i[u]$ is also used, rather than fixing the value at a specific time. In Ref.~\cite{Khairnar:2026mtm}, $G_i[u]$ is specifically fixed to agree with its PN value in some time interval, which ensures that the system was in the center-of-mass frame in the infinite past. In both the SXS scheme and ours, time translations are fixed by aligning waveforms (e.g., aligning NR and PN waveforms).
    \item Our scheme fixes the supertranslations by eliminating the even-parity shear on some time slice. The SXS scheme achieves this indirectly by using the $\l\geq 2$ modes of the Moreschi supermomentum $\P_M$, \cref{eq:MoreschiSupermomentum}. To fix to the superrest frame, the supermomentum is fixed to (minus) the Bondi mass at late times. To fix to the PN frame, where the shear vanished in the infinite past, the supermomentum is fixed to its PN value, $\P^{PN}_M$, in a common time window, where $\P^{PN}_M$ is given in Eq.~(A1) of Ref.~\cite{mitman2022fixing} to 3PN order without spins and 2PN order with spins.
    \item In our scheme, we fix the rotation frame by eliminating the $x$ and $y$ components (i.e., the $\l=1,\m=\pm1$ modes) of the perturbed Wald--Zoupas angular momentum aspect~\eqref{eq:JT} on a given slice, using \cref{eq:BMSRotationconst}. This aligns our frame with the total Bondi angular momentum on that slice. We can then fix the axial rotation by aligning the waveform phase. In the SXS scheme, one instead uses an angular velocity that can be extracted directly from the waveform and which approximates the direction of a binary's orbital angular momentum~\cite{Boyle:2013nka}. One then aligns the $z$ axis with the direction of this angular velocity at a given time, or else one aligns the angular velocity with its PN direction at that time.
\end{itemize}

\section{\label{sec:conclusions}Conclusion}


In this paper, we constructed a perturbative treatment of the Bondi--Sachs gauge on Kerr spacetime, accompanied by a BMS frame-fixing scheme and an analysis of BMS symmetries in the context of the multiscale expansion used in self-force calculations. Our motivations are threefold: being able to extract BMS charges from BHPT calculations and unambiguously specify the BMS frames of BHPT waveforms; working in an asymptotically flat gauge which avoids infrared divergences in second-order calculations; and fixing the first-order gauge to make the Teukolsky variable $\psi^{(2)}_{4L}$ into an invariant quantity near future null infinity. All of these should allow sharper comparisons between different self-force calculations and between self-force, PN, PM, ringdown, and NR results. 

Our derivation of the perturbative BS gauge is built on the Bondi--Sachs coordinates describing the Kerr metric as derived by Ref.~\cite{bai2007KerrBS}. The transformation from Boyer--Lindquist to BS coordinates in Kerr spacetime is given by \cref{eq:BStoBLtransform}. The Kerr BS coordinates are expressed using an asymptotic expansion, which can be truncated to a given order in $1/\bsr$. While the asymptotic expansion can be extended to higher orders, we have specifically limited ourselves to the number of orders needed to extract the perturbed BMS charges from the asymptotic metric. 
Hence, this description is sufficient to analyse the BMS frame of both the background Kerr BS coordinates and of perturbations. 

In \cref{sec:pertBSgauge}, we described the perturbative BS gauge, defining the gauge conditions on the first-order metric perturbation. The perturbative BS gauge can be enforced in any background coordinates, but it is most easily expressed in background BS coordinates, as we have done in \cref{eq:pertBSconditions}. Using the BMS balance laws, we were immediately able to conclude, in \cref{constant Poincare}, that all the Poincaré charges are constant in first-order perturbation theory (since the fluxes are quadratic); at least, this is the case in \emph{regular} perturbation theory, if not in a multiscale expansion. In the remainder of Sec.~\ref{sec:BSgaugevector}, we found the gauge vector that takes one from any asymptotically flat gauge to a perturbative BS gauge. The gauge vector satisfies a hierarchical set of first-order radial ODEs along null rays extending to $\mathscr{I}^+$.  By introducing an asymptotic expansion of the metric perturbation and truncating to a given order in $1/\bsr$, we solved the ODEs to obtain the gauge vector as a large-$\bsr$ expansion, given in Eqs.~\eqref{eq:xiu3}--\eqref{eq:xir3}. However, these results are all expressed in the background Kerr BS coordinates, limiting their practical utility. To make our formalism more practical, in Appendix~\ref{App:pBSinBL} we convert the gauge vector from Kerr BS coordinates to outgoing Kerr--Newman coordinates and express the gauge vector in terms of its Newman--Penrose tetrad components; see \cref{eq:xil,eq:xin,eq:xim}. We expect these latter expressions to be the ones used in practice.

Our transformation to a BS gauge in Sec.~\ref{sec:BSgaugevector} leaves residual gauge freedom: integration constants that, in regular perturbation theory, are precisely the generators of first-order BMS transformations. In Sec.~\ref{sec:ConstrInfBMS}, we investigated how to use this residual freedom to fix the BMS frame at first perturbative order. 
Our frame-fixing scheme, summarized in Sec.~\ref{sec:BMS fixing summary}, specializes to a center-of-mass frame by eliminating the Bondi linear momentum $P^{(1)}_i$ and boost charge $K^{(1)}_i$ (\cref{eq:Pt,eq:KT}). We align the $z$ axis with the total Bondi angular momentum (\cref{eq:JT}), and we eliminate the even-parity part of the shear on a chosen time slice. This fixes all the BMS freedom up to the time translation and axial rotation, which are Killing symmetries of the background spacetime. These last freedoms can be fixed when comparing two waveforms by aligning the waveform time and phase. 

A key upshot of our scheme is that it fully fixes the first-order gauge in a neighbourhood of $\mathscr{I}^+$. This allows us to construct second-order gauge invariant scalar variables near \scri using any of $\{\psi^{(2)}_{4L},\psi^{(2)}_{4},\psi^{(2)}_{0L},\psi^{(2)}_{0}\}$ (though we only emphasized $\psi^{(2)}_{4L}$ here). These variables only depend on first-order gauge transformations; they are invariant under second-order gauge transformations. Hence, once the first-order gauge is fully fixed, they become gauge-invariant observables in our specified asymptotic frame. 

We believe our scheme has the greatest significance in the context of self-force theory. As we stressed in the Introduction, there has been very little work on understanding the asymptotic frame of self-force calculations, and infrared gauge singularities appear at second order in these calculations. In Sec.~\ref{sec:BS gauge in SF theory}, we explained how these issues are resolved by extending our scheme to self-force applications. This involved two major aspects: 

First, we formulated an iterative scheme, similar in spirit to the ``MPM$\to$radiative'' scheme in Refs.~\cite{Blanchet:1986dk,Trestini:2023wwg}, for transforming to the BS gauge. In this scheme, one solves for $h^{(1)}_{\alpha\beta}$ in any desired gauge, transforms to BS gauge (in our fixed BMS frame), then solves for $h^{(2)}_{\alpha\beta}$ in any desired gauge. Putting $h^{(1)}_{\alpha\beta}$ in the BS gauge ensures that the source for $h^{(2)}_{\alpha\beta}$ decays rapidly enough toward $\mathscr{I}^+$ to avoid infrared divergences. Hence, this scheme provides the flexibility of using whatever gauge is convenient for the left-hand side of the field equations, while ensuring that the right-hand side is well behaved. With this scheme, the distinction between near and far zone is eliminated: the multiscale expansion used in the near zone extends all the way to $\mathscr{I}^+$. There is then no need for the SF-MPM far-zone expansion that has been used to date in second-order self-force calculations~\cite{PoundLargeScales,Cunningham:2024dog}.

Second, we found several nontrivial features of the extension from regular perturbation theory to the multiscale expansion used in self-force theory: 
        \begin{itemize}
            \item In regular perturbation theory, the linear perturbations to the Poincaré charges are constant in time, and they can therefore be set to zero for \emph{all} time with a frame choice. In the multiscale expansion, on the other hand, these charges can slowly evolve (with vanishing oscillatory part), meaning they can only be set to zero on a single slice.
            \item In regular perturbation theory, the integration constants that arise when transforming to BS gauge are precisely the generators of a first-order BMS transformation, meaning they involve only time-independent parameters. In the multiscale expansion, these parameters instead must be allowed to slowly evolve; the transformation from a Lorenz gauge, for example, to a BS gauge must generically involve slowly evolving Poincaré transformations and slowly evolving supertranslations.
            \item  In regular perturbation theory, gravitational-wave memory only enters the waveform at second order; in the multiscale expansion, due to long accumulation, it enters at first order (a point already stressed in Ref.~\cite{Cunningham:2024dog}). However, in the gauges typically used in self-force calculations, signatures of memory do not enter the near zone until second order, and memory effects such as the memory distortion found in Ref.~\cite{Cunningham:2024dog} are only discovered through the SF-MPM scheme. This led us to call the gauges traditionally used in self-force theory ``forgetful gauges''. Our analysis here revealed that the slowly evolving supertranslations from a forgetful gauge to a BS gauge directly introduces the gravitational-wave memory into the first-order solution, and moreover, it extends the first-order memory into the near zone, adding soft hair to the system. In our slowly evolving supertranslation scheme, the memory distortion appears naturally in the process of solving the second-order field equations. 
        \end{itemize}
    
Finally, in Sec.~\ref{sec:CompareNR}, we compared our frame-fixing scheme to the one employed by the SXS collaboration. The two schemes are conceptually similar, particularly in the multiscale case where the frame alignment can only happen at a particular time. But they have a pronounced difference in how they align the frame with the spin axis: we align the $z$ axis with the system's total Bondi angular momentum, while the SXS method uses an angular velocity vector defined directly from the waveform~\cite{mitman2024review,mitman2022fixing}.

We envision two immediate applications of our work: applying our iterative scheme to second-order self-force calculations; and employing our frame-fixing strategy, together with our sharpened understanding of perturbative BMS frames more broadly, to analyse the asymptotic frame of self-force calculations. In future work we will also include the previously neglected memory distortion effect in 1PA waveform models.


\begin{acknowledgments}
AS and AP thank Keefe Mitman for helpful discussions and Ben Leather for collaboration on a related project that helped inform this one. AP thanks Sarp Akcay, Geoffrey Compère, Kevin Cunningham, Josh Mathews, J\o rgen Musaeus, Ariadna Ribes Metidieri, Ali Seraj, Jonathan Thompson, and especially Maarten van de Meent and Alex Grant for additional helpful discussions. 
AP acknowledges the support of a Royal Society University Research Fellowship and the ERC Consolidator/UKRI Frontier Research Grant GWModels (selected by the ERC and funded by UKRI [grant number EP/Y008251/1]), and AP and AS acknowledge the support of a Royal Society University Research Fellowship Enhancement Award. AS acknowledges the partial support
from the STFC Consolidated Grant no. ST/V005596/1. AS~acknowledges funding from the European Union's Horizon Europe research and innovation programme under the Marie Sklodowska-Curie grant agreement no. 101199153. Google Gemini was used to help typeset \cref{eq:xil,eq:xin,eq:xim}.

\end{acknowledgments}

\appendix

\section{Expansion of the Bondi--Sachs determinant condition}\label{sec:perturbative det condition}

The BS gauge conditions~\eqref{eq:BSconditions} include a condition on the determinant of the angular piece of the metric: ${\rm det}[\gamma_{AB}]={\rm det}[\Omega_{AB}]$. When the metric is perturbatively expanded, this translates into a condition on the trace of the angular part of the metric perturbation, as enforced in Eq.~\eqref{eq:pertBSconditions}. We derive that condition at first and second order here.

Concretely, we write the angular metric as
\begin{equation}
    \gamma_{AB} = \gamma^{(0)}_{AB}+\varepsilon\gamma^{(1)}_{AB} + \varepsilon^2\gamma^{(2)}_{AB} + {\cal O}(\varepsilon^3).
\end{equation}
Using bold symbols to denote matrices, we write this compactly as $\bm{\gamma}=\bm{\gamma}_{(0)}+\bm{\delta\gamma}=\bm{\gamma}_{(0)}\Bigl(\bm{1}+\bm{\gamma}_{(0)}^{-1}\bm{\delta\gamma}\Bigr)$. We  then see the condition ${\rm det}[\bm{\gamma}]={\rm det}[\bm{\Omega}]$ implies ${\rm det}\Bigl[\bm{1}+\bm{\gamma}_{(0)}^{-1}\bm{\delta\gamma}\Bigr]=1$. Using the Jacobi formula for the determinant, we expand in powers of $\bm{\gamma}$:
\begin{align}
    {\rm det}\Bigl[\bm{1}+\bm{\gamma}_{(0)}^{-1}\bm{\delta\gamma}\Bigr]
    &= 1 + {\rm Tr}\bigl[\bm{\gamma}_{(0)}^{-1}\bm{\delta\gamma}\bigr]  + \frac{1}{2}\left({\rm Tr}\bigl[\bm{\gamma}_{(0)}^{-1}\bm{\delta\gamma}\bigr]\right)^2 - \frac{1}{2}{\rm Tr}\bigl[\bm{\gamma}_{(0)}^{-1}\bm{\delta\gamma}\,\bm{\gamma}_{(0)}^{-1}\bm{\delta\gamma}\bigr] +\ldots
\end{align}
Writing $\delta\gamma_{AB}=\varepsilon\gamma^{(1)}_{AB}+\varepsilon^2\gamma^{(2)}_{AB}+{\cal O}(\varepsilon^2)$ and appealing to 
\begin{equation}
\gamma^{(0)}_{AB} = \Omega_{AB} 
+ \order{\bsr^{-3}} \quad\text{and}\quad  \gamma^{(1)}_{AB}={\cal O}(\bsr^{-1})=\gamma^{(2)}_{AB},
\end{equation}
we can now read off 
\begin{align}
    \Omega^{AB}\gamma^{(1)}_{AB} &= \order{\bsr^{-4}},\\
    \Omega^{AB}\gamma^{(2)}_{AB} &=  -\frac{1}{2} \gamma_{(1)}^{AB}\gamma^{(1)}_{AB} + \order{\bsr^{-4}},
\end{align}
where 
we recall that angular indices are raised with $\Omega^{AB}$.

\section{Comparing Bondi--Sachs coordinates to Kerr--Newman coordinates}
\label{app:Kerr-Newman}

In the body of the paper, we work exclusively with the BS coordinates defined by Eq.~\eqref{eq:BStoBLtransform}. But most practical calculations will be performed in the more traditional coordinates of Kerr spacetime, such as Boyer--Lindquist or retarded Kerr--Newman coordinates. Here we relate BS coordinates to Kerr--Newman coordinates. Here, as in Sec.~\ref{sec:KerrBS}, we use $M$ to denote the background mass parameter, omitting a ``(0)'' label.

The Kerr--Newman coordinates are related to Boyer--Lindquist coordinates by
\begin{align}
    u_{\mathrm{KN}}&=t-r_*=t-r-\frac{r_+^2+a^2}{r_+-r_-} \ln\bigg[ \frac{r-r_+}{r_+}\bigg] +\frac{r_-^2+a^2}{r_+-r_-} \ln\bigg[ \frac{r-r_-}{r_-} \bigg],\notag \\
    r_{\mathrm{KN}}&=r, \notag \\
    \theta_{\mathrm{KN}} &= \theta, \notag \\
    \phi_* &= \phi -\frac{a}{r_+-r_-} \ln\bigg[ \frac{r-r_+}{r-r_-} \bigg], \label{eq:KNcoords}
\end{align}
where $r_\pm=M\pm\sqrt{M^2-a^2}$. These coordinates have the useful property that the outgoing principal null direction in Kerr is purely radial; in the Kinnersley tetrad,
\begin{align}
    l^\mu_{\mathrm{KN}}&=(  0, 1, 0 , 0 ),\label{eq:KN-l}  \\
    n^\mu_{\mathrm{KN}}&=\frac{1}{\Sigma}\Bigl(   r^2+a^2, -\frac{\Delta}{2}, 0 , a  \Bigr),\label{eq:KN-n} \\
    m^\mu_{\mathrm{KN}}&=\frac{1}{\sqrt{2}\bar\zeta}\biggl(   ia\sin\theta, 0, 1 , \frac{i}{\sin\theta}  \bigg), \label{eq:KN-m}\\
    \bar m^\mu_{\mathrm{KN}}&=\frac{1}{\sqrt{2}\zeta}\biggl(   -ia\sin\theta, 0, 1 , \frac{-i}{\sin\theta}  \biggr).    \label{eq:KN-mbar} 
\end{align}
Here $\Delta:=r^2-2Mr+a^2$, $\Sigma:=r^2+a^2\cos^2\theta$, and $\zeta:=r-ia\cos\theta$ are the standard functions in the Kerr geometry~\cite{Pound-Wardell:2021}.

Compare $u_{\rm KN}$ to the retarded time in BS coordinates from \cref{eq:BStoBLtransform}, expressed as large-$r$ expansions:
\begin{align}
    \bsu&=t-r-2M\ln\left[\frac{r}{2M} \right]+\frac{4M^2-\frac{1}{2}a^2\sin^2\theta}{r}  +\frac{4M^3-Ma^2}{r^2} +\mathcal{O}(r^{-3}), \notag \\
    u_{\mathrm{KN}}&=t-r -2M\ln\left[ \frac{r}{M+\sqrt{M^2-a^2}} \right] +\frac{4M^2}{r} +\frac{4M^3-Ma^2}{r^2}+\mathcal{O}(r^{-3}).\label{eq:uu}
\end{align} 
We see that in Schwarzschild, $\bsu$ and $u_{\mathrm{KN}}$ both reduce to the Eddington--Finkelstein retarded time. However, for $a\neq 0$, $\bsu$ and $u_{\mathrm{KN}}$ disagree from their logarithm term onwards. The disagreement in the logarithm is equivalent to a constant shift in retarded time,
\begin{align}
    (\bsu - u_{\mathrm{KN}})_{r\rightarrow\infty}= 2M\ln\left[ \frac{2M}{M+\sqrt{M^2-a^2}} \right]:=\Delta u.\label{eq:uminusu}
\end{align}
On the other hand, the differences at higher orders in $1/r$ are angle dependent, as the coordinate $\bsu$ is dependent on $\theta$ while $u_{\mathrm{KN}}$ has no angular dependence. In the limit $M\rightarrow 0$ (while $a$ remains constant), we see $\bsu\rightarrow t-\Big(r +\frac{a^2\sin^2\theta}{2r} +\ldots\Big)$, similar to the coordinates of an oblate spheroid in flat space~\cite{bai2007KerrBS}. On the other hand, $u_{\mathrm{KN}} \rightarrow t-r$ in this limit. 

Similarly, the large-$r$ expansions of $\bsp$ and $\phi_*$ clearly differ,
\begin{align}
    \hat{\phi} &=\phi+\frac{aM}{r^2}+\frac{4M^2a}{3r^3}+\mathcal{O}(r^{-4}),\notag \\
        \phi_* &= \phi + \frac{a}{r} +\frac{aM}{r^2} -\frac{a(a^2-4M^2)}{3r^3}+\mathcal{O}(r^{-4}),\label{eq:phistar}
\end{align}
though the difference is angle-independent at these orders.

\section{\label{App:pBSinBL}Tetrad components of the Bondi--Sachs gauge vector in Kerr--Newman coordinates}

The BS coordinates we have used to express the gauge vector in \cref{eq:xiu3,eq:xiA3,eq:xir3} are not generally used for BHPT calculations; here, we transform our results to the commonly used Kerr--Newman coordinates reviewed in Appendix~\ref{app:Kerr-Newman}. Additionally, for convenience in practical implementations, we express the components of the gauge vector and the components of the metric perturbation in terms of the Kinnersley tetrad. 

\subsection{Tetrad components}

The radial expansion~\eqref{eq:metricPertExpansion} of the BS coordinate components of the metric perturbation equates to an equivalent expansion of the Kinnersley tetrad components, with coefficients of powers of $1/\bsr$ related by
\begingroup%
\allowdisplaybreaks%
\begin{align}
    h_{\bsr\bsr}^{\ro}&= h_{ll}^{\ro}, \nonumber\\ 
    h_{\bsr\bsr}^{\rr} &= h_{ll}^{\rr}+ i a\sqrt{2}   \sin[\bst] \Big(h_{lm}^{\ro} -h_{l\mb}^{\ro}\Big), \notag \\
    h_\bssub{\bsr\bst}^{\ro}&= \frac{h_{lm}^{\ro}+h_{l\mb}^{\ro}}{\sqrt2}, \notag\\
    h_\bssub{\bsr\bst}^{\rr}&= \frac{1}{\sqrt{2}} \bigg\{ h_{lm}^{\rr} +h_{l\mb}^{\rr} +i 2 a \cos[\bst] \Big( h_{lm}^{\ro} 
 -h_{l\mb}^{\ro} \Big) \notag \\*
 &\quad +i \sqrt2 a \sin[\bst] \Big( h_{mm}^{\ro} 
 -h_{\mb\mb}^{\ro} \Big) \bigg\},
    \notag \\
    h_\bssub{\bsr\bsp}^{\ro} &= -i\sin[\bst]\frac{h_{lm}^{\ro}-h_{l\mb}^{\ro}}{\sqrt2} ,\notag \\
    h_\bssub{\bsr\bsp}^{\rr}&=\frac{\sin[\bst]}{\sqrt2}\bigg\{ i\Big( h_{l\mb}^{\rr} - h_{lm}^{\rr}\Big) + a \cos[\bst]\Big( h_{lm}^{\ro}+h_{lm}^{\ro} \Big)\notag \\*
  &\quad + \frac{a \sin[\bst]}{\sqrt2}\Big( h_{\mb\mb}^{\ro}+  h_{mm}^{\ro} -2 h_{m\mb}^{\ro} -h_{ll}^{\ro} -2h_{ln}^{\ro} \Big) \bigg\},
    \notag \\
    h_\bssub{\bst\bst}^{\{1\}} +\csc^2[\bst] h_\bssub{\bsp\bsp}^{\{1\}}&= 2h_{m\mb}^\ro, \notag \\
     h_\bssub{\bst\bst}^{\{2\}} +\csc^2[\bst] h_\bssub{\bsp\bsp}^{\{2\}}&= 2h_{m\mb}^\rr+\frac{ia\sin[\bst]}{\sqrt2}\Big(h_{lm}^\ro+h_{l\bar{m}}^\ro+2h_{nm}^\ro+h_{n\bar{m}}^\ro\Big).\label{eq:BStoKinnersley}
\end{align}%
\endgroup%

We assume an $m$-mode azimuthal decomposition and a Fourier decomposition with respect to retarded time:
\begin{align}
     h_\bssub{\mu\nu}^{\{a\}}[\bsu,\bst,\bsp]=e^{-i\omega u}e^{i m \bsp}h_\bssub{\mu\nu}^{\{a\}{\rm BS}\omega m}[\bst],\\
     h_\bssub{\mu\nu}^{\{a\}}[\bsu,\theta,\KNp]=e^{-i\omega u_{\rm KN}}e^{i m \KNp}h_\bssub{\mu\nu}^{\{a\}{\rm KN}\omega m}[\theta],
\end{align}
where the BS and KN labels indicate the coordinates used for the Fourier, azimuthal, and radial expansions, and we have omitted the integral over $\omega$ and the sum over $m$ for brevity. In the conversions $u\rightarrow u_{KN}$ and $\bsp\rightarrow \KNp$, we relate the coefficients of metric perturbation in BS coordinates and KN coordinates,
\begin{align}
    e^{-i\omega\bsu+im\bsp}h_{\mu\nu}^{{\rm BS} \omega m}&=e^{-i\omega\KNu+im\KNp}h_{\mu\nu}^{{\rm KN} \omega m},\\
    h_{\mu\nu}^{{\rm BS} \omega m}&=e^{-i\omega (u_{\rm KN}-\bsu)}e^{im(\KNp-\bsp)}h_{\mu\nu}^{{\rm KN} \omega m},\\
    \Rightarrow h_\bssub{\mu\nu}^{\{1\}{\rm BS}\omega m}&=P_0 h_\bssub{\mu\nu}^{\{1\}{\rm KN}\omega m},\\
    h_\bssub{\mu\nu}^{\{2\}{\rm BS}\omega m}&=P_0 h_\bssub{\mu\nu}^{\{2\}{\rm KN}\omega m}+P_1[\theta] h_\bssub{\mu\nu}^{\{1\}{\rm KN}\omega m},\label{eq:hKNtoBS}
\end{align}
where
\begin{align}
    P_0=\bigg(\frac{2M}{M+\sqrt{M^2-a^2}} \bigg)^{2i M\omega}, \ \ \ P_1[\theta]= -\frac{ia(a\omega\sin^2[\theta]-2m)}{2}P_0\label{eq:P0P1}.
\end{align}
Note that $P_0$ is a phase factor as it has a modulus of 1. Additionally, we consider the gauge vector to be admit a similar $m$-mode azimuthal decomposition and a Fourier decomposition in KN coordinates,
\begin{align}
     \xi^\mu[\bsu,r,\theta,\KNp]=e^{-i\omega u_{\rm KN}}e^{i m \KNp}\xi^\mu_{{\rm KN}\omega m}[r,\theta].
\end{align}
Inverting \cref{eq:hKNtoBS} we find
\begin{align}
    \xi^{\rm KN\omega m}_\mu&= e^{-i\omega (\bsr^*-r^*_{\rm KN})}e^{-im(\KNp-\bsp)}\xi^{{\rm BS} \omega m}\label{eq:xiBStoKN}.
\end{align}
The coordinate changes from \cref{eq:hKNtoBS,eq:xiBStoKN} cancel eachother except when a radial or polar angle derivative acts on the metric perturbation in \cref{eq:BStoKinnersley}. Using \cref{eq:BStoKinnersley,eq:BStoBLtransform,eq:KNcoords,eq:hKNtoBS,eq:xiBStoKN}, we can construct tetrad components of the gauge vector, \cref{eq:xiu3,eq:xiA3,eq:xir3}, expressed in KN coordinates: 
\begingroup%
\allowdisplaybreaks%
\begin{align}
    \xi_l &= -\frac{ h_{ll}^\ro\ln[r]}{2} + a\sin^2[\theta]\xi_\circl^\bsp - \xi_\circl^\bsu \notag  - \frac{i a m  h_{ll}^\ro \ln[r]}{2r}+ \frac{1}{8r}\Big( 4h_{ll}^\rr \notag \\
    &\quad+ a\big( -4im h_{ll}^\ro - 3a\sin[2\theta]\xi_\circl^\bst + 2a\sin^2[\theta]\big( \xi_{\circl,\theta}^\bst + \xi_{\circl,\KNp}^\bsp \big) - 8\xi_{\circl,\KNp}^\bsu \big) \Big)+\O\left(\frac{1}{r^2}\right), \label{eq:xil}\\
    \xi_n&= \frac12\Big( \cot[\theta]\xi_\circl^\bst + \partial_{\theta}\xi_{\circl}^\bst + \partial_{\phi}\xi_{\circl}^\bsp \Big)r +\frac{1}{4}\ln[r] \Big( (-1 + m^2 \csc^2[\theta]) h_{ll}^\ro - \cot[\theta] h_{ll,\theta}^\ro - h_{ll,\theta\theta}^\ro \Big)\notag\\*
&\quad +\frac{1}{4} \Big(  m^2 \csc^2[\theta] h_{ll}^\ro + \sqrt{2}(m + \cos[\theta])\csc[\theta] h_{lm}^\ro + 2h_{m\mb}^\ro + 2a\sin^2[\theta]\xi_\circl^\bsp - 2\xi_\circl^\bsu \notag \\
    &\quad + \sqrt{2}h_{lm,\theta}^\ro + \sqrt{2}h_{l\mb,\theta}^\ro + \cot[\theta]\big( \sqrt{2}h_{l\mb}^\ro - h_{ll,\theta}^\ro - 2\xi_{\circl,\theta}^\bsu \big) - h_{ll,\theta\theta}^\ro \notag \\
    &\quad - 2\xi_{\circl,\theta\theta}^\bsu - \csc[\theta]\big( \sqrt{2}m h_{l\mb}^\ro + 2\csc[\theta]\xi_{\circl,\KNp\KNp}^\bsu \big) \Big)  +\frac{1}{8r}\ln[r] \Big(  \big( 4M \notag \\ 
&\quad+ ia(-2m + a(1+P_0)\omega) + 3ia^2(1+P_0)\omega\cos[2\theta] \big) h_{ll}^\ro  + 4ia^2\omega\sin[2\theta] h_{ll,\theta}^\ro \Big)\notag \\ 
&\quad+ \frac{1}{16r}\bigg(  2ia(-3m + a(1+P_0)\omega + 3a(1+P_0)\omega\cos[2\theta])h_{ll}^\ro - 4iam h_{ln}^\ro \notag \\
    &\quad + 2\Big(-m^2\csc^2[\theta]h_{ll}^\rr + \sqrt{2}(m+\cos[\theta])\csc[\theta]h_{lm}^\rr - \sqrt{2}m\csc[\theta]h_{l\mb}^\rr  + 8M\xi_\circl^\bsu \notag \\
    &\quad+ \cot[\theta]\big(\sqrt{2}h_{l\mb}^\rr + h_{ll,\theta}^\rr\big)\Big)  + 2\Big(2h_{ll}^\rr + 4h_{m\mb}^\rr + \sqrt{2}h_{lm,\theta}^\rr + \sqrt{2}h_{l\mb,\theta}^\rr + h_{ll,\theta\theta}^\rr\Big) \notag \\
    &\quad + a^2\Big(-4\cot[\theta]\xi_\circl^\bst - 4(1+3\cos[2\theta])\xi_{\circl,\KNu}^\bsu + (5-9\cos[2\theta])\xi_{\circl,\theta}^\bst \notag \\
    &\quad + \sin[2\theta]\big(9\xi_\circl^\bst - 4\xi_{\circl,\theta \KNu}^\bsu - 2i\omega(\sqrt{2}(1+P_0)h_{lm}^\ro + \sqrt{2}(1+P_0)h_{l\mb}^\ro - 4h_{ll,\theta}^\ro) \notag \\
    &\quad - 4\xi_{\circl,\KNu\theta}^\bsu\big) + 10\sin^2[\theta]\xi_{\circl,\KNp}^\bsp\Big) + ia\Big(\sqrt{2}(1-2m^2+2m\cos[\theta]+3\cos[2\theta])\csc[\theta]h_{lm}^\ro \notag \\
    &\quad + \sqrt{2}(-1+2m^2+2m\cos[\theta]-3\cos[2\theta])\csc[\theta]h_{l\mb}^\ro + 2\Big((m+2\cos[\theta])h_{mm}^\ro \notag \\
    &\quad- 2mh_{m\mb}^\ro  + (m-2\cos[\theta])h_{\mb\mb}^\ro + 4\sqrt{2}\cos[\theta](h_{lm,\theta}^\ro - h_{l\mb,\theta}^\ro) \notag \\
    &\quad + \sin[\theta]\big(2\sqrt{2}h_{nm}^\ro + 2\sqrt{2}h_{n\mb}^\ro + 8iM\sin[\theta]\xi_\circl^\bsp + h_{mm,\theta}^\ro - h_{\mb\mb,\theta}^\ro + \sqrt{2}(h_{lm,\theta\theta}^\ro - h_{l\mb,\theta\theta}^\ro)\big) \notag \\
    &\quad + 4i\xi_{\circl,\KNp}^\bsu\Big)\Big) \bigg)
    + \O\bigg(\frac{1}{r^2}\bigg), \label{eq:xin}
\end{align}
and
\begin{align}
    \xi_m&= \frac{\big(i \sin[\theta]\xi_\circl^\bsp +\xi_\circl^\bst  \big)r}{\sqrt2} +\frac{\ln[r]}{2\sqrt{2}} \Big( m\csc[\theta]h_{ll}^\ro - h_{ll,\theta}^\ro \Big) +\frac{1}{4} \Big(  \sqrt{2}m\csc[\theta] h_{ll}^\ro + 4h_{lm}^\ro \notag \\
    &\quad + \sqrt{2}\big( a\sin[2\theta]\xi_\circl^\bsp - ia\cos[\theta]\xi_\circl^\bst - h_{ll,\theta}^\ro - 2\xi_{\circl,\theta}^\bsu  + ia\sin[\theta]\big( \xi_{\circl,\theta}^\bst + \xi_{\circl,\KNp}^\bsp \big) - 2i\csc[\theta]\xi_{\circl,\KNp}^\bsu \big) \Big)\notag \\
    &\quad
    +\frac{1}{4\sqrt{2}r}ia\ln[r]\sin[\theta] \Big(  \big(-2 + 2a(1+P_0)\omega\cos[\theta] + m\csc[\theta](-2\cot[\theta] + m\csc[\theta])\big) h_{ll}^\ro \notag \\
    &\quad + \cot[\theta] h_{ll,\theta}^\ro - h_{ll,\theta\theta}^\ro \Big) + \frac{1}{8r}\bigg\{  4h_{lm}^\rr + i\sqrt{2}a\csc[\theta]h_{ll}^\ro\big( m^2 - 2m\cos[\theta] - \sin^2[\theta] \notag \\
    &\quad+ a(1+P_0)\omega\sin[\theta]\sin[2\theta] \big)  + 2\sqrt{2}a^2\big( 2i\sin^3[\theta]\xi_\circl^\bsp - \cos[2\theta]\xi_\circl^\bst - \sin[2\theta]\xi_{\circl,\KNu}^\bsu \big) + \sqrt{2}h_{ll,\theta}^\rr \notag \\
    &\quad - ia\sin[\theta]\Big( 2\sqrt{2}h_{ln}^\ro - 2\sqrt{2}h_{mm}^\ro - 4h_{lm,\theta}^\ro  + \sqrt{2}\big( 4\xi_\circl^\bsu + h_{ll,\theta\theta}^\ro + 2\xi_{\circl,\theta\theta}^\bsu \big) \notag \\
    &\quad - \sqrt{2}\cot[\theta]\big( h_{ll,\theta}^\ro + 2\xi_{\circl,\theta}^\bsu + 4i\csc[\theta]\xi_{\circl,\KNp}^\bsu \big) \Big)  - \sqrt{2}\csc[\theta]\big( m h_{ll}^\rr + 2ia\xi_{\circl,\KNp\KNp}^\bsu \big) \bigg\}\notag \\
    &\quad + \O\bigg(\frac{1}{r^2}\bigg), \label{eq:xim}
\end{align}%
\endgroup%
where we have dropped the ${\rm KN}\omega m$ labels on $h_{\mu\nu}^{\{c\}{\rm KN}\omega m}$, absorbed the $e^{-i(\omega \KNu -\m \KNp)}$ dependence, and omitted the integral over $\omega$ and sum over $m$ for brevity. Equations~\eqref{eq:xil}, \eqref{eq:xin}, and \eqref{eq:xim} are available in the accompanying Mathematica notebook~\cite{BS-Mathematica-notebook}.

\subsection{Expressing the BMS gauge vector in Boyer--Lindquist coordinates}

In the previous section, we left the tetrad components of $\xi^\alpha$ in terms of the BMS generator $\xi_\circl^\alpha[\bsu,\bst^A]$, which is a function of BS coordinates. To complete the work of this appendix, we re-express the BMS generator in terms of KN coordinates. 

To consistently neglect only $\O(\frac{1}{r^2})$ terms in \cref{eq:xil,eq:xin,eq:xim}, the coefficients $\xi_\circl^\bst[\bst^A]$ and $\xi_\circl^\bst[\bst^A]$ need to be calculated up to and including $\O(1/r^2)$. Therefore, we need to include up to and including the $\O(1/r^2)$ parts of the transformation of $\bst\rightarrow\theta$ and $\bsp\rightarrow\KNp$ (see \cref{eq:KNcoords} and \cref{eq:BStoBLtransform}); therefore, 
\begin{align}
    \delta\theta^A[r,\theta]&=\hat\theta^A - \theta^A_{\rm KN} = (\delta\theta,\delta\phi_*),\notag \\
    \delta\theta[r,\theta]&=\frac{a^2\sin^2\theta}{2r^2} +\mathcal{O}(r^{-3}),\notag \\
    \delta\phi_*[r,\theta]&=-\frac{a}{r}+\mathcal{O}(r^{-3}).
\end{align}
We also need to include up to and including the $\O(1/r)$ parts of the transformation $\bsu\rightarrow\KNu$, as $\xi_\circl^\bsu[u,\bst^A]$ needs to be calculated up to and including $\O(1/r)$. From \cref{eq:uu}, we define 
\begin{align}
    \bsu=\KNu+\Delta u+\delta u+\O(1/r^2),
\end{align}
where the transformation of retarded time includes a non-zero constant shift,
\begin{align}
    \Delta u=2M\ln\left[ \frac{2M}{M+\sqrt{M^2-a^2}} \right] ,
\end{align}
and a small radial dependent shift,
\begin{align}
    \delta u=-\frac{a^2\sin^2[\theta]}{2r}.
\end{align}

We begin with the expressions for $\xi_\circl^\alpha$, \cref{eq:xiuBMS,eq:xiABMS}, in BS coordinates. The coefficients $ \MB[{(1)}]_{\!\!\!\!(1,\m)}$, $ \hat{N}^{{\rm e}(1,\m)}_{(1)}$, $ \hat{N}^{(1)}_{{\rm o}(1, \m)}$, and $\Phi^{(1)}_{{\rm e}(l,m)}[u_0]$ in \cref{eq:xiuBMS} are determined from radial expansions of the metric perturbation in Kerr BS coordinates, \cref{eq:BStoBLtransform}. Here we find expressions for these quantities in terms of the metric perturbation in KN coordinates. The coefficients $ \MB[{(1)}]_{\!\!\!\!(1,\m)}$, $ \hat{N}^{{\rm e}(1,\m)}_{(1)}$, $ \hat{N}^{(1)}_{{\rm o}(1, \m)}$, and $\Phi^{(1)}_{{\rm e}(l,m)}[u_0]$ are determined from the components $\hat{h}^{\rm BS}_{uu}$, $\hat{h}^{\rm BS}_{uA}$, and $\hat{h}^{\rm BS}_{AB}$ of the metric perturbation in the perturbative BS gauge, expressed in BS coordinates. Applying the KN to BS coordinate transformation using \cref{eq:BStoBLtransform,eq:KNcoords}, gives, 
\begin{align}
\hat{h}^{\rm BS}_{uu}&=\hat{h}^{\rm KN}_{\KNu\KNu}+\order{r^{{\rm M}-3}}, \notag \\
\hat{h}^{\rm BS}_{uA}&=\hat{h}^{\rm KN}_{\KNu A}+\order{r^{{\rm M}-3}}, \notag \\
\hat{h}^{\rm BS}_{AB}&=\hat{h}^{\rm KN}_{AB}+\order{r^{{\rm M}-2}},\label{eq:hcomponenttransform}
\end{align}
where $\hat{h}^{\rm KN}_{\mu\nu}$ is the metric perturbation in the perturbative BS gauge, expressed in KN coordinates. 

Assuming the asymptotic radial expansion of the metric perturbation in KN coordinates is given by
\begin{align}
\hat{h}^{\rm KN}_{\KNu\KNu}&=\frac{M^{(1)}_{\rm KN}}{r}+\order{r^{-2}}, \label{eq:MKN} \\
\hat{h}^{\rm KN}_{\KNu A}&=U^{(1)\rm KN}_A-\frac{2N^{(1)\rm KN}_A}{3r}+\order{r^{-2}}, \label{eq:NKN} \\
    \hat{h}^{\rm KN}_{AB}&=rC^{(1)\rm KN}_{AB}+\order{r^{0}}.\label{eq:CKN}
\end{align}
Equations~\eqref{eq:MKN} and \eqref{eq:CKN} immediately provide
\begin{align}
    \MB^{(1)}&=M^{(1)}_{\rm KN}[\theta^A_{\rm KN}] , \label{eq:MBMKN}\\
    C^{(1)}_{AB}&=C^{(1)\rm KN}_{AB}[\KNu+\Delta u,\theta^A_{\rm KN}].\label{eq:CCKN}
\end{align}
Equation~\eqref{eq:MKN} must be treated with more subtlety, as we are interested in the second-order in $\frac1r$ coefficient. Equating the BS radial expansion to the KN radial expansion gives
\begin{align}
    U^{(1)}_A&=U^{(1)\rm KN}_A[\KNu+\Delta u,\theta^A_{\rm KN}]=-\frac{D^BC_{AB}[\KNu+\Delta u,\theta^A_{\rm KN}]}{2}, \notag \\
     N^{(1)}_A&= N^{(1)\rm KN}_A[\KNu+\Delta u,\theta^A_{\rm KN}]+\frac{3}{2}\big(\delta\KNp\partial_\KNp+\delta u\partial_\KNu\big)U_A[\KNu+\Delta u,\theta^A_{\rm KN}]\notag\\
     \quad &= N^{(1)\rm KN}_A[\KNu+\Delta u,\theta^A_{\rm KN}] +\frac{3a}{8r} \Big( 2\partial_\phi +a\sin^2[\theta]\partial_\KNu \Big)D^CC_{AC}[\KNu+\Delta u,\theta^A_{\rm KN}],
\end{align}
where we have used \cref{eq:pertBSfalloffs}.
We can then use \cref{eq:Nhat}, giving
\begin{align}
     \hat N^{(1)}_A&= N^{(1)\rm KN}_A[\KNu+\Delta u,\theta^A_{\rm KN}] -(\KNu+\Delta u)D_A M^{(1)}_{\rm KN}[\theta^A_{\rm KN}]  \notag \\
     &\quad +\frac{3a}{8r} \Big( 2\partial_\phi +a\sin^2[\theta]\partial_\KNu \Big)D^CC_{AC}[\KNu+\Delta u,\theta^A_{\rm KN}] \label{eq:NhatKN}.
\end{align}
Note that \cref{eq:MBMKN,eq:NhatKN} are time independent due to \cref{constant Poincare}. As $\bst^A|_{\mathcal{I}^+}=\theta^A_{\rm KN}|_{\mathcal{I}^+}$, we can decompose \cref{eq:MBMKN,eq:NhatKN,eq:CCKN} in terms of $\theta^A_{\rm KN}$ scalar, vector, and tensor harmonics, respectively, to find the coefficients $\MB[(1)]_{\!\!\!\!(1,\m)}$, $\hat{N}^{(1)}_{{\rm o}(1,\pm 1)} $, $N^{{\rm e}(1,\m )}_{(1)}$, and $\Phi^{(1)}_{{\rm e}(l,m)}[u_0]$ (using a consistent $u_0$ whether in BS or KN coordinates considering $u|_{\mathcal{I}^+}=u_{\rm KN}|_{\mathcal{I}^+}+\Delta u$). These coefficients provide the inputs for \cref{eq:boost-transform,eq:BMSRotationconst,eq:translation-transform,eq:supertranslation-transform} for $\chi^{(1)}_{(1,\m)}$, $\kappa^{(1)}_{(1,\pm 1)}$, $\alpha^{(1)}_{(1,\m)}$, and $\alpha^{(1)}_{(l,m)}$, respectively.


Next, we need to convert $\hat{\theta}^A$ in \cref{eq:xiuBMS,eq:xiABMS} to $\theta^A_{\rm KN}=\{\theta,\phi_*\}$ using \cref{eq:BStoBLtransform,eq:KNcoords}, introducing a radial dependence in KN coordinates. The conversion involves a radial expansion; hence, a dependence on $r$ is introduced:
\begin{align}
    Y_{lm}[\hat\theta^A]&=Y_{lm}[\theta^A_{\rm KN} + \delta\theta^A_{\rm KN}]\notag \\
    &=Y_{lm}[\theta^A_{\rm KN}]+\delta\theta^B[r,\theta^A_{\rm KN}]Y^{lm}_B[\theta^A_{\rm KN}]+\order{r^{-3}}.\label{eq:DY}
\end{align}
We define
\begin{align}
    \tilde Y_{lm}[r, \theta^A_{\rm KN}]&:=Y_{lm}[\theta^A_{\rm KN}]+\delta\theta^B[r,\theta]Y^{lm}_B[\theta^A_{\rm KN}].\label{eq:Ytilde}
\end{align}

Additionally, we will require the transformation of the $\l=1$ even and odd vector harmonics,
\begin{align}
    Z^{A_{\rm BS}}_{lm} &= Z^{A}_{lm} +{\cal L}_{\delta\vec\theta}Z^{A}_{lm} + \order{r^{-3}},\\
    X^{A_{\rm BS}}_{lm} &= X^{A}_{lm} +{\cal L}_{\delta\vec\theta}X^{A}_{lm}+\order{r^{-3}},\label{eq:XA expansion}
\end{align}
and we define
\begin{align}
    \tilde Z^{A_{\rm BS}}_{lm}[r, \theta^A_{\rm KN}] &= Z^{A}_{lm}[\theta^A_{\rm KN}] +{\cal L}_{\delta\vec\theta[r, \theta^A_{\rm KN}]}Z^{A}_{lm}[\theta^A_{\rm KN}] ,\\
    \tilde X^{A_{\rm BS}}_{lm}[r, \theta^A_{\rm KN}] &= X^{A}_{lm}[\theta^A_{\rm KN}] +{\cal L}_{\delta\vec\theta[r, \theta^A_{\rm KN}]}X^{A}_{lm}[\theta^A_{\rm KN}].\label{eq:XA expansion def}
\end{align}

Hence, Eqs.~\eqref{eq:xiuBMS} and \eqref{eq:xiABMS} can then be written in Kerr--Newman coordinates as
\begin{align}
    \xi^\bsu_\circl[u_0,\KNu,r,\theta,\KNp]
    &= \alpha^{(1)}_{(0,0)}Y_{00} + \frac{1}{3M}\sum_{m=-1}^{+1}\Big\{-\hat{N}^{{\rm e}(1,\m)}_{(1)} \notag\\
    &\quad+ \big(a\m i  -  (u_{\rm KN}+\Delta u+\delta u[r])\big)\MB[(1)]_{\!\!\!\!(1,\m)}\Big\}\tilde Y_{1,\m}[r,\theta^A_{\rm KN}]\notag\\
    &\quad +\frac{1}{2}\sum_{l\geq2}^\infty\Phi^{(1)}_{{\rm e}(l,m)}[u_0]\tilde Y_{\ell m}[r,\theta^A_{\rm KN}]+\order{r^{-3}},\label{eq:xiuBMSKN}
\end{align}
and 
\begin{align}
    \xi^A_\circl [u_0,r,\theta^B_{\rm KN}]
    &= \kappa^{(1)}_{(1,0)} \tilde X^A_{1,0}[r,\theta^B_{\rm KN}]+ \frac{1}{3M}\sum_{m=-1}^1\Big\{ \MB[(1)]_{\!\!\!\!(1,\m)}\tilde Z^A_{1,\m}[r,\theta^B_{\rm KN}] \notag\\
    &\quad+ \frac{i \m}{a}\hat{N}^{(1)}_{{\rm o}(1, \m)} \tilde X^A_{1,\m}[r,\theta^B_{\rm KN}]\Big\}+\order{r^{-2}}.
    \label{eq:xiABMSKN}
\end{align}

\bibliographystyle{JHEP}
\bibliography{bib.bib}

\end{document}